\documentclass[11pt]{article}
\usepackage{eurosym}
\usepackage{amssymb,amsmath,amsfonts,amsthm,enumerate}
\usepackage{setspace, natbib,graphicx,color,caption,subcaption,float,relsize}
\usepackage{natbib,url}
\usepackage{hyperref}
\usepackage[nottoc,notlof,notlot]{tocbibind}
\usepackage{sectsty}

\hypersetup{colorlinks = true, urlcolor = blue, linkcolor = blue, citecolor = magenta }
\def\maketag@@@#1{\hbox{\m@th\normalfont\normalsize#1}}
\makeatother
\renewenvironment{abstract}
 {\small
  \begin{center}
  \bfseries \abstractname\vspace{-.0em}\vspace{0pt}
  \end{center}
  \list{}{    \setlength{\leftmargin}{0mm}
    \setlength{\rightmargin}{\leftmargin}  }  \item\relax}
 {\endlist}
\makeatletter
\def\maketag@@@#1{\hbox{\m@th\normalfont\normalsize#1}}
\makeatother

\DeclareMathOperator{\argmax}{argmax}
\DeclareMathOperator{\argmin}{argmin}
\allowdisplaybreaks
\newtheorem {theorem}{Theorem}[section]
\newtheorem {assumption}{Assumption}

\newtheorem{example}[theorem]{Example}

\newtheorem{lemma}[theorem]{Lemma}

\newtheorem{remark}{Remark}

\begin{document}

\title{A Smoothed P-Value Test When There is a Nuisance Parameter under the
Alternative}
\author{Jonathan B.~Hill\thanks{
Dept. of Economics, University of North Carolina, Chapel Hill;
jbhill@email.unc.edu; https://jbhill.web.unc.edu. \ This article benefited
from expert commentary from two referees.} \\
%EndAName
University of North Carolina -- Chapel Hill}
\date{{\normalsize \today}}
\maketitle

\begin{abstract}
\setstretch{1}We present a new test when there is a nuisance parameter $%
\lambda $ under the alternative hypothesis. The test exploits the p-value
occupation time [PVOT], the measure of the subset of $\lambda $ on which a
p-value test based on a test statistic $\mathcal{T}_{n}(\lambda )$ rejects
the null hypothesis. Key contributions are: (\textit{i}) An asymptotic
critical value upper bound for our test is the significance level $\alpha $,
making inference easy. (\textit{ii}) We only require $\mathcal{T}%
_{n}(\lambda )$ to have a known or bootstrappable limit distribution, hence
we do not require $\sqrt{n}$-Gaussian asymptotics, allowing for weak or
non-identification, boundary values, heavy tails, infill asymptotics, and so
on. (\textit{iii}) A test based on the transformed p-value $\sup_{\lambda
\in \Lambda }p_{n}(\lambda )$\ may be conservative and in some cases have
trivial power, while the PVOT naturally controls for this by smoothing over
the nuisance parameter space. Finally, (iv) the PVOT uniquely allows for
bootstrap inference in the presence of nuisance parameters when some
estimated parameters may not be identified.\bigskip \newline
\end{abstract}

\setstretch{1.4}

\section{Introduction\label{sec:intro}}

This paper develops a test for cases when a nuisance parameter $\lambda $ $%
\in $ $\mathbb{R}^{k}$\ is present under the alternative hypothesis $H_{1}$,
where $k$ $\geq $ $1$ is finite. Let $\mathcal{S}_{n}$ $\equiv $ $%
\{z_{t}\}_{t=1}^{n}$ be the observed sample of random variables $z_{t}$ $\in
$ $\mathbb{R}^{q},$ $q$ $\geq $ $1$, with sample size $n$ $\geq $ $1$ and
joint distribution $P$ $\subset $ $\mathcal{P}$ from some collection of
distributions $\mathcal{P}$. We want to test the hypothesis $H_{0}$ $:$ $P$ $%
\in $ $\mathcal{P}_{0}$ against $H_{1}$ $:$ $P$ $\notin $ $\mathcal{P}_{0}$
for some subset $\mathcal{P}_{0}$ $\subset $ $\mathcal{P}$. Let $\mathcal{T}%
_{n}(\lambda )$ $\equiv $ $\mathcal{T}(\mathcal{S}_{n},\lambda )$ be a test
statistic function of $\lambda $ for testing a $H_{0}$. We assume $\mathcal{T%
}_{n}(\lambda )$ $\geq $ $0$, and that large values are indicative of $H_{1}$%
.\ We present a simple smoothed p-value test based on the Lebesgue measure
of the subset of $\lambda ^{\prime }s$ on which we reject $H_{0}$ based on $%
\mathcal{T}_{n}(\lambda )$, defined as the \textit{P-Value Occupation Time}
[PVOT]. In order to focus ideas, we ignore cases where $\lambda $ may be a
set, interval, or function, or infinite dimensional as in nonparametric
estimation problems.

The PVOT was originally explored in \cite{HillAguilar13} and \cite%
{Hill_white_2012} as a way to gain inference in the presence of a trimming
tuning parameter. We extend the idea to test problems where $\lambda $ is a
nuisance parameter under $H_{1}$, and provide a complete asymptotic theory
for the first time.

Nuisance parameters under $H_{1}$ arise in two over-lapping cases. First, $%
\lambda $ is part of the data generating process under $H_{1}$, e.g. ARMA
models with common roots \citep{AndrewsCheng2012}; tests of no GARCH effects
\citep{Engle_etal87,
Andrews2001}; tests for common factors \citep{AndrewsPloberger1994}; tests
for a Box-Cox transformation \citep{AT_Gallant1983}; and structural change
tests \citep{Andrews1993}, to name a few. A standard example is the
regression of scalar $y_{t}$ $=$ $\beta _{0}^{\prime }x_{t}$ $+$ $\gamma
_{0}h(\lambda ,x_{t})$ $+$ $\epsilon _{t}$ where $x_{t}$ are covariates, $h$
is a known function, and $E[\epsilon _{t}|x_{t}]$ $=$ $0$ $a.s.$ for unique $%
(\beta _{0},\gamma _{0})$. If $H_{0}$ $:$ $\gamma _{0}$ $=$ $0$ is true then
$\lambda $ is not identified. In this case $z_{t}$ $=$ $[x_{t}^{\prime
},y_{t}]^{\prime }$ and the null distribution subset $\mathcal{P}_{0}$
contains all joint distributions of $\{x_{t},y_{t}\}_{t=1}^{n}$ such that $%
E[y_{t}|x_{t}]$ $=$ $\beta _{0}^{\prime }x_{t}$ $a.s.$, and under $H_{1}$
the joint distribution $P$ depends on $\lambda $. This test class includes
the Box-Cox transform, neural networks, flexible functional forms, and
regime switching models. See, e.g., \citet{Gallant1981,Gallant1984}, \cite%
{Gallant_Golub1984}, \cite{White1989}, \cite{AndrewsPloberger1994}, \cite%
{Terasvirta1994}, \cite{Hansen1996}, \cite{LiLi2011}, \cite{AndrewsCheng2012}
and \cite{Goracci_etal2021}.

Second, $\lambda $ is used to compute an estimator, or perform a general
model specification test, and therefore need not appear in the joint
distribution $P$ under either hypothesis. This includes tests of omitted
nonlinearity against general alternatives
\citep[see][amongst many
others]{White1989,Bierens1990,BierensPloberger1997,StinchWhite1998,Hill_white_2012}%
; and tests of marginal effects in models with mixed frequency data where $%
\lambda $ is used to reduce regressor dimensionality %
\citep{GhyselsHillMotegi2016}. An example is the regression $y_{t}$ $=$ $%
\beta _{0}^{\prime }x_{t}$ $+$ $\epsilon _{t}$ where we want to test $H_{0}$
$:$ $E[\epsilon _{t}|x_{t}]$ $=$ $0$ $a.s.$ We again have $z_{t}$ $=$ $%
[x_{t}^{\prime },y_{t}]^{\prime }$ and $\mathcal{P}_{0}$ $=$ $\{P$ $:$ $%
E[y_{t}|x_{t}]$ $=$ $\beta _{0}^{\prime }x_{t}$ $a.s.\}$. This is
fundamentally different from the preceding example where $E[\epsilon
_{t}|x_{t}]$ $=$ $0$ $a.s.$ was \textit{assumed}. A parametric test
statistic can be based on the fact that $E[\epsilon _{t}F(\lambda ^{\prime
}x_{t})]$ $\neq $ $0$ \textit{if and only if} $E[\epsilon _{t}|x_{t}]$ $=$ $%
0 $ $a.s.$ is false, for all $\lambda $ in any compact set $\Lambda $
outside of a measure zero subset, provided $F$ $:$ $\mathbb{R}$ $\rightarrow
$ $\mathbb{R}$ is exponential \citep{Bierens1990}, logistic \citep{White1989}%
, or any real analytic non-polynomial \citep{StinchWhite1998}, or
multinomials of $x_{t}$ \citep{Bierens1982}. Notice that $\lambda $ need not
be part of the data generating process since $E[y_{t}|x_{t}]$ $=$ $\beta
_{0}^{\prime }x_{t}$ $+$ $\gamma _{0}F(\lambda ^{\prime }x_{t})$ $a.s.$ may
not be true under $H_{1}$. Detailed examples involving a test of function
form where weak identification is possible, and a test of no GARCH effects,
are presented in Sections \ref{sec:ex_start}, \ref{ex:omitted_nl} and \ref%
{sec:examples}.

A classic approach for handling nuisance parameters in the broad sense is to
compute a p-value $p_{n}(\lambda )$ $\equiv $ $p(\mathcal{S}_{n},\lambda )$.
and use $\sup_{\lambda \in \Lambda }p_{n}(\lambda )$ for some compact subset
$\Lambda $ of $\mathbb{R}^{k}$\citep[see][Chap. 3.1]{Lehmann1994}. This may
lead to a conservative test, although it promotes a test with the correct
asymptotic level.\footnote{%
Let $\tau _{n}$ $\in $ $[0,1]$ be a test statistic, and suppose we reject a
null hypothesis at nominal significance level $\alpha $ when $\tau _{n}$ $>$
$\alpha $. Recall that the asymptotic \textit{level} of the test is $\alpha $
if $\lim_{n\rightarrow \infty }P(\tau _{n}$ $>$ $\alpha |H_{0})$ $\leq $ $%
\alpha $, and if $\lim_{n\rightarrow \infty }P(\tau _{n}$ $>$ $\alpha
|H_{0}) $ $=$ $\alpha $ then $\alpha $ is the asymptotic size %
\citep[cf.][]{Lehmann1994}.} Further, $\sup_{\lambda \in \Lambda
}p_{n}(\lambda )$ may not promote a consistent test even when $\mathcal{T}%
_{n}(\lambda )$ and its transforms like $\sup_{\lambda \in \Lambda }\mathcal{%
T}_{n}(\lambda )$ do. An example is a \cite{Bierens1990}-type test which is
known to be consistent $\forall \lambda $ $\in $ $\Lambda /S$ where $S$ has
measure zero. This means $\sup_{\lambda \in \Lambda }p_{n}(\lambda )$ $%
\overset{p}{\rightarrow }$ $(0,1)$ under $H_{1}$\ is possible despite $%
p_{n}(\lambda )$ $\overset{p}{\rightarrow }$ $0$ $\forall \lambda $ $\in $ $%
\Lambda /S$. We find the test where $H_{0}$ is rejected at nominal level $%
\alpha $ when $\sup_{\lambda \in \Lambda }p_{n}(\lambda )$ $<$ $\alpha $
leads to profound size distortions and trivial power for a test of no GARCH
effects, and is relatively conservative as a test of omitted nonlinearity.
In the case where $\lambda $ is identified under either hypothesis, \cite%
{Silvapule1996} proposes an improvement with better size and power
properties.

The challenge of constructing valid tests in the presence of nuisance
parameters under $H_{1}$ dates at least to \cite{ChernoffZacks1964} and %
\citet{Davies77,Davies87}. Recent contributions include \cite{Nyblom1989},
\cite{Andrews1993}, \cite{Dufour1997}, %
\citet{AndrewsPloberger1994,AndrewsPloberger1995}, \cite{Hansen1996}, and %
\citet{AndrewsCheng2012,AndrewsCheng2013,AndrewsCheng2014} to name but a
few. Nuisance parameters that are not identified under $H_{1}$ are either
chosen at random, thereby sacrificing power \citep[e.g.][]{White1989}; or $%
\mathcal{T}_{n}(\lambda )$ is smoothed over $\Lambda $, resulting in a
non-standard limit distribution and in general the necessity of a bootstrap
step \citep[e.g.][]{ChernoffZacks1964,Davies77,AndrewsPloberger1994}.
Examples are the average $\int_{\Lambda }\mathcal{T}_{n}(\lambda )\mu
(d\lambda )$ and supremum $\sup_{\lambda \in \Lambda }\mathcal{T}%
_{n}(\lambda )$, where $\mu (\lambda )$ is an absolutely continuous
probability measure \citep{ChernoffZacks1964,Davies77,AndrewsPloberger1994}.
The non-standard limit distribution, moreover, cannot be bootstrapped using
conventional methods when some parameters may be weakly or non-identified.
See \cite{Hill2021_weak}, and see below for discussion. Further, even if
bootstrapping is valid, it adds significant computation time due to the many
repeated generated bootstrap samples.

Let $p_{n}(\lambda )$ be a p-value or asymptotic p-value based on $\mathcal{T%
}_{n}(\lambda )$: $p_{n}(\lambda )$ may be based on a known limit
distribution, or if the limit distribution is non-standard then a bootstrap
or simulation method is assumed available for computing an asymptotically
valid approximation to $p_{n}(\lambda )$. Assume that $p_{n}(\lambda )$
leads to an asymptotically correctly sized test, uniformly on compact $%
\Lambda $ $\subset $ $\mathbb{R}^{k}$:%
\begin{equation}
\sup_{\lambda \in \Lambda }\left\vert P\left( p_{n}(\lambda )<\alpha
|H_{0}\right) -\alpha \right\vert \rightarrow 0\text{ for any }\alpha \in
\left( 0,1\right) .  \label{pn}
\end{equation}%
If $p_{n}(\lambda )$ is uniformly distributed then $\alpha $ is the size of
the test, else by (\ref{pn}) $\alpha $ is the asymptotic size. The terms
"asymptotic p-value" and "asymptotic size" are correct when convergence in (%
\ref{pn}) is uniform over the null distribution subset $\mathcal{P}_{0}$.
The latter is not possible here because for generality we do not specify a
model or $H_{0}$. If $p_{n}(\lambda )$ is asymptotically free of any other
nuisance parameters then uniform convergence over $H_{0}$ is immediate given
that (\ref{pn}) is uniform over $\Lambda $ \citep[e.g.][p.
417]{Hansen1996}. Since this problem is common, we will not focus on it, and
will simply call $p_{n}(\lambda )$ a "p-value" for brevity.

The p-value [PV] test with nominal level $\alpha $ for a chosen value of $%
\lambda $ is (\ref{pn}):%
\begin{equation}
\text{\textbf{PV Test:} reject }H_{0}\text{ if }p_{n}(\lambda )<\alpha \text{%
, otherwise fail to reject }H_{0}.  \label{T_test}
\end{equation}%
Now assume $\Lambda $ has unit Lebesgue measure $\int_{\Lambda }d\lambda $ $%
= $ $1$, and compute the \textit{p-value occupation time} [PVOT] of $%
p_{n}(\lambda )$ below the nominal level $\alpha $ $\in $ $(0,1)$:%
\begin{equation}
\text{\textbf{PVOT:} }\mathcal{P}_{n}^{\ast }(\alpha )\equiv \int_{\Lambda
}I\left( p_{n}(\lambda )<\alpha \right) d\lambda ,  \label{PVOT}
\end{equation}%
where $I(\cdot )$ is the indicator function. If $\int_{\Lambda }d\lambda $ $%
\neq $ $1$ then we use $\mathcal{P}_{n}^{\ast }(\alpha )$ $\equiv $ $%
\int_{\Lambda }I(p_{n}(\lambda )$ $<$ $\alpha )d\lambda /\int_{\Lambda
}d\lambda $. $\mathcal{P}_{n}^{\ast }(\alpha )$ is just the Lebesgue measure
of the subset of $\lambda ^{\prime }s$ on which we reject $H_{0}$. Thus, a
large occupation time in the rejection region asymptotically indicates $%
H_{0} $ is false.

As long as $\{\mathcal{T}_{n}(\lambda )$ $:$ $\lambda $ $\in $ $\Lambda \}$
converges weakly under $H_{0}$ to a stochastic process $\{\mathcal{T}%
(\lambda )$ $:$ $\lambda $ $\in $ $\Lambda \}$ on a space endowed with,
e.g., the uniform metric (sup-norm), and $\mathcal{T}(\lambda )$ has a
continuous distribution for all $\lambda $ outside a set of measure zero,
then asymptotically $\mathcal{P}_{n}^{\ast }(\alpha )$ has a mean $\alpha $\
and the probability that $\mathcal{P}_{n}^{\ast }(\alpha )$ $>$ $\alpha $ is
not greater than $\alpha $. Evidence against $H_{0}$ is therefore simply $%
\mathcal{P}_{n}^{\ast }(\alpha )$ $>$ $\alpha $. Further, if asymptotically
with probability approaching one the PV test (\ref{T_test}) rejects $H_{0}$
for each $\lambda $ in a subset of $\Lambda $ that has Lebesgue measure
greater than $\alpha $, then $\mathcal{P}_{n}^{\ast }(\alpha )$ $>$ $\alpha $
asymptotically with probability one. The PVOT test at the chosen level $%
\alpha $ is then:%
\begin{equation}
\text{\textbf{PVOT Test}: reject }H_{0}\text{ if }\mathcal{P}_{n}^{\ast
}(\alpha )>\alpha \text{, otherwise fail to reject }H_{0}.  \label{PVOT_test}
\end{equation}%
These results are formally derived in Section \ref{sec:asym_theory}. Thus,
an asymptotic level $\alpha $ critical value is simply $\alpha $, a useful
simplification over transforms with non-standard asymptotic distributions,
like $\int_{\Lambda }\mathcal{T}_{n}(\lambda )\mu (d\lambda )$ and $%
\sup_{\lambda \in \Lambda }\mathcal{T}_{n}(\lambda )$. A simulation study in
Section \ref{sec:sim} suggests the critical value $\alpha $ leads to an
asymptotically correctly sized test for tests of omitted nonlinearity and
GARCH effects, and strong power in each case. We may therefore expect that
similar tests have this property.

The PVOT yields several useful innovations. First, when $\mathcal{T}%
_{n}(\lambda )$ is derived from a regression model in which some parameters
may be weakly or non-identified, there is no known valid standard bootstrap
or simulation approach for approximating the limit distribution of smoothed
test statistics in the class considered in \cite{AndrewsPloberger1994},
including $\int_{\Lambda }\mathcal{T}_{n}(\lambda )\mu (d\lambda )$ and $%
\sup_{\lambda \in \Lambda }\mathcal{T}_{n}(\lambda )$. This is because a
valid bootstrap, for example, must approximate the covariance structure of
the limit process $\{\mathcal{T}(\lambda )$ $:$ $\lambda $ $\in $ $\Lambda
\} $ which generally requires consistent estimates of model parameters. If
some parameters are weakly or non-identified, then they cannot be
consistently estimated \citep[see, e.g.,][]{Gallant1977,AndrewsCheng2012}.
\cite{Hill2021_weak} presents an asymptotically valid bootstrap method for
the non-smoothed $\mathcal{T}_{n}(\lambda )$ for any degree of
(non)identification. The resulting bootstrapped p-value leads to a valid
smoothed p-value test, even though smoothed test statistics \textit{cannot}
be consistently bootstrapped. See Example \ref{sec:ex_start} in Section \ref%
{sec:ex_start} below.

Second, since the PVOT critical value upper bound is simply $\alpha $ under
any asymptotic theory for $\mathcal{T}_{n}(\lambda )$, we only require $%
\mathcal{T}_{n}(\lambda )$ to have a known or bootstrappable limit
distribution. Thus, $\sqrt{n}$-Gaussian asymptotics is not required as is
nearly always assumed %
\citep[e.g.][]{AndrewsPloberger1994,Hansen1996,AndrewsCheng2012}.
Non-standard asymptotics are therefore allowed, including inference
concerning parameters that lie on the parameter space boundary %
\citep{Andrews2001,CavaliereNielsenRahbek2017}, test statistics when a
parameter is weakly identified, GARCH tests \citep[e.g.][]{Andrews2001};
inference under heavy tails; and non-$\sqrt{n}$ asymptotics are covered, as
in heavy tail robust tests %
\citep[e.g.][]{Hill_white_2012,HillAguilar13,AguilarHill15}, or when infill
asymptotics or nonparametric estimators are involved %
\citep[e.g.][]{BandiPhillips2007}, or in high dimensional settings when a
regularized estimator is used.

Third, the local power properties of specific PVOT tests appear to be on par
with the power optimal exponential class developed in \cite%
{AndrewsPloberger1994}. We derive general results, and apply them to a test
of omitted nonlinearity. We show in a numerical experiment that the PVOT
test achieves local power on par with the highest achievable (weighted
average) power. In view of the general result, the local power merits of the
PVOT test appear to extend to any consistent test on $\Lambda $, but any
such claim requires a specific test statistic and numerical exercise to
verify.

Fourth, although we focus on the PVOT test, in Appendix B of the
supplemental material \cite{Hill_supp_ot} [SM] we show the PVOT naturally
arises as a measure of test optimality when $\lambda $ is part of the true
data generating process under $H_{1}$. This requires Andrews and Ploberger's %
\citeyear{AndrewsPloberger1994} notion of weighted average local power of a
test based on $\mathcal{T}_{n}(\lambda )$, where the average is computed
over $\lambda $ and a drift parameter \citep[cf.][]{Wald1943}. In this
environment, the PVOT is just a point estimate of the weighted average
probability of PV test rejection evaluated under $H_{0}$. Since that
probability is asymptotically no larger than $\alpha $ when the null is
true, the PVOT test rejects $H_{0}$ when the PVOT is larger than $\alpha $.
Thus, the PVOT is a natural way to transform a test statistic.

Fifth, when $\mathcal{T}_{n}(\lambda )$ has a known distribution limit (e.g.
standard normal, chi-squared) then performing the PVOT test is significantly
computationally faster that bootstrapping a smoothed test statistic (e.g. $%
\sup_{\lambda \in \Lambda }\mathcal{T}_{n}(\lambda )$). Indeed, if $\mathcal{%
M}$ bootstrap samples are required then the PVOT test is trivially $\mathcal{%
M}$-times faster.

The relevant literature also includes \cite{King_Shiv93} whose
re-parameterization leads to a conventional, but not general, test. \cite%
{Hansen1996} presents a wild bootstrap for computing the p-value for a
smoothed LM statistic when $\lambda $ is part of the data generating
process, extending ideas in \cite{Wu1986} and \cite{Liu1988}. The method
implicitly requires strong identification of regression model parameters.
Our simulation study for tests of functional form and GARCH effects show the
PVOT test performs on par with, or is better than, the average and supremum
test. Moreover, when model parameters are weakly or non-identified, a PVOT
test of functional form substantially dominates $p_{n}(\lambda ^{\ast })$
with randomly selected $\lambda ^{\ast }$, $\sup_{\lambda \in \Lambda
}p_{n}(\lambda )$, and bootstrapped $\sup_{\lambda \in \Lambda }\mathcal{T}%
_{n}(\lambda )$ and $\int_{\Lambda }\mathcal{T}_{n}(\lambda )\mu (d\lambda )$%
. Indeed, the latter two fail to be valid for the reasons explained above.

\cite{Bierens1990} creatively compares supremum and pointwise statistics to
achieve standard asymptotics for a functional form test, while \cite%
{BierensPloberger1997} compute a critical value upper bound for their
integrated conditional moment statistic. We show that the latter upper bound
leads to an under-sized test and potentially low power in a local power
numerical exercise and a simulation study presented below.\medskip

The remainder of the paper is as follows. In Section \ref{sec:ex_start} we
introduce examples showcasing uses of the PVOT test: tests of omitted
nonlinearity (with possibly weakly identified parameters), and GARCH
effects. We then present the formal list of assumptions and the main results
for the PVOT test in Section \ref{sec:asym_theory}. Local power is analyzed
in general Section \ref{app:locpow}, and applied to a test of function form
with a numerical exercise. Section \ref{sec:examples} continues the Section %
\ref{sec:ex_start} examples with complete theory verifying the main
assumptions. We perform a simulation study\ in Section \ref{sec:sim}
involving tests of omitted nonlinearity and GARCH effects. Concluding
remarks are left for Section \ref{sec:conclusion}.

\subsection{Examples\label{sec:ex_start}}

We discuss examples showcasing the use of the PVOT test.

\begin{example}[\textbf{Test of Functional Form with Possible Weak
Identification}]
\label{ex_weak}\normalfont This example showcases a unique advantage of the
PVOT test: it allows for robust bootstrap inference when weak identification
is possible and a nuisance parameter is present, \textit{and} it promotes a
consistent test. Conversely, test statistic functionals like the supremum $%
\sup_{\lambda \in \Lambda }\mathcal{T}_{n}(\lambda )$ and average $%
\int_{\Lambda }\mathcal{T}_{n}(\lambda )\mu (d\lambda )$ cannot be validly
bootstrapped asymptotically when weak identification is possible %
\citep[see][]{Hill2021_weak}, and $\sup_{\lambda \in \Lambda }p_{n}(\lambda
) $ with a weak identification robust $p_{n}(\lambda )$ need not be
consistent. The following is based on ideas developed in \cite{Hill2021_weak}%
; consult that source for more details and references.

We work with the following model:%
\begin{equation}
y_{t}=\zeta _{0}^{\prime }x_{t}+\beta _{0}^{\prime }g(x_{t},\pi
_{0})+\epsilon _{t}=f(\theta _{0},x_{t})+\epsilon _{t}\text{ where }x_{t}\in
\mathbb{R}^{k_{x}}\text{ and }\theta _{0}=\left[ \zeta _{0}^{\prime },\beta
_{0}^{\prime },\pi _{0}^{\prime }\right] ^{\prime }\in \Theta \text{,}
\end{equation}%
where $g$ is a known function, and $E[\epsilon _{t}]$ $=$ $0$ and $%
E[\epsilon _{t}^{2}]$ $\in $ $\left( 0,\infty \right) $ for unique $\theta
_{0}$ $\in $ $\Theta $ and compact $\Theta $. We want to test $H_{0}$ $:$ $%
E[y_{t}|x_{t}]$ $=$ $f(\theta _{0},x_{t})$ $a.s$. against $%
H_{1}:\sup_{\theta \in \Theta }P(E[y_{t}|x_{t}]$ $=$ $f(\theta ,x_{t}))$ $<$
$1$. Let $\{x_{t},y_{t}\}_{t=1}^{n}$ have joint distribution $P$ $\in $ $%
\mathcal{P}$, a collection of joint distributions, and let $\mathcal{P}_{0}$
$\subset $ $\mathcal{P}$ be all distributions consistent with $%
E[y_{t}|x_{t}] $ $=$ $f(\theta _{0},x_{t})$ $a.s.$ The null coincides with $P
$ $\in $ $\mathcal{P}_{0}$.

Let $\Psi $ be a $1$-$1$ bounded mapping from $\mathbb{R}^{k}$ to $\mathbb{R}%
^{k}$, let $\mathcal{F}$ $:$ $\mathbb{R}$ $\rightarrow $ $\mathbb{R}$ be
analytic and non-polynomial (e.g. exponential or logistic), and assume $%
\lambda $ $\in $ $\Lambda $, a compact subset of $\mathbb{R}^{k}$.
Mis-specification \linebreak $\sup_{\theta \in \Theta }P(E[y_{t}|x_{t}]$ $=$
$f(\theta ,x_{t}))$ $<$ $1$ implies $E[e_{t}\mathcal{F}(\lambda ^{\prime
}\Psi (x_{t}))]$ $\neq $ $0$ $\forall \lambda $ $\in $ $\Lambda /\mathcal{S}$%
, where $\mathcal{S}$ has Lebesgue measure zero. See \cite{White1989}, \cite%
{Bierens1990} and \cite{StinchWhite1998} for seminal results for iid data,
and see \cite{deJong1996} and \cite{Hill2008} for dependent data. Thus a
LM-type statistic based on a sample version of $E[e_{t}\mathcal{F}(\lambda
^{\prime }\Psi (x_{t}))]$ can be used to test $H_{0}.$

If $\beta _{0}$ $=$ $0$ then $\pi _{0}$ is not identified. Or if there is
local drift $\beta _{0}$ $=$ $\beta _{n}$ $\rightarrow $ $0$ with $\sqrt{n}%
||\beta _{n}||$ $\rightarrow $ $[0,\infty )$, then estimators of $\pi _{0}$
have random probability limits, and estimators for $\theta _{0}$ have
nonstandard limit distributions \citep{AndrewsCheng2012}. In either case we
say $\pi _{0}$ is \emph{weakly identified}. The literature on consistent
specification testing generally assumes strong identification %
\citep[e.g.][]{Bierens1982,White1989,Bierens1990,HongWhite1995,deJong1996,BierensPloberger1997,Hill2008}%
, while the weak identification literature presumes model correctness $%
E[y_{t}|x_{t}]$ $=$ $f(\theta _{0},x_{t})$ $a.s.$ %
\citep[e.g.][]{AndrewsCheng2012,AndrewsCheng2013,AndrewsCheng2014}. \cite%
{Hill2021_weak} allows for both weak identification \textit{and} model
mis-specification. There a modified Conditional Moment [CM] test statistic
and bootstrap procedure, both to account for possible weak identification.
Let $\hat{\theta}_{n}$ be the nonlinear least squares estimator of $\theta
_{0}$. The CM statistic is:
\begin{equation*}
\mathcal{T}_{n}(\lambda )\equiv \left( \frac{1}{\hat{v}_{n}(\hat{\theta}%
_{n},\lambda )}\frac{1}{\sqrt{n}}\sum_{t=1}^{n}\epsilon _{t}(\hat{\theta}%
_{n})F\left( \lambda ^{\prime }\Psi (x_{t})\right) \right) ^{2}
\end{equation*}%
where $\hat{v}_{n}(\theta ,\lambda )$\ is a scale estimator.

Under strong identification, $\{\mathcal{T}_{n}(\lambda )$ $:$ $\lambda $ $%
\in $ $\Lambda \}$ converges weakly to a chi-squared process. Under weak
identification the limit process is non-standard with nuisance parameter $%
\lambda $, and other nuisance parameters $h$ containing distribution
parameters (e.g. $\pi _{0}$ and $E[\epsilon _{t}^{2}]$). Let $\{\mathcal{T}%
(\lambda )$ $:$ $\lambda $ $\in $ $\Lambda \}$ denote either limit process.

Test statistic transforms like $\sup_{\lambda \in \Lambda }\mathcal{T}%
_{n}(\lambda )$ and $\int_{\Lambda }\mathcal{T}_{n}(\lambda )\mu (d\lambda )$
cannot be consistently bootstrapped or simulated if weak identification is
possible. The reason is a consistent estimate of the covariance kernel for $%
\{\mathcal{T}(\lambda )$ $:$ $\lambda $ $\in $ $\Lambda \}$ is required,
which depends on $\pi _{0}$. The latter cannot be consistently estimated
under weak identification \citep{AndrewsCheng2012}. Invalidity of the
bootstrap is easily demonstrated by simulation: see \cite{Hill2021_weak},
and see Section \ref{sec:sim_weak} below.

\citet[see][Sections 5 and 6]{Hill2021_weak} therefore takes a different
approach by bootstrapping a p-value $p_{n}(\lambda )$ for $\mathcal{T}%
_{n}(\lambda )$ that is consistent for the asymptotic p-value, under any
degree of (non)identification. The key steps involve computing (or
bootstrapping) the asymptotic p-value under strong identification, wild
bootstrapping the p-value under weak identification, and then combining the
two in a way that promotes valid inference asymptotically under any degree
of identification.\footnote{\cite{Hill2021_weak} uses the \textit{least
favorable} and \textit{identification category selection} constructions from
\cite{AndrewsCheng2012} as the basis for p-value combinations. \cite%
{AndrewsCheng2012} use those notions for critical value combinations under
assumed model correctness and without a nuisance parameter under a specific
hypothesis.} Let $\hat{p}_{n,\mathcal{M}}(\lambda )$\ be the resulting
combined wild bootstrapped p-value based on $\mathcal{M}$\ independently
drawn bootstrap samples, hence the PVOT is $\mathcal{\hat{P}}_{n,\mathcal{M}%
}(\alpha )$ $\equiv $ $\int_{\Lambda }I(\hat{p}_{n,\mathcal{M}}(\lambda )$ $%
< $ $\alpha )d\lambda $. The test rejects $H_{0}$ when $\mathcal{\hat{P}}_{n,%
\mathcal{M}}(\alpha )$ $>$ $\alpha $.
\end{example}

\begin{example}[\textbf{Test of GARCH Effects}]
\label{ex_GARCH}\normalfont We want to test the hypothesis that a random
variable $y_{t}$ is not governed by a GARCH process. Consider a stationary
GARCH(1,1) model \citep{Boller86, Nelson90}:%
\begin{eqnarray}
&&y_{t}=\sigma _{t}\epsilon _{t}\text{ where }\epsilon _{t}\text{ is iid, }%
E[\epsilon _{t}]=0\text{, }E[\epsilon _{t}^{2}]=1\text{, and }E\left\vert
\epsilon _{t}\right\vert ^{r}<\infty \text{ for }r>4  \label{garch} \\
&&\sigma _{t}^{2}=\omega _{0}+\delta _{0}y_{t-1}^{2}+\lambda _{0}\sigma
_{t-1}^{2}\text{ where }\omega _{0}>0\text{, }\delta _{0},\lambda _{0}\in
\lbrack 0,1)\text{, and }E\left[ \ln \left( \delta _{0}\epsilon
_{t}^{2}+\lambda _{0}\right) \right] <0.  \notag
\end{eqnarray}%
Under $H_{0}$: $\delta _{0}$ $=$ $0$ if the starting value is $\sigma
_{0}^{2}$ $=$ $\tilde{\omega}$ $=$ $\omega _{0}/(1$ $-$ $\lambda _{0})$ $>$ $%
0$ then $\sigma _{1}^{2}$ $=$ $\omega _{0}$ $+$ $\lambda _{0}\omega _{0}/(1$
$-$ $\lambda _{0})$ $=$ $\tilde{\omega}$\ and so on, hence $\sigma _{t}^{2}$
$=$ $\tilde{\omega}$ $\forall t$ $\geq $ $0$ which means there are no GARCH
effects. In this case the $\sigma _{t-1}^{2}$ marginal effect $\lambda _{0}$
is not identified. Further, $\delta _{0},\lambda _{0}$ $\geq $ $0$ must be
maintained during estimation to ensure a positive conditional variance, and
because this includes a boundary value, QML asymptotics are non-standard %
\citep{Andrews1999,Andrews2001}.

Let $\theta $ $=$ $[\omega ,\delta ,\lambda ]$, and define the parameter
subset $\pi $ $=$ $[\omega ,\delta ]^{\prime }$ $\in $ $\Pi $ $\equiv $ $%
[\iota _{\omega },u_{\omega }]$ $\times $ $[0,1$ $-$ $\iota _{\delta }]$ for
tiny $(\iota _{\omega },\iota _{\delta })$ $>$ $0$ and some $u_{\omega }$ $>$
$0$. Express the volatility process as $\sigma _{t}^{2}(\pi ,\lambda )$ $=$ $%
\omega $ $+$ $\delta y_{t-1}^{2}$ $+$ $\lambda \sigma _{t-1}^{2}(\pi
,\lambda )$ for an imputed $\lambda $ $\in $ $\Lambda $ $\equiv $ $%
[0,1-\iota _{\lambda }]$ and tiny $\iota _{\lambda }$ $>$ $0$, with initial
condition $\sigma _{0}^{2}(\pi ,\lambda )$ $=$ $\omega /(1$ $-$ $\lambda )$.
Denote the unrestricted QML estimator of $\pi _{0}$ for a given $\lambda $ $%
\in $ $\Lambda $: $\hat{\pi}_{n}(\lambda )$ $=$ $[\hat{\omega}_{n}(\lambda ),%
\hat{\delta}_{n}(\lambda )]^{\prime }$ $\equiv $ $\arg \min_{\pi \in \Pi
}1/n\sum_{t=1}^{n}\{\ln (\sigma _{t}^{2}(\pi ,\lambda ))$ $+$ $%
y_{t}^{2}/\sigma _{t}^{2}(\pi ,\lambda )\}$. Andrews' (\citeyear{Andrews2001}%
) test statistic is:%
\begin{equation}
\mathcal{T}_{n}(\lambda )=n\hat{\delta}_{n}^{2}(\lambda ).
\label{T_no_garch}
\end{equation}%
The process $\{\mathcal{T}_{n}(\lambda )$ $:$ $\lambda $ $\in $ $\Lambda \}$
has a well defined limit that can be easily simulated resulting in a
simulation-based p-value approximation $\hat{p}(\lambda )$. The PVOT is
therefore $\int_{\Lambda }I(\hat{p}(\lambda )$ $<$ $\alpha )d\lambda $.
\end{example}

\section{Asymptotic Theory\label{sec:asym_theory}}

The following notation is used. $[z]$ rounds $z$ to the nearest integer. $%
|\cdot |$ is the $l_{1}$-matrix norm, and $||\cdot ||$ is the Euclidean
norm, unless otherwise noted. Assume the sample $\mathcal{S}_{n}$ $\equiv $ $%
\{z_{t}\}_{t=1}^{n}$ lies in $\mathbb{R}^{n\times q}$ for some $q$ $\in $ $%
\mathbb{N}$.

We require a notion of weak convergence and accompanying metric that can
handle a range of applications. A fundamental concern is that the test
statistic and p-value mappings $\mathcal{T}$ $:$ $\Lambda $ $\times $ $%
\mathbb{R}^{n\times q}$ $\rightarrow $ $[0,\infty )$ and $p$ $:$ $\Lambda $ $%
\times $ $\mathbb{R}^{n\times q}$ $\rightarrow $ $[0,1]$\ are not here
defined, making measurability a challenge for their sample paths $\{\mathcal{%
T}_{n}(\lambda )$ $:$ $\lambda $ $\in $ $\Lambda \}$ and $\{p_{n}(\lambda )$
$:$ $\lambda $ $\in $ $\Lambda \}$\ and their transforms like $\sup_{\lambda
\in \Lambda }p_{n}(\lambda )$ and $\int_{\Lambda }I(p_{n}(\lambda )$ $<$ $%
\alpha )d\lambda $. Even with $\mathcal{T}$ and $p$ in hand, measurability
may need to be assumed due to iterative estimation algorithms (e.g. GARCH
test). Let $\mathcal{B}(\mathcal{A})$ be the Borel $\sigma $-field on $%
\mathcal{A}$. We therefore assume $\mathcal{T}(\mathcal{S}_{n},\lambda )$
and $p(\mathcal{S}_{n},\lambda )$\ are $\sigma (\mathcal{S}_{n})$ $\otimes $
$\mathcal{B}(\Lambda )$ measurable and exist on a complete measure space.%
\footnote{%
Completeness is not trivial because $\mathcal{B}(\Lambda )$ is not complete
for any $\sigma $-finite measure, and even if extended to be complete under
Lebesque measure, the product $\sigma (\mathcal{S}_{n})$ $\otimes $ $%
\mathcal{B}(\Lambda )$ need not be complete under, e.g., any $\sigma $%
-finite measure. Thus $\sigma (\mathcal{S}_{n})$ $\otimes $ $\mathcal{B}%
(\Lambda )$ measurability and completeness implies we operate on the
completed $\sigma (\mathcal{S}_{n})$ $\otimes $ $\mathcal{B}(\Lambda )$ and
associated product measure.} Now majorants and integrals over uncountable
families of measurable functions like $\{p_{n}(\lambda )$ $:$ $\lambda $ $%
\in $ $\Lambda \}$ are measurable, and probabilities where applicable are
outer probability measures. See especially Pollard's (\citeyear{Pollard1984}%
: Appendix C) \textit{permissibility} criteria based on the notion of
analytic sets in \cite{DellacherieMeyer1978}. Under completeness,
permissibility necessarily holds (e.g. Dellacherie and Meyer %
\citeyear{DellacherieMeyer1978}, Section 33; cf. Pollard %
\citeyear{Pollard1984}, p. 195-196). See also \citet[Section
3]{Dudley1978} for the closely related \textit{Souslin} measurability %
\citep[cf.][Section 16]{DellacherieMeyer1978}.

We use weak convergence on $l_{\infty }(\Lambda )$, the space of bounded
functions on $\Lambda $ with sup-norm topology, in the sense of %
\citet{HoffJorg1991}:%
\begin{equation*}
\left\{ \mathcal{T}_{n}(\lambda )\right\} \Rightarrow ^{\ast }\left\{
\mathcal{T}(\lambda )\right\} \text{ in }l_{\infty }(\Lambda )\text{, where }%
\left\{ \mathcal{T}_{n}(\lambda )\right\} =\left\{ \mathcal{T}_{n}(\lambda
):\lambda \in \Lambda \right\} \text{, etc.}
\end{equation*}%
If, for instance, the sample is $\mathcal{S}_{n}$ $\equiv $ $%
\{x_{t},y_{t}\}_{t=1}^{n}$ $\in $ $\mathbb{R}^{n\times q}$, and $\mathcal{T}%
_{n}(\lambda )$ is a measurable mapping $h(\mathcal{Z}(\mathcal{S}%
_{n},\lambda ))$ of a function $\mathcal{Z}$ $:$ $\mathbb{R}^{n\times q}$ $%
\times $ $\Lambda $ $\rightarrow $ $\mathbb{R}$, then $h(\mathcal{Z}%
(s,\lambda ))$ $\in $ $l_{\infty }(\Lambda )$ requires the uniform bound $%
\sup_{\lambda \in \Lambda }|h(\mathcal{Z}(s,\lambda ))|$ $<$ $\infty $ for
each $s$ $\in $ $\mathbb{R}^{n\times q}$.\footnote{%
If more details are available, then boundedness can be refined. For example,
if $\mathcal{T}_{n}(\lambda )$ $=$ $(1/\sqrt{n}\sum_{t=1}^{n}z(y_{t},\lambda
))^{2}$ where $z$ $:$ $\mathbb{R}$ $\times $ $\Lambda $ $\rightarrow $ $%
\mathbb{R}$, then we need $\sup_{\lambda \in \Lambda }|z(y,\lambda )|$ $<$ $%
\infty $ for each $y$.} Sufficient conditions for weak convergence to a
Gaussian process, for example, are convergence in finite dimensional
distributions,\ and stochastic equicontinuity: $\forall \epsilon $ $>$ $0$
and $\eta $ $>$ $0$ there exists $\delta $ $>$ $0$ such that $%
\lim_{n\rightarrow \infty }P(\sup_{||\lambda -\tilde{\lambda}||\leq \delta }|%
\mathcal{T}_{n}(\lambda )$ $-$ $\mathcal{T}_{n}(\tilde{\lambda})|$ $>$ $\eta
)$ $<$ $\epsilon $. Consult, e.g., \cite{Dudley1978}, \cite{GineZinn84}, and
\cite{Pollard1984}.

A large variety of test statistics are known to converge weakly under
regularity conditions. In many cases $\mathcal{T}_{n}(\lambda )$ is a
continuous function $h(\mathcal{Z}_{n}(\lambda ))$ of a sequence of sample
mappings $\{\mathcal{Z}_{n}(\lambda )\}_{n\geq 1}$ such that $\sup_{x\in
A}|h(x)|$ $<$ $\infty $ on every compact subset $A$ $\subset $ $\mathbb{R}$,
and $\{\mathcal{Z}_{n}(\lambda )\}$ $\Rightarrow ^{\ast }$ $\{\mathcal{Z}%
(\lambda )\}$ a Gaussian process. Two examples of $h$ are $h(x)$ $=$ $x^{2}$
for pointwise asymptotic chi-squared tests of functional form or structural
change; or $h(x)$ $=$ $\max \{0,x\}$ for a GARCH test \citep{Andrews2001}.

A \textit{version} is a process with the same finite dimensional
distributions. If $\{\mathcal{Z}(\lambda )\}$ is Gaussian, then any other
Gaussian process with the same mean $E[\mathcal{Z}(\lambda )]$ and
covariance kernel $E[\mathcal{Z}(\lambda _{1})\mathcal{Z}(\lambda _{2})]$ is
a version of $\{\mathcal{Z}(\lambda )\}$.\footnote{%
Even in the Gaussian case it is not true that all versions have continuous
sample paths, but if a version of $\{\mathcal{Z}(\lambda )\}$ has continuous
paths then this is enough to apply the continuous mapping theorem to
transforms of $\mathcal{Z}_{n}(\lambda )$ over $\Lambda $. See %
\citet{Dudley67, Dudley1978}.}

\begin{assumption}[weak convergence]
\label{assum:main}Let $H_{0}$ be true.\medskip \newline
$a.$ $\{\mathcal{T}_{n}(\lambda )\}$ $\Rightarrow ^{\ast }$ $\{\mathcal{T}%
(\lambda )\}$, a process with a version that has \emph{almost surely}
bounded uniformly continuous sample paths (with respect to the sup-norm). $%
\mathcal{T}(\lambda )$ $\geq $ $0$ $a.s.$, $\sup_{\lambda \in \Lambda }%
\mathcal{T}(\lambda )$ $<$ $\infty $ $a.s$., and $\mathcal{T}(\lambda )$\
has an absolutely continuous distribution function $F_{0}(c)$ $\equiv $ $P(%
\mathcal{T}(\lambda )$ $\leq $ $c)$ that is not a function of $\lambda $%
.\medskip \newline
$b.$ $\sup_{\lambda \in \Lambda }|p_{n}(\lambda )$ $-$ $\bar{F}_{0}(\mathcal{%
T}_{n}(\lambda ))|$ $\overset{p}{\rightarrow }$ $0$, where $\bar{F}_{0}(c)$ $%
\equiv $ $P(\mathcal{T}(\lambda )$ $>$ $c)$.
\end{assumption}

\begin{remark}
\normalfont$(a)$ is broadly applicable (see Section \ref{sec:examples}).
Continuity of the distribution of $\mathcal{T}(\lambda )$ and $(b)$ ensure $%
p_{n}(\lambda )$ has asymptotically a uniform limit distribution under $H_{0}
$. This is mild since often $\mathcal{T}_{n}(\lambda )$ is a continuous
transformation of a standardized sample analogue to a population moment. In
a great variety of settings a standardized sample moment has a Gaussian or
stable distribution limit, or converges to a function of a Gaussian or
stable law. See \cite{GineZinn84} and \cite{Pollard1984} for weak
convergence to stochastic processes, exemplified with Gaussian functional
asymptotics, and see \cite{Bartkiewicz10} for weak convergence to a Stable
process for a (possibly dependent) heavy tailed process.
\end{remark}

\begin{remark}
\normalfont$(b)$ is required when $p_{n}(\lambda )$ is not computed as the
asymptotic p-value $\bar{F}_{0}(\mathcal{T}_{n}(\lambda ))$, for example
when a simulation or bootstrap method is used because $\bar{F}_{0}$\ is
unknown or a better small sample approximation is desired. Thus, in order to
obtain lower level conditions we need to know how $p_{n}(\lambda )$ was
computed. In Section \ref{sec:func_form_weak}, for example, we use Hill's (%
\citeyear{Hill2021_weak}) weak identification robust bootstrap method for
p-value computation; and in Section \ref{sec:garch} we use Andrews' (%
\citeyear{Andrews2001}) simulation method for p-value computation for a
GARCH test.
\end{remark}

All proofs are presented in Appendix \ref{sec:proofs}.

\begin{theorem}
\label{th:main}Let Assumption \ref{assum:main} hold.\medskip \newline
$a.$ In general $\lim_{n\rightarrow \infty }P(\mathcal{P}_{n}^{\ast }(\alpha
)$ $>$ $\alpha )\leq \alpha $.$\medskip $\newline
$b.$ The asymptotic size is exactly $\lim_{n\rightarrow \infty }P(\mathcal{P}%
_{n}^{\ast }(\alpha )$ $>$ $\alpha )$ $=$ $\alpha $ when $\mathcal{T}%
(\lambda )$ $=$ $\mathcal{T}(\lambda ^{\ast })$ $=$ $a.s.$ $\forall \lambda $
$\in $ $\Lambda $ and some $\lambda ^{\ast }$ $\in $ $\Lambda .\medskip
\newline
c.\lim_{n\rightarrow \infty }P(\mathcal{P}_{n}^{\ast }(\alpha )$ $>$ $\alpha
)$ $>0$ under the following condition: $\{\bar{F}_{0}(\mathcal{T}(\lambda
))\}$ is weakly dependent in the sense that $P(\bar{F}_{0}(\mathcal{T}%
(\lambda ))$ $<$ $\alpha ,\bar{F}_{0}(\mathcal{T}(\tilde{\lambda}))$ $<$ $%
\alpha )$ $>$ $\alpha ^{2}$ for each couplet $\{\lambda ,\tilde{\lambda}\}$\
on a subset of $\Lambda $ $\times $ $\Lambda $ with positive measure.
\end{theorem}

\begin{remark}
\normalfont Under $H_{0}$ the pointwise PV test rejects $H_{0}$
asymptotically with probability $\alpha $. The above theorem proves this
implies asymptotically no more than an $\alpha $ portion of all $\lambda
^{\prime }s$ lead to a rejection.
\end{remark}

\begin{remark}
\normalfont In general the asymptotic \emph{level} of the test is $\alpha $
when the critical value is itself $\alpha $
\citep[cf.][eq.
(3.1)]{Lehmann1994}. The proof reveals if $\mathcal{T}(\lambda )$ $=$ $%
\mathcal{T}(\lambda ^{\ast })$ $a.s.$ for some $\lambda ^{\ast }$ and all $%
\lambda $ such that they are perfectly dependent, then $\lim_{n\rightarrow
\infty }P(\mathcal{P}_{n}^{\ast }(\alpha )$ $>$ $\alpha )$ $=$ $\alpha $ and
the asymptotic size is $\alpha $. This occurs when $\lambda $ is a tuning
parameter since these do not appear in the limit process %
\citep[see][]{HillAguilar13}.
\end{remark}

Next, asymptotic global power of PV test (\ref{T_test}) translates to global
power for PVOT test (\ref{PVOT_test}).

\begin{theorem}
\label{th:main_h1} $\ \ $\newline
$a$. $\lim_{n\rightarrow \infty }P(\mathcal{P}_{n}^{\ast }(\alpha )$ $>$ $%
\alpha )$ $>$ $0$ \emph{if and only if} there exists a subset $\tilde{\Lambda%
}$ $\subset $ $\Lambda $ with Lebesgue measure \emph{greater} than $\alpha $
($\int_{\tilde{\Lambda}}1d\lambda $ $>$ $\alpha $) such that $\lim
\inf_{n\rightarrow \infty }P(p_{n}(\lambda )$ $<$ $\alpha )$ $>$ $0$.$%
\medskip $\newline
$b$. The PVOT test is consistent $P(\mathcal{P}_{n}^{\ast }(\alpha )$ $>$ $%
\alpha )$ $\rightarrow $ $1$ if the PV test is consistent $P(p_{n}(\lambda )$
$<$ $\alpha )$ $\rightarrow $ $1$\ on a subset of $\Lambda $ with Lebesgue
measure greater than $\alpha $.
\end{theorem}

\begin{remark}
\normalfont As long as the PV test is consistent on a subset of $\Lambda $
with measure greater than $\alpha $, then the PVOT test is consistent. In
view of $\int_{\Lambda }d\lambda $ $=$ $1$\ this trivially holds when the PV
test is consistent for any $\lambda $ outside a set with measure zero,
including Andrews' (\citeyear{Andrews2001}) GARCH test which is consistent
on a known compact $\Lambda $; \cite{White1989}, \cite{Bierens1990} and \cite%
{BierensPloberger1997} tests (and others) of omitted nonlinearity; Andrews' (%
\citeyear{Andrews1993}) structural break test; and a test of an omitted
Box-Cox transformation. See Section \ref{sec:examples}. At the risk of
abusing terminology, we will say a test based on $\mathcal{T}_{n}(\lambda )$
is \emph{randomized} when $\lambda $ is drawn from a uniform distribution on
$\Lambda $ independent of the data. The randomized test is consistent only
if the PV test is consistent for every $\lambda $ outside a set with measure
zero.\footnote{%
Here and elsewhere we refer to a test based on $\mathcal{T}_{n}(\lambda
_{\ast })$ as a \textit{randomized test}, which is generally different from
the classical definition of a randomized test \citep[cf.][]{Lehmann1994}.}
The transforms $\int_{\Lambda }\mathcal{T}_{n}(\lambda )\mu (d\lambda )$ and
$\sup_{\lambda \in \Lambda }\mathcal{T}_{n}(\lambda )$, however, are
consistent if the PV test is consistent on a subset of $\Lambda $ with
positive measure. Thus, the PVOT test ranks above the randomized test, but
below average and supremum tests in terms of required PV test asymptotic
power over $\Lambda $. As we discussed in Section \ref{sec:intro}, it is
difficult to find a relevant example in which this matters, outside a toy
example. We give such an example below.
\end{remark}

The following shows how PV test power transfers to the PVOT test.

\begin{example}
\label{ex:consist_test}\normalfont Let $\lambda _{\ast }$ be a random draw
from a uniform distribution on $\Lambda $. The parameter space is $\Lambda $
$=$ $[0,1]$, $\mathcal{T}_{n}(\lambda )$ $\overset{p}{\rightarrow }$ $\infty
$ for $\lambda $ $\in $ $[.5,.56]$ such that the PV test is consistent on a
subset with measure $\beta $ $=$ $.06$, and $\{\mathcal{T}_{n}(\lambda )$ $:$
$\lambda $ $\in $ $\Lambda /[.5,.56]\}$ $\Rightarrow ^{\ast }$ $\{\mathcal{T}%
(\lambda )$ $:$ $\lambda $ $\in $ $\Lambda /[.5,.56]\}$ such that there is
only trivial power. Thus, $\int_{\Lambda }\mathcal{T}_{n}(\lambda )\mu
(d\lambda )$ and $\sup_{\lambda \in \Lambda }\mathcal{T}_{n}(\lambda )$ have
asymptotic power of one. A uniformly randomized PV test is not consistent at
any level, and at level $\alpha $ $<$ $.06$ has trivial power.

In the PVOT case, however, by applying arguments in the proof of Theorem \ref%
{th:main}, we can show $\lim_{n\rightarrow \infty }P(\mathcal{P}_{n}^{\ast
}(\alpha )$ $>$ $\alpha )$ is identically%
\begin{equation*}
P\left( \int_{\lambda \in \lbrack .5,.56]}d\lambda +\int_{\lambda \notin
\lbrack .5,.56]}I\left( \mathcal{U}(\lambda )<\alpha \right) d\lambda
>\alpha \right) =P\left( \int_{\lambda \notin \lbrack .5,.56]}I\left(
\mathcal{U}(\lambda )<\alpha \right) d\lambda >\alpha -.06\right)
\end{equation*}%
for some process $\{\mathcal{U}(\lambda )$ $:$ $\lambda $ $\in $ $\Lambda
/[.5,.56]\}$ where $\mathcal{U}(\lambda )$ is uniform on $[0,1]$. This
implies the PVOT test is consistent at level $\alpha $ $\leq $ $.06$ since $%
\int_{\lambda \notin \lbrack .5,.56]}I(\mathcal{U}(\lambda )$ $<$ $\alpha
)d\lambda $ $>$ $0$ $a.s.$
\end{example}

\section{Local Power \label{app:locpow}}

A characterization of local power requires an explicit hypothesis and some
information on the construction of $\mathcal{T}_{n}(\lambda )$. Assume an
observed sequence $\{y_{t}\}_{t=1}^{n}$ has a parametric joint distribution $%
f(y;\theta _{0})$, where $\theta _{0}$ $=$ $[\beta _{0}^{\prime },\delta
_{0}^{\prime },]$ and $\beta _{0}$ $\in $ $\mathbb{R}^{r}$, $r$ $\geq $ $1$.
Consider testing whether the subvector $\beta _{0}$ $=$ $0$, while $\delta
_{0}$ may contain other distribution parameters. If some additional
parameter $\lambda $ is part of $\delta _{0}$ only when $\beta _{0}$ $\neq $
$0$, and therefore not identified under $H_{0}$, then we have Andrews and
Ploberger's (\citeyear{AndrewsPloberger1994}) setting, but in general $%
\lambda $ need not be part of the true data generating process.

We first treat a general environment that includes each test example
mentioned in this paper. We then study a test of omitted nonlinearity, and
perform a numerical experiment in order to compare local power.

\subsection{Local Power : General Case}

The sequence of local alternatives we consider is:%
\begin{equation}
H_{1}^{L}:\beta _{0}=\mathcal{N}_{n}^{-1}b\text{ for some }b\in \mathbb{R}%
^{r},  \label{H1L}
\end{equation}%
where $(\mathcal{N}_{n}\}$ is a sequence of diagonal matrices $[\mathcal{N}%
_{n,i,j}]_{i,j=1}^{r}$, $\mathcal{N}_{n,i,i}$ $\rightarrow $ $\infty $. The
test statistic is $\mathcal{T}_{n}(\lambda )$ $\equiv $ $h(\mathcal{Z}%
_{n}(\lambda ))$ for a sequence of random functions $\{\mathcal{Z}%
_{n}(\lambda )\}$ on $\mathbb{R}^{q}$, $q$ $\geq $ $1$, and a measurable
function $h$ $:$ $\mathbb{R}^{q}$ $\rightarrow $ $[0,\infty )$ where $h(x)$
is monotonically increasing in $||x||$, and $h(x)$ $\rightarrow $ $\infty $
as $||x||$ $\rightarrow $ $\infty $. This covers LM and Wald statistics, and
each test statistic discussed in this paper.

We assume regularity conditions apply such that under $H_{1}^{L}$
\begin{equation}
\left\{ \mathcal{Z}_{n}(\lambda ):\lambda \in \Lambda \right\} \Rightarrow
^{\ast }\left\{ \mathcal{Z}(\lambda )+c(\lambda )b:\lambda \in \Lambda
\right\} ,  \label{Zn_weak}
\end{equation}%
for some matrix $c(\lambda )$ $\in $ $\mathbb{R}^{q\times r}$, and $\{%
\mathcal{Z}(\lambda )\}$ is a zero mean process on $\mathbb{R}^{q}$ with a
version that has \textit{almost surely} uniformly continuous sample paths
(with respect to some norm $||\cdot ||$). In many cases in the literature $\{%
\mathcal{Z}(\lambda )\}$ is a Gaussian process with $E[\mathcal{Z}(\lambda )%
\mathcal{Z}(\lambda )^{\prime }]$ $=$ $I_{q}$.

Combine (\ref{Zn_weak}) and the continuous mapping theorem to deduce under $%
H_{0}$ the limiting distribution function $F_{0}(x)$ $\equiv $ $P(h(\mathcal{%
Z}(\lambda ))$ $\leq $ $x)$ for $\mathcal{T}_{n}(\lambda )$, cf. %
\citet[Theorem 2.7]{Billingsley1999}. An asymptotic p-value is $%
p_{n}(\lambda )$ $=$ $\bar{F}_{0}(\mathcal{T}_{n}(\lambda ))$ $\equiv $ $1$ $%
-$ $F_{0}(\mathcal{T}_{n}(\lambda ))$, hence $\int_{\Lambda }I(p_{n}(\lambda
)$ $<$ $\alpha )d\lambda $ $\overset{d}{\rightarrow }$ $\int_{\Lambda }I(%
\bar{F}_{0}(h(\mathcal{Z}(\lambda ))$ $+$ $c(\lambda )b))$ $<$ $\alpha )$
under $H_{1}^{L}$. Similarly, any continuous mapping $g$\ over $\Lambda $\
satisfies $g(\mathcal{T}_{n}(\lambda ))$ $\overset{d}{\rightarrow }$ $g(h(%
\mathcal{Z}(\lambda )$ $+$ $c(\lambda )b))$, including $\int_{\Lambda }%
\mathcal{T}_{n}(\lambda )\mu (d\lambda )$ and $\sup_{\lambda \in \Lambda }%
\mathcal{T}_{n}(\lambda )$. Obviously if $c(\lambda )b$ $=$ $0$ when $b$ $%
\neq $ $0$\ then local power is trivial at $\lambda $. Whether any of the
above tests has non-trivial asymptotic local power depends on the measure of
the subset of $\Lambda $ on which $\inf_{\xi ^{\prime }\xi =1}||\xi ^{\prime
}c(\lambda )||$ $>$ $0$.

In order to make a fair comparison across tests, we assume each is
asymptotically correctly sized for a nominal level $\alpha $ test. The next
result follows from the preceding properties, hence a proof is omitted.

\begin{theorem}
\label{th:local_pow}Let (\ref{H1L}), (\ref{Zn_weak}) and $b$ $\neq $ $0$
hold, and write $\inf_{\xi ^{\prime }\xi =1}||\xi ^{\prime }c(\lambda )||$.
Assume the randomized statistic $\mathcal{T}_{n}(\lambda ^{\ast })$ uses a
draw $\lambda ^{\ast }$ from a uniform distribution on $\Lambda $.
Asymptotic local power is non-trivial for (i) the PVOT test when $\inf_{\xi
^{\prime }\xi =1}||\xi ^{\prime }c(\lambda )||$ $>$ $0$ on a subset of $%
\Lambda $ with measure greater than $\alpha $; and (ii) the uniformly
randomized, average and supremum tests when $\inf_{\xi ^{\prime }\xi
=1}||\xi ^{\prime }c(\lambda )||$ $>$ $0$ on a subset of $\Lambda $ with
positive measure.\newline
$b$. Under cases (i) and (ii), asymptotic local power is monotonically
increasing in $|b|$ and converges to one as $|b|$ $\rightarrow $ $\infty $.
\end{theorem}

\begin{remark}
\normalfont The PVOT test ranks lower than randomized, average and supremum
tests because it rejects only when the PV tests rejects on a subset of $%
\Lambda $ with measure greater than $\alpha $. Indeed, the PVOT test cannot
asymptotically distinguish between PV tests that are consistent on a subset
with measure less than $\alpha $ and have trivial power otherwise, or have
trivial power everywhere. This cost is slight since a meaningful example in
which it matters is difficult to find. The previously cited tests of omitted
nonlinearity and GARCH effects all have randomized, PVOT, average and
supremum versions with non-trivial local power, although we only give
complete details for a test of omitted nonlinearity below.
\end{remark}

\subsection{Example : Test of Omitted Nonlinearity\label{ex:omitted_nl}}

The proposed model to be tested is%
\begin{equation*}
y_{t}=f\left( x_{t},\zeta _{0}\right) +e_{t},
\end{equation*}%
where $\zeta _{0}$ lies in the interior of $\mathfrak{Z}$, a compact subset
of $\mathbb{R}^{q}$, $x_{t}$ $\in $ $\mathbb{R}^{k}$ contains a constant
term and may contain lags of $y_{t}$, and $f:\mathbb{R}^{k}\times \mathfrak{Z%
}$ $\rightarrow $ $\mathbb{R}$ is a known response function. The null is $%
H_{0}$ $:$ $E[y_{t}|x_{t}]$ $=$ $f(x_{t},\zeta _{0})$ $a.s.$

Assume $\{e_{t},x_{t},y_{t}\}$ are stationary for simplicity. Let $\Psi $ be
a $1$-$1$ bounded mapping from $\mathbb{R}^{k}$ to $\mathbb{R}^{k}$, let $%
\mathcal{F}$ $:$ $\mathbb{R}$ $\rightarrow $ $\mathbb{R}$ be analytic and
non-polynomial (e.g. exponential or logistic), and assume $\lambda $ $\in $ $%
\Lambda $, a compact subset of $\mathbb{R}^{k}$. Mis-specification $%
\sup_{\zeta \in \mathbb{R}^{q}}P(E[y_{t}|x_{t}]$ $=$ $f(x_{t},\zeta ))$ $<$ $%
1$ implies $E[e_{t}\mathcal{F}(\lambda ^{\prime }\Psi (x_{t}))]$ $\neq $ $0$
$\forall \lambda $ $\in $ $\Lambda /\mathcal{S}$, where $\mathcal{S}$ has
Lebesgue measure zero. See \cite{White1989}, \cite{Bierens1990} and \cite%
{StinchWhite1998} for seminal results for iid data, and see \cite{deJong1996}
and \cite{Hill2008} for dependent data. The test statistic for a test of the
hypothesis $H_{0}$ $:$ $E[y_{t}|x_{t}]$ $=$ $f(x_{t},\zeta _{0})$ $a.s.$ is%
\begin{equation}
\mathcal{T}_{n}(\lambda )=\left( \frac{1}{\hat{v}_{n}(\lambda )}\frac{1}{%
\sqrt{n}}\sum_{t=1}^{n}e_{t}(\hat{\zeta}_{n})\mathcal{F}\left( \lambda
^{\prime }\Psi (x_{t})\right) \right) ^{2}\text{ where }e_{t}(\zeta )\equiv
y_{t}-f(x_{t},\zeta ).  \label{Tn_CM}
\end{equation}%
The estimator $\hat{\zeta}_{n}$ is assumed $\sqrt{n}$-consistent for a
strongly identified $\zeta _{0}$, and $\hat{v}_{n}^{2}(\lambda )$ is a
consistent estimator of $E[\{1/\sqrt{n}\sum_{t=1}^{n}e_{t}(\hat{\zeta}_{n})%
\mathcal{F}(\lambda ^{\prime }\Psi (x_{t}))\}^{2}]$. By application of
Theorem \ref{lm:locpow_suff}, below, the asymptotic p-value is $%
p_{n}(\lambda )$ $\equiv $ $1$ $-$ $F_{0}\left( \mathcal{T}_{n}(\lambda
)\right) $ $\equiv $ $\bar{F}_{0}\left( \mathcal{T}_{n}(\lambda )\right) $
where $F_{0}$ is the $\chi ^{2}(1)$ distribution function.

In view of $\sqrt{n}$-asymptotics, a sequence of local-to-null alternatives
is
\begin{equation}
H_{1}^{L}:\beta _{0}=b/n^{1/2}\text{ for }b\in \mathbb{R}.  \label{H1_L}
\end{equation}%
We assume for now that regularity conditions apply such that, for some
sequence of positive finite non-random numbers $\{c(\lambda )\}$ :%
\begin{equation}
\text{under }H_{1}^{L}\text{ : }\left\{ \mathcal{T}_{n}(\lambda ):\lambda
\in \Lambda \right\} \Rightarrow ^{\ast }\{\left( \mathcal{Z}(\lambda
)+bc(\lambda )\right) ^{2}:\lambda \in \Lambda \},  \label{TZc}
\end{equation}%
where $\{\mathcal{Z}(\lambda )$ $+$ $c(\lambda )b\}$ is a Gaussian process
with mean $\{c(\lambda )b\}$, and \emph{almost surely} uniformly continuous
sample paths. See below for low level assumptions that imply (\ref{TZc}).
The latter implies by Theorem \ref{th:main} that the PVOT asymptotic
probability of rejection $\lim_{n\rightarrow \infty }P(\mathcal{P}_{n}^{\ast
}(\alpha )$ $>$ $\alpha )$, under $H_{0}$, is between $(0,\alpha ]$.

Let $F_{J,\nu }(c)$ denote a noncentral $\chi ^{2}(J)$ law with
noncentrality $\nu $, hence $(\mathcal{Z}(\lambda )$ $+$ $c(\lambda )b)^{2}$
is distributed $F_{1,b^{2}c(\lambda )^{2}}$. Under the null $b$ $=$ $0$ by
construction $p_{n}(\lambda )$ $\overset{d}{\rightarrow }$ $\bar{F}_{1,0}((%
\mathcal{Z}(\lambda )$ $+$ $c(\lambda )b)^{2})$ $=$ $\bar{F}_{1,0}(\mathcal{Z%
}(\lambda )^{2})$ is uniformly distributed on $[0,1]$. Under the global
alternative $\sup_{\zeta \in \mathbb{R}^{q}}P(E[y_{t}|x_{t}]$ $=$ $%
f(x_{t},\zeta ))$ $<$ $1$ notice $\mathcal{T}_{n}(\lambda )$ $\overset{p}{%
\rightarrow }$ $\infty $ $\forall \lambda $ $\in $ $\Lambda /S$ implies $%
p_{n}(\lambda )$ $\overset{p}{\rightarrow }$ $0$ $\forall \lambda $ $\in $ $%
\Lambda /S$, hence $\mathcal{P}_{n}^{\ast }(\alpha )$ $\overset{p}{%
\rightarrow }$ $1$ by Theorem \ref{th:main_h1}. The latter implies the PVOT
test of $E[y_{t}|x_{t}]$ $=$ $f(x_{t},\zeta _{0})$ $a.s.$ is consistent. The
following contains the result under $H_{1}^{L}$.

\begin{theorem}
\label{th:local_pow_nl}Under (\ref{TZc}), asymptotic local power of the PVOT
test is $P(\int_{\Lambda }I(\bar{F}_{1,0}(\{\mathcal{Z}(\lambda )$ $+$ $%
c(\lambda )b\}^{2})$ $<$ $\alpha )d\lambda >\alpha )$. Hence, under $%
H_{1}^{L}$ the probability the PVOT test rejects $H_{0}$ increases to unity
monotonically as the drift parameter $|b|$ $\rightarrow $ $\infty $, for any
nominal level $\alpha $ $\in $ $[0,1)$.{}
\end{theorem}

The following assumptions detail sufficient conditions leading to (\ref{TZc}%
). These are not the most general possible, but are fairly compact for the
sake of brevity.

\begin{assumption}[nonlinear regression and functional form test]
\label{assum:locpow_suff}\ \ \medskip \newline
a. \emph{Memory and Moments}: All random variables lie on the same complete
measure space. $\{y_{t},x_{t},\epsilon _{t}\}$ are stationary; $%
E|y_{t}|^{4+\iota }$ $<$ $\infty $ and $E|\epsilon _{t}|^{4+\iota }$\ for
tiny $\iota $ $>$ $0$; $E[\epsilon _{t}|x_{t}]$ $=$ $0$ $a.s.$ under $%
H_{1}^{L}$; $E[\inf_{\lambda \in \Lambda }w_{t}^{2}(\lambda )]$ $>$ $0$, $%
E[\epsilon _{t}^{2}\inf_{\lambda \in \Lambda }w_{t}^{2}(\lambda )]$ $>$ $0$,
and $\inf_{\lambda \in \Lambda }||(\partial /\partial \lambda )E[\epsilon
_{t}^{2}F(\lambda ^{\prime }\Psi (x_{t}))^{2}]||$ $>$ $0$; $\{x_{t},\epsilon
_{t}\}$ are $\beta $-mixing with mixing coefficients $\beta _{h}$ $=$ $%
O(h^{-4-\delta })$ for tiny $\delta $ $>$ $0$.\medskip \newline
b. \emph{Response Function}: $f:\mathbb{R}^{k}\times \mathfrak{Z}$ $%
\rightarrow $ $\mathbb{R}$; $f(\cdot ,\zeta )$ is twice continuously
differentiable; $(\partial /\partial \zeta )^{i}f(x,\zeta )$\ are Borel
measurable for each $\zeta $ $\in $ $\mathfrak{Z}$ and $i$ $=$ $0,1,2$;
write $h_{t}^{(i)}(\zeta )$ $=$ $(\partial /\partial \zeta
)^{i}f(x_{t},\cdot )$ for $i$ $=$ $0,1,2$: $E[\sup_{\zeta \in \mathfrak{Z}%
}|h_{t}^{(i)}(\zeta )|^{4+\delta }]$ $<$ $\infty $ for tiny $\delta $ $>$ $0$
and each $i$; $(\partial /\partial \zeta )f(x_{t},\zeta _{0})$ has full
column rank.\medskip \newline
c. \emph{Test Weight}: $F(\cdot )$ is analytic, nonpolynomial, and $%
(\partial /\partial c)^{i}F(c)$ is bounded for $i$ $=$ $0,1,2$\ uniformly on
any compact subset; $\Psi $ is one-to-one and bounded.\medskip \newline
d. \emph{Variance Estimator}: $\hat{v}_{n}^{2}(\lambda )$ $\equiv $ $%
1/n\sum_{s,t=1}^{n}\mathcal{K}((s$ $-$ $t)/\gamma _{n})e_{s}(\hat{\zeta}%
_{n})e_{t}(\hat{\zeta}_{n})\hat{w}_{n,s}(\lambda ,\hat{\zeta}_{n})\hat{w}%
_{n,t}(\lambda ,\hat{\zeta}_{n})$ with kernel $\mathcal{K}$ and bandwidth $%
\gamma _{n}$ $\rightarrow $ $\infty $ \textit{and }$\gamma _{n}$\textit{\ }$%
= $\textit{\ }$o(\sqrt{n})$. $\mathcal{K}$\textit{\ is continuous at }$0$%
\textit{\ and all but a finite number of points}, $\mathcal{K}$ $:$ $\mathbb{%
R}$ $\rightarrow $ $[-1,1]$, $\mathcal{K}(0)$ $=$ $1,$ $\mathcal{K}(x)$ $=$ $%
\mathcal{K}(-x)$ $\forall x$ $\in $ $\mathbb{R},$ $\int_{-\infty }^{\infty }|%
\mathcal{K}(x)|dx$ $<$ $\infty $; and there exists $\{\delta _{n}\}$, $%
\delta _{n}$ $>$ $0$, $\delta _{n}/\sqrt{n}$ $\rightarrow $ $\infty $, such
that $\int_{\delta _{n}}^{\infty }\{|\mathcal{K}(x)|$ $+$ $|\mathcal{K}%
(-x)|\}dx$ $=$ $o(1/\sqrt{n})$.\medskip \newline
e.\emph{\ Plug-In}: $\zeta _{0}$ is an interior point of $\mathfrak{Z}$, and
$\hat{\zeta}_{n}$ $\equiv $ $\argmin_{\zeta \in \mathfrak{Z}%
}\{1/n\sum_{t=1}^{n}(y_{t}$ $-$ $f(x,\zeta ))^{2}\}.$
\end{assumption}

\begin{remark}
\normalfont The kernel variance $\hat{v}_{n}^{2}(\lambda )$ form follows
from a standard expansion of \linebreak $1/\sqrt{n}\sum_{t=1}^{n}e_{t}(\hat{%
\zeta}_{n})\mathcal{F}(\lambda ^{\prime }\Psi (x_{t}))$ around $\zeta _{0}$
under $H_{0}$. We exploit a kernel estimator in order to prove uniform
convergence of $\hat{v}_{n}^{2}(\lambda )$ without the assumption that $%
H_{0} $ is true, a generality that may be of separate interest. See Lemma
C.1 in \citet[Appendix C]{Hill_supp_ot}.
\end{remark}

\begin{remark}
\normalfont Property (d), other than the requirement that $\mathcal{I}_{n}$ $%
\equiv $ $\int_{\delta _{n}}^{\infty }\{|\mathcal{K}(x)|$ $+$ $|\mathcal{K}%
(-x)|\}dx$ $=$ $o(1/\sqrt{n})$ for $\delta _{n}/\sqrt{n}$ $\rightarrow $ $%
\infty $, is similar to properties in \cite{AndrewsHAC91} and elsewhere,
covering Bartlett, Parzen, Tukey-Hanning and Quadratic-Spectral kernels. We
use $\mathcal{I}_{n}$ $=$ $o(1/\sqrt{n})$ with $\delta _{n}/\sqrt{n}$ $%
\rightarrow $ $\infty $ to prove uniform convergence $\sup_{\lambda \in
\Lambda }|\hat{v}_{n}^{2}(\lambda )$ $-$ $v^{2}(\lambda )|$ $\overset{p}{%
\rightarrow }$ $0$. The bound $\mathcal{I}_{n}$ $=$ $o(1/\sqrt{n})$ is
trivially satisfied for any $\delta _{n}$ $\geq $ $K$ and some finite $K$ $>$
$0$ for Bartlett, Parzen, and Tukey-Hanning kernels, while the
Quadratic-Spectral kernel obtains $\mathcal{I}_{n}\leq K\int_{\delta
_{n}}^{\infty }x^{-2}dx=K\delta _{n}^{-3}$ hence $\mathcal{I}_{n}$ $=$ $o(1/%
\sqrt{n})$ for any $\delta _{n}/n^{1/6}$ $\rightarrow $ $\infty $.
\end{remark}

The next claim is proven in Appendix C of the SM since it follows from
standard arguments.

\begin{theorem}
\label{lm:locpow_suff} $\ \ \ \medskip $\newline
$a.$ Assumption \ref{assum:locpow_suff} implies Assumption \ref{assum:main}.
In particular, under $H_{0}$ we have $\{\mathcal{T}_{n}(\lambda )$ $:$ $%
\lambda $ $\in $ $\Lambda \}$ $\Rightarrow ^{\ast }$ $\{\mathcal{Z}(\lambda
)^{2}$ $:$ $\lambda $ $\in $ $\Lambda \}$ where $\{\mathcal{Z}(\lambda )$ $:$
$\lambda $ $\in $ $\Lambda \}$ is a zero mean Gaussian process with a
version that has \emph{almost surely} uniformly continuous sample paths, and
covariance kernel
\begin{equation}
E\left[ \mathcal{\tilde{Z}}_{n}(\lambda )\mathcal{\tilde{Z}}_{n}(\tilde{%
\lambda})\right] =\frac{E\left[ \epsilon _{t}^{2}w_{t}(\lambda )w_{t}(\tilde{%
\lambda})\right] }{\left( E[\epsilon _{t}^{2}w_{t}^{2}(\lambda )]E[\epsilon
_{t}^{2}w_{t}^{2}(\tilde{\lambda})]\right) ^{1/2}}.  \label{cov_kern}
\end{equation}%
$b.$ Under $H_{1}^{L}$\ weak convergence (\ref{TZc}) is valid with $%
c(\lambda )$ $=$ $E[w_{t}^{2}(\lambda )]/(E[\epsilon
_{t}^{2}w_{t}^{2}(\lambda )])^{1/2}$ $>$ $0$ where $w_{t}(\lambda )$ $\equiv
$ $F_{t}(\lambda )$ $-$ $E[F_{t}(\lambda )g_{t}(\zeta _{0})^{\prime }]$ $%
\times $ $(E[g_{t}(\zeta _{0})g_{t}(\zeta _{0})^{\prime }])^{-1}g_{t}(\zeta
_{0})$.
\end{theorem}

Theorem \ref{lm:locpow_suff}.a implies under $H_{0}$ the test statistic
converges weakly $\{\mathcal{T}_{n}(\lambda )$ $:$ $\lambda $ $\in $ $%
\Lambda \}$ $\Rightarrow ^{\ast }$ $\{\mathcal{Z}(\lambda )^{2}$ $:$ $%
\lambda $ $\in $ $\Lambda \}$, where $\{\mathcal{Z}(\lambda )\}$ is weakly
dependent in the sense of Theorem \ref{th:main}: $P(\bar{F}_{0}(\mathcal{T}%
(\lambda ))$ $<$ $\alpha ,\bar{F}_{0}(\mathcal{T}(\tilde{\lambda}))$ $<$ $%
\alpha )$ $>$ $\alpha ^{2}$ on a subset of $\Lambda $ $\times $ $\Lambda $
with positive measure. This follows instantly from Gaussianicity of $\{%
\mathcal{Z}(\lambda )\}$ and its continuous covariance kernel (\ref{cov_kern}%
). This in turn implies by Theorem \ref{th:main} that the PVOT $\mathcal{P}%
_{n}^{\ast }(\alpha )$ $\equiv $ $\int_{\Lambda }I(p_{n}(\lambda )$ $<$ $%
\alpha )d\lambda $ does not have a degenerate limit distribution, which
yields the following result by invoking Theorems \ref{th:main} and \ref%
{lm:locpow_suff}.a.

\begin{theorem}
\label{lm:omitted_nl_null}Let Assumption \ref{assum:locpow_suff} and $H_{0}$
hold. Then $\lim_{n\rightarrow \infty }P(\mathcal{P}_{n}^{\ast }(\alpha )$ $%
> $ $\alpha )$ $\in $ $(0,\alpha ].$
\end{theorem}

\subsection{Numerical Experiment : Test of Omitted Nonlinearity\label%
{sec:local_num}}

Our final goal in this section is to compare asymptotic local power for
tests based on the PVOT, average $\int_{\Lambda }\mathcal{T}_{n}(\lambda
)\mu (d\lambda )$ with uniform measure $\mu (\lambda )$, supremum $%
\sup_{\lambda \in \Lambda }\mathcal{T}_{n}(\lambda )$, and Bierens and
Ploberger's (\citeyear{BierensPloberger1997}) Integrated Conditional Moment
[ICM] statistics. We work with a simple model $y_{t}$ $=$ $\zeta _{0}x_{t}$ $%
+$ $\beta _{0}\exp \{\lambda x_{t}\}+\epsilon _{t}$, where $\zeta _{0}$ $=$ $%
1$, $\beta _{0}$ $=$ $b/\sqrt{n}$, and $\{\epsilon _{t},x_{t}\}$ are iid $%
N(0,1)$ distributed. We omit a constant term entirely for simplicity. In
order to abstract from the impact of sampling error on asymptotics, we
assume $\zeta _{0}$ $=$ $1$\ is known, hence the test statistic is%
\begin{equation*}
\mathcal{T}_{n}(\lambda )\equiv \frac{\hat{z}_{n}^{2}(\lambda )}{\hat{v}%
_{n}^{2}(\lambda )}\text{\ where }\hat{z}_{n}(\lambda )\equiv \frac{1}{\sqrt{%
n}}\sum_{t=1}^{n}\left( y_{t}-\zeta _{0}x_{t}\right) \exp \{\lambda x_{t}\}%
\text{, }\hat{v}_{n}^{2}(\lambda )\equiv \frac{1}{n}\sum_{t=1}^{n}\left(
y_{t}-\zeta _{0}x_{t}\right) ^{2}\exp \{2\lambda x_{t}\}.
\end{equation*}%
The nuisance parameter space is $\Lambda $ $=$ $[0,1]$. A Gaussian setting
implies the main results of \cite{AndrewsPloberger1994} apply: the average $%
\int_{\Lambda }\mathcal{T}_{n}(\lambda )\mu (d\lambda )$ has the highest
weighted average local power for alternatives close to the null.

In view of Gaussianicity, and Theorem \ref{lm:locpow_suff}, it can be shown $%
\{\mathcal{T}_{n}(\lambda )\}$ $\Rightarrow ^{\ast }$ $\{(\mathcal{Z}%
(\lambda )$ $+$ $c(\lambda )b)^{2}\}$, where $c(\lambda )$ $=$ $E[\exp
\{2\lambda x_{t}\}]/(E[\epsilon _{t}^{2}\exp \{2\lambda x_{t}\}])^{1/2}$ $=$
$(E[\exp \{2\lambda x_{t}\}])^{1/2}$ $=$ $\exp \{\lambda ^{2}\}$, and $\{%
\mathcal{Z}(\lambda )\}$ is a zero mean Gaussian process with \emph{almost
surely} uniformly continuous sample paths, and covariance function $E[%
\mathcal{Z}(\lambda )\mathcal{Z}(\tilde{\lambda})]$ $=$ $\exp \{-.5(\lambda $
$-$ $\tilde{\lambda})^{2}\}$. Local asymptotic power is therefore:
\begin{eqnarray*}
&&\text{PVOT}\text{: }P\left( \int_{0}^{1}I\left( \bar{F}_{1,0}\left(
\left\{ \mathcal{Z}(\lambda )+b\exp \{\lambda ^{2}\}\right\} ^{2}\right)
<\alpha \right) d\lambda >c_{\alpha }^{(pvot)}\right) \\
&&\text{randomized: }P\left( \left\{ \mathcal{Z}(\lambda _{\ast })+b\exp
\{\lambda _{\ast }^{2}\}\right\} ^{2}>c_{\alpha }^{(rand)}\right) \\
&&\text{average: }P\left( \int_{0}^{1}\left\{ \mathcal{Z}(\lambda )+b\exp
\{\lambda ^{2}\}\right\} ^{2}d\lambda >c_{\alpha }^{(ave)}\right) \\
&&\text{supremum}\text{: }P\left( \sup_{\lambda \in \lbrack 0,1]}\left\{
\mathcal{Z}(\lambda )+b\exp \{\lambda ^{2}\}\right\} ^{2}>c_{\alpha }^{(\sup
)}\right) ,
\end{eqnarray*}%
where $\bar{F}_{1,0}$ is the upper tail probability of a $\chi ^{2}(1)$
distribution; $\lambda _{\ast }$ is a uniform random variable on $\Lambda $,
independent of $\{\epsilon _{t},x_{t}\}$; and $c_{\alpha }^{(\cdot )}$ are
level $\alpha $ asymptotic critical values under the null: $c_{\alpha
}^{(pvot)}$ $\equiv $ $\alpha $, and $c_{\alpha }^{rand)}$ is the $1$ $-$ $%
\alpha $ quantile from a $\chi ^{2}(1)$ distribution. See below for
approximating $\{c_{\alpha }^{(ave)},c_{\alpha }^{(\sup )}\}$.

Local power for Bierens and Ploberger's (\citeyear{BierensPloberger1997})
ICM statistic $\widehat{\mathcal{I}}_{n}$ $\equiv $ $\int_{0}^{1}\hat{z}%
_{n}^{2}(\lambda )\mu (d\lambda )$ is based on their Theorem 7 critical
value upper bound $\lim_{n\rightarrow \infty }P(\widehat{\mathcal{I}}_{n}$ $%
\geq $ $u_{\alpha }\int_{0}^{1}v_{n}^{2}(\lambda )\mu (d\lambda ))$ $\leq $ $%
\alpha $, where $v_{n}^{2}(\lambda )$ $=$ $\exp \{2\lambda ^{2}\}$ satisfies
$\sup_{\lambda \in \lbrack 0,1]}|\hat{v}_{n}^{2}(\lambda )$ $-$ $%
v_{n}^{2}(\lambda )|$ $\overset{p}{\rightarrow }$ $0$, and $%
\{u_{.01},u_{.05},u_{.10}\}$ $=$ $\{6.81$, $4.26$, $3.23\}$. We use a
uniform measure $\mu (\lambda )$ $=$ $\lambda $ since this promotes the
highest weighted average local power for alternatives near $H_{0}$ %
\citep{AndrewsPloberger1994,Boning_Sowell_99}. Under $H_{1}^{L}$ we have $\{%
\hat{z}_{n}(\lambda )\}$ $\Rightarrow ^{\ast }$ $\{z(\lambda )$ $+$ $b\exp
\{\lambda ^{2}\}\}$ for some zero mean Gaussian process $\{z(\lambda )\}$
with \emph{almost surely} uniformly continuous sample paths, and $%
\int_{0}^{1}v_{n}^{2}(\lambda )d\lambda $ $=$ $\int_{0}^{1}\exp \{2\lambda
^{2}\}d\lambda $ $=$ $2.3645$. This yields local asymptotic power:
\begin{equation*}
\text{ICM: }P\left( \int_{0}^{1}\left\{ z(\lambda )+b\exp \{\lambda
^{2}\})\right\} ^{2}d\lambda >c_{\alpha }^{(icm)}\right) \text{ where }%
c_{\alpha }^{(icm)}\equiv 2.3645\times u_{\alpha }.
\end{equation*}%
Asymptotically valid critical values can be easily computed for the present
experiment by mimicking the steps below, in which case PVOT, average,
supremum, and ICM tests are essentially identical. We are, however,
interested in how well Bierens and Ploberger's (%
\citeyear{BierensPloberger1997}) solution to the problem of non-standard
inference compares to existing methods.

Local power is computed as follows. We draw $R$ samples $\{\epsilon
_{i,t},x_{i,t}\}_{t=1}^{T}$, $i$ $=$ $1,...,R$, of iid random variables $%
(\epsilon _{i,t},x_{i,t})$ from $N(0,1)$, and draw iid $\lambda _{\ast ,i}$,
$i$ $=$ $1,...,R$, from a uniform distribution on $\Lambda $. Then $\{%
\mathcal{Z}_{T,i}(\lambda )\}$ $\equiv $ $\{1/\sqrt{T}\sum_{t=1}^{T}\epsilon
_{i,t}\exp \{\lambda x_{i,t}$ $-$ $\lambda ^{2}\}\}$ becomes a draw from the
limit process $\{\mathcal{Z}(\lambda )\}$ as $T$ $\rightarrow $ $\infty $.
We draw $R$ $=$ $100,000$ samples of size $T$ $=$ $100,000$, and compute $%
\mathcal{T}_{T,i}^{(PVOT)}(b)$ $\equiv $ $\int_{0}^{1}I(\bar{F}_{1,0}(\{%
\mathcal{Z}_{T,i}(\lambda )+b\exp \{\lambda ^{2}\}\}^{2})$ $<$ $\alpha
)d\lambda $, $\mathcal{T}_{T,i}^{(ave)}(b)$ $\equiv $ $\int_{0}^{1}\{%
\mathcal{Z}_{T,i}$ $+$ $b\exp \{\lambda ^{2}\}\}^{2}d\lambda $ and $\mathcal{%
T}_{T,i}^{(\sup )}(b)$ $\equiv $ $\sup_{\lambda \in \lbrack 0,1]}\{\mathcal{Z%
}_{T,i}(\lambda )$ $+$ $b\exp \{\lambda ^{2}\}\}^{2}$ and $\mathcal{T}%
_{T,i}^{(rand)}(b)$ $\equiv $ $\{\mathcal{Z}_{T,i}(\lambda _{\ast ,i})$ $+$ $%
b\exp \{\lambda _{\ast ,i}^{2}\}\}^{2}$. The critical values $\{c_{\alpha
}^{(ave)},c_{\alpha }^{(\sup )}$ $\}$ are the $1$ $-$ $a$ quantiles of $\{%
\mathcal{T}_{T,i}^{(ave)}(0),\mathcal{T}_{T,i}^{(\sup )}(0)\}_{i=1}^{R}$. In
the ICM case $\{z_{T,i}(\lambda )\}$ $\equiv $ $\{1/\sqrt{T}%
\sum_{t=1}^{T}\epsilon _{i,t}\exp \{\lambda x_{i,t}\}\}$ becomes a draw from
$\{z(\lambda )\}$ as $T$ $\rightarrow $ $\infty $, hence we compute $%
\mathcal{T}_{T,i}^{(icm)}(b)$ $\equiv $ $\int_{0}^{1}\{z_{T,i}$ $+$ $b\exp
\{\lambda ^{2}\}\}^{2}d\lambda $. Local power is $1/R\sum_{i=1}^{R}I(%
\mathcal{T}_{T,i}^{(\cdot )}(b)$ $>$ $c_{\alpha }^{(\cdot )})$. Integrals
are computed by the midpoint method based on the discretization $\lambda $ $%
\in $ $\{.001,.002,...,.999,1\}$, hence there are $1000$ points ($\lambda $ $%
=$ $0$ is excluded because power is trivial in that case).

Figure E.1 in the SM contains local power plots at level $\alpha $ $=$ $.05$
over drift parameters $b$ $\in $ $[0,2]$ and $b$ $\in $ $[0,7]$. Notice that
under the null $b$ $=$ $0$ each test, except ICM, achieves power of nearly
exactly $.05$ (PVOT, average and supremum are $.0499,$ and randomized is $%
.0511$), providing numerical verification that the correct critical value
for the PVOT test at level $\alpha $ is simply $\alpha $. The ICM critical
value upper bound leads to an under sized test with asymptotic size $.0365$.

Second, local power is virtually identical across PVOT, random, average and
supremum tests. This is logical since the underlying PV test is consistent
on any compact $\Lambda $ outside of a measure zero subset, it has
non-trivial local power, and local power is asymptotic. Since the average
test has the highest weighted average power aimed at alternatives near the
null \citep[eq. (2.5)]{AndrewsPloberger1994}, we have evidence that PVOT
test power is at the highest possible level. The randomized test has
slightly lower power for deviations far from the null $b$ $\geq $ $2.5$
ostensibly because for large $b$ larger values of $\lambda $ lead to a
higher power test, while the randomized $\lambda $ may be small. Finally,
ICM power is lower near the null $b$ $\in $ $(0,1.5]$ since these
alternatives are most difficult to detect, and the test is conservative, but
power is essentially identical to the remaining tests for drift $b$ $\geq $ $%
1.5$.

\section{Examples \protect\ref{ex_weak} and \protect\ref{ex_GARCH} Continued
\label{sec:examples}}

We complete the Section \ref{sec:ex_start} examples by providing relevant
theory results that verify Assumption \ref{assum:main}.

\subsection{Example \protect\ref{ex_weak}: Test of Functional Form with
Possible Weak Identification \label{sec:func_form_weak}}

Recall the regression model is $y_{t}$ $=$ $\zeta ^{\prime }x_{t}$ $+$ $%
\beta ^{\prime }g(x_{t},\pi )$ $+$ $\epsilon _{t}$ $=$ $f(\theta ,x_{t})$ $+$
$\epsilon _{t}$. We want to test $H_{0}$ $:$ $E[y_{t}|x_{t}]$ $=$ $f(\theta
_{0},x_{t})$ $a.s$. for unique $\theta _{0}$ $\in $ $\Theta $\ against $%
H_{1}:\sup_{\theta \in \Theta }P(E[y_{t}|x_{t}]$ $=$ $f(\theta ,x_{t}))$ $<$
$1$. If $\beta _{0}$ $\neq $ $0$ then $\pi _{0}$ is not identified. If there
is local drift $\beta _{0}$ $=$ $\beta _{n}$ $\rightarrow $ $0$ with $\sqrt{n%
}||\beta _{n}||$ $\rightarrow $ $[0,\infty )$, then estimators of $\pi _{0}$
have random probability limits, and estimators for $\theta _{0}$ have
nonstandard limit distributions \citep{AndrewsCheng2012}. Let $\hat{\theta}%
_{n}$ be the nonlinear least squares estimator of $\theta _{0}$ and define%
\begin{eqnarray*}
&&d_{\theta ,t}(\omega ,\pi )\equiv \left[ g(x_{t},\pi )^{\prime
},x_{t}^{\prime },\omega ^{\prime }\frac{\partial }{\partial \pi }%
g(x_{t},\pi )\right] ^{\prime }\text{ and }\mathfrak{\hat{b}}_{\theta
,n}(\omega ,\pi ,\lambda )\equiv \frac{1}{n}\sum_{t=1}^{n}F\left( \lambda
^{\prime }\Psi (x_{t})\right) d_{\theta ,t}(\omega ,\pi ) \\
&&\widehat{\mathcal{H}}_{n}=\frac{1}{n}\sum_{t=1}^{n}d_{\theta ,t}(\omega (%
\hat{\beta}_{n}),\hat{\pi}_{n})d_{\theta ,t}(\omega (\hat{\beta}_{n}),\hat{%
\pi}_{n})^{\prime }\text{ where }\omega (\beta )\equiv \left\{
\begin{array}{ll}
\beta /\left\Vert \beta \right\Vert  & \text{if }\beta \neq 0 \\
1_{k_{\beta }}/\left\Vert 1_{k_{\beta }}\right\Vert  & \text{if }\beta =0%
\end{array}%
\right.  \\
&&\hat{v}_{n}^{2}(\hat{\theta}_{n},\lambda )\equiv \frac{1}{n}%
\sum_{t=1}^{n}\epsilon _{t}^{2}(\hat{\theta}_{n})\left\{ F\left( \lambda
^{\prime }\Psi (x_{t})\right) -\mathfrak{\hat{b}}_{\theta ,n}(\omega (\hat{%
\beta}_{n}),\hat{\pi}_{n},\lambda )^{\prime }\widehat{\mathcal{H}}%
_{n}^{-1}d_{\theta ,t}(\omega (\hat{\beta}_{n}),\hat{\pi}_{n})\right\} ^{2}.
\end{eqnarray*}%
The CM statistic is $\mathcal{T}_{n}(\lambda )$ $\equiv $ $\{\hat{v}%
_{n}^{-1}(\hat{\theta}_{n},\lambda )\sum_{t=1}^{n}\epsilon _{t}(\hat{\theta}%
_{n})F\left( \lambda ^{\prime }\Psi (x_{t})\right) /\sqrt{n}\}^{2}$, which
is similar to statistics in \cite{Bierens1990} and \cite{StinchWhite1998}.
The scale $\hat{v}_{n}(\hat{\theta}_{n},\lambda )$, however, has been
altered by dividing by $||\beta ||$ in order to avoid a singular Hessian
matrix under semi-strong identification $\beta _{0}$ $=$ $0$ and $\sqrt{n}%
||\beta _{n}||$ $\rightarrow $ $\infty $
\citep[cf.][Section
3.5]{AndrewsCheng2012}.

Technical results are derived under two overlapping identification cases:
under case $\mathcal{C}(i,b)$ there is $\beta _{n}\rightarrow \beta _{0}=0$
and $\sqrt{n}\beta _{n}\rightarrow b$ where $b\in (\mathbb{R}\cup \{\pm
\infty \})^{k_{\beta }}$; and under case $\mathcal{C}(ii,\omega _{0})$, $%
\beta _{n}\rightarrow \beta _{0}$ where $\beta _{0}\gtreqless 0$, $\sqrt{n}%
\left\Vert \beta _{n}\right\Vert \rightarrow \infty ,$ and $\beta
_{n}/\left\Vert \beta _{n}\right\Vert \rightarrow \omega _{0}$ where $%
\left\Vert \omega _{0}\right\Vert =1$. Case $\mathcal{C}(i,b)$ contains
sequences $\beta _{n}$ close to zero, and when $||b||$ $<$ $\infty $ then $%
\pi _{0}$ is either weakly or non-identified. Case $\mathcal{C}(ii,\omega
_{0})$ contains sequences $\beta _{n}$ farther from zero, covering
semi-strong ($\beta _{0}$ $=$ $0$ and $\sqrt{n}||\beta _{n}||$ $\rightarrow $
$\infty $) and strong ($\beta _{0}$ $\neq $ $0$) identification for $\pi
_{0} $. Cf. \cite{AndrewsCheng2012}.

Let $\hat{p}_{n,\mathcal{M}}(\lambda )$\ be the weak identification robust
bootstrapped p-value in \cite{Hill2021_weak} based on $\mathcal{M}$\
independently drawn bootstrap samples. The PVOT is $\mathcal{\hat{P}}_{n,%
\mathcal{M}}(\alpha )$ $\equiv $ $\int_{\Lambda }I(\hat{p}_{n,\mathcal{M}%
}(\lambda )$ $<$ $\alpha )d\lambda $. The PVOT test has the correct
asymptotic level and is consistent. See \citet[Theorem
6.3]{Hill2021_weak} for a proof of the following result.

\begin{theorem}
\label{th:pvot}Let $\mathcal{M}$ $=$ $\mathcal{M}_{n}$ $\rightarrow $ $%
\infty $ as $n$ $\rightarrow $ $\infty $. Under regularity conditions
presented in \citet[Theorem 6.3]{Hill2021_weak}, if $H_{0}$ is true then $%
\lim_{n\rightarrow \infty }P(\mathcal{\hat{P}}_{n,\mathcal{M}}(\alpha )$ $>$
$\alpha )$ $\leq $ $\alpha $, and otherwise $P(\mathcal{\hat{P}}_{n,\mathcal{%
M}}(\alpha )$ $>$ $\alpha )$ $\rightarrow $ $1$.
\end{theorem}

\begin{remark}
\normalfont As stated above, there does not exist a valid bootstrap method
for handling \emph{test statistic} functionals like the average and
supremum. The bootstrap method developed in \cite{Hill2021_weak} is only
valid for computing an approximate p-value for the non-smoothed $\mathcal{T}%
_{n}(\lambda )$ that is asymptotically consistent for the asymptotic p-value %
\citep[Theorem 6.2]{Hill2021_weak}. The practitioner is therefore left with
smoothing such a p-value approximation $\hat{p}_{n,\mathcal{M}}(\lambda )$.
The supremum $\sup_{\lambda \in \Lambda }\hat{p}_{n,\mathcal{M}}(\lambda )$,
however, promotes a conservative test that is not consistent. Even though $%
\hat{p}_{n,\mathcal{M}}(\lambda )$ $\overset{p}{\rightarrow }$ $0$ $\forall
\lambda $ $\in $ $\Lambda /S$ where $S$ has Lebesgue measure zero, as long
as there exists $\lambda $ $\in $ $\Lambda $ such that a Type II error
occurs, i.e. $\hat{p}_{n,\mathcal{M}}(\lambda )$ $\overset{p}{\rightarrow }$
$(0,1]$, then $\sup_{\lambda \in \Lambda }\hat{p}_{n,\mathcal{M}}(\lambda )$
$\overset{p}{\rightarrow }$ $(0,1]$ and the sup-p-value test is
inconsistent. Conversely, the PVOT test with $\hat{p}_{n,\mathcal{M}%
}(\lambda )$ is both consistent and immune to weak identification,
asymptotically with probability approaching one.
\end{remark}

\subsection{Example \protect\ref{ex_GARCH}: Test of GARCH Effects \label%
{sec:garch}}

Recall the GARCH process $y_{t}$ $=$ $\sigma _{t}\epsilon _{t}$ where $%
\sigma _{t}^{2}$ $=$ $\omega _{0}$ $+$ $\delta _{0}y_{t-1}^{2}$ $+$ $\lambda
_{0}\sigma _{t-1}^{2}$, $\omega _{0}$ $>$ $0$, and $\delta _{0},\lambda
_{0}\in \lbrack 0,1)$. The unrestricted QML estimator of $\delta _{0}$ for a
given $\lambda $ $\in $ $\Lambda $ is $\hat{\delta}_{n}(\lambda )$, and the
test statistic is $\mathcal{T}_{n}(\lambda )=n\hat{\delta}_{n}^{2}(\lambda )$
\citep{Andrews1999}. We first show the limit distribution for $\mathcal{T}%
_{n}(\lambda )$ is a one-sided normal.

\begin{theorem}
\label{th:garch}Let $\{y_{t}\}$ be generated by process (\ref{garch}).
Assumption \ref{assum:main} applies where $\mathcal{T}(\lambda )$ $=$ $(\max
\{0,\mathcal{Z}(\lambda )\})^{2}$, and $\{\mathcal{Z}(\lambda )\}$ is a zero
mean Gaussian process with a version that has \emph{almost surely} uniformly
continuous sample paths, and covariance function $E[\mathcal{Z(}\lambda _{1})%
\mathcal{Z(}\lambda _{2})]$ $=$ $(1$ $-$ $\lambda _{1}^{2})(1$ $-$ $\lambda
_{2}^{2})/(1$ $-$ $\lambda _{1}\lambda _{2})$.
\end{theorem}

A simulation procedure can be used to approximate the asymptotic p-value %
\citep[cf.][]{Andrews2001}. Draw $\widetilde{\mathcal{M}}$ $\in $ $\mathbb{N}
$ samples of iid standard normal random variables $\{Z_{j,i}\}_{j=1}^{%
\widetilde{\mathcal{R}}}$, $i$ $=$ $1,...,\widetilde{\mathcal{M}}$, and
compute $\mathfrak{Z}_{\widetilde{\mathcal{R}},i}(\lambda )$ $\equiv $ $(1$ $%
-$ $\lambda ^{2})\sum_{j=0}^{\widetilde{\mathcal{R}}}\lambda ^{j}Z_{j,i}$
and $\mathcal{T}_{\widetilde{\mathcal{R}},i}(\lambda )$ $\equiv $ $(\max \{0,%
\mathfrak{Z}_{\widetilde{\mathcal{R}},i}(\lambda )\})^{2}$. Notice $%
\mathfrak{Z}_{\widetilde{\mathcal{R}}}(\lambda )$ $\equiv $ $(1$ $-$ $%
\lambda ^{2})\sum_{j=0}^{\widetilde{\mathcal{R}}}\lambda ^{j}Z_{j}$ is zero
mean Gaussian with the same covariance function as $\mathcal{Z}(\lambda )$
when $\widetilde{\mathcal{R}}$ $=$ $\infty $, hence $\{\mathcal{T}_{\infty
,i}(\lambda )$ $:$ $\lambda $ $\in $ $\Lambda \}$ is an independent draw
from the limit process $\{\mathcal{T}(\lambda )$ $:$ $\lambda $ $\in $ $%
\Lambda \}$. The p-value approximation is $\hat{p}_{\widetilde{\mathcal{R}},%
\widetilde{\mathcal{M}},n}(\lambda )$ $\equiv $ $1/\widetilde{\mathcal{M}}%
\sum_{i=1}^{\widetilde{\mathcal{M}}}I(\mathcal{T}_{\widetilde{\mathcal{R}}%
,i}(\lambda )$ $>$ $\mathcal{T}_{n}(\lambda ))$. Since we can choose $%
\widetilde{\mathcal{M}}$ and $\widetilde{\mathcal{R}}$ to be arbitrarily
large, we can make $\hat{p}_{\widetilde{\mathcal{R}},\widetilde{\mathcal{M}}%
,n}(\lambda )$ arbitrarily close (in probability) to the asymptotic p-value
by the Glivenko-Cantelli theorem. Now compute the PVOT $\mathcal{P}_{%
\widetilde{\mathcal{R}},\widetilde{\mathcal{M}},n}^{\ast }(\alpha )$ $\equiv
$ $\int_{\Lambda }I(\hat{p}_{\widetilde{\mathcal{R}},\widetilde{\mathcal{M}}%
,n}(\lambda )$ $<$ $\alpha )d\lambda $.

\begin{theorem}
\label{th:garch_p_val}Let $\{y_{t}\}$ be generated by (\ref{garch}), and let
$\{\widetilde{\mathcal{R}}_{n},\widetilde{\mathcal{M}}_{n}\}_{n\geq 1}\ $be
sequences of positive integers, $\widetilde{\mathcal{R}}_{n}$ $\rightarrow $
$\infty $ and $\widetilde{\mathcal{M}}_{n}$ $\rightarrow $ $\infty $. If $%
H_{0}$: $\delta _{0}$ $=$ $0$ is true then $\lim_{n\rightarrow \infty }P(%
\mathcal{P}_{\widetilde{\mathcal{R}}_{n},\widetilde{\mathcal{M}}%
_{n},n}^{\ast }(\alpha )$ $>$ $\alpha )$ $\in $ $(0,\alpha ]$. Otherwise if $%
\delta _{0}$ $>$ $0$ then $P(\mathcal{P}_{\widetilde{\mathcal{R}}_{n},%
\widetilde{\mathcal{M}}_{n},n}^{\ast }(\alpha )$ $>$ $\alpha )$ $\rightarrow
$ $1$.
\end{theorem}

\begin{remark}
\normalfont\label{rm:p_boot_h}Under $H_{0}$, $h(\mathcal{T}_{n}(\lambda ))$ $%
\overset{d}{\rightarrow }$ $h(\mathcal{T}(\lambda ))$ for mappings $h$ $:$ $%
\mathbb{R}$ $\rightarrow $ $\mathbb{R}$, continuous \emph{a.e.}, by
exploiting theory in \citet[Section 4]{Andrews2001}. The relevant simulated
p-value is $\hat{p}_{\widetilde{\mathcal{R}},\widetilde{\mathcal{M}}%
,n}^{(h)} $ $\equiv $ $1/\widetilde{\mathcal{M}}\sum_{i=1}^{\widetilde{%
\mathcal{M}}}I(h(\mathcal{T}_{\widetilde{\mathcal{R}},i}(\lambda ))$ $>$ $%
h\left( \mathcal{T}_{n}(\lambda )\right) )$. Arguments used to prove Theorem %
\ref{th:garch_p_val} easily lead to a proof that $\hat{p}_{\widetilde{%
\mathcal{R}},\widetilde{\mathcal{M}},n}^{(h)}$ is consistent for the
corresponding asymptotic p-value.
\end{remark}

\section{Simulation Study\label{sec:sim}}

We perform three Monte Carlo experiments concerning tests of functional form
with and without the possibility of weak identification, and GARCH effects.
The same discretized $\Lambda $ is used for PVOT and bootstrap p-value
tests, and integrals are discretized using the midpoint method. Wild
bootstrapped p-values are computed with $R$ $=$ $1000$ samples of iid
standard normal random variables $\{z_{t,i}\}_{t=1}^{n}$. Sample sizes are $%
n $ $\in $ $\{100,250,500\}$ and $10,000$ samples $\{y_{t}\}_{t=1}^{n}$ are
independently drawn in each case. Nominal levels are $\alpha $ $\in $ $%
\{.01,.05.,.10\}$.

\subsection{Test of Functional Form}

We work with a threshold process in which all parameters are strongly
identified.

\subparagraph{Step-Up}

Samples $\{y_{t}\}_{t=1}^{n}$ are drawn from one of four data generating
processes: linear $y_{t}$ $=$ $2x_{t}$ $+$ $\epsilon _{t}$\ or quadratic $%
y_{t}$ $=$ $2x_{t}$ $+$ $.1x_{t}^{2}$ $+$ $\epsilon _{t}$, where $%
\{x_{t},\epsilon _{t}\}$ are iid standard normal random variables; and AR(1)
$y_{t}$ $=$ $.9x_{t}$ $+$ $\epsilon _{t}$\ or Self-Exciting Threshold AR(1) $%
y_{t}$ $=$ $.9x_{t}$ $-$ $.4x_{t}I(x_{t}$ $>$ $0)$ $+$ $\epsilon _{t}$,
where $x_{t}$ $=$ $y_{t-1}$ and $\epsilon _{t}$ is iid standard normal. In
the time series cases we draw $2n$ observations with starting values $y_{1}$
$=$ $\epsilon _{1}$ and retain the last $n$ observations. Now write $\sum $
for sample summations: for iid data $\sum $ $=$ $\sum_{t=1}^{n}$ and for
time series $\sum $ $=$ $\sum_{t=2}^{n}$. The estimated model is $y_{t}$ $=$
$\beta x_{t}$ $+$ $\epsilon _{t}$, and we test $H_{0}$ $:$ $E[y_{t}|x_{t}]$ $%
=$ $\beta _{0}x_{t}$ $a.s.$ for some $\beta _{0}$.

We compute $\mathcal{T}_{n}(\lambda )$ in (\ref{Tn_CM}) with logistic $%
F(\Psi (x_{t}))$ $=$ $(1$ $+$ $\exp \{\Psi (x_{t})\})^{-1}$ and $\Psi
(x_{t}) $ $=$ $\arctan (x_{t}^{\ast })$, where $x_{t}^{\ast }$ $\equiv $ $%
x_{t}$ $-$ $1/n\sum x_{t}$. Write $F_{t}(\lambda )$ $=$ $F(\lambda \Psi
(x_{t}))$, let $\hat{\beta}_{n}$ be the least squares estimator, and define $%
\hat{z}_{n}(\lambda )$ $\equiv $ $1/n^{1/2}\sum (y_{t}$ $-$ $\hat{\beta}%
_{n}x_{t})F_{t}(\lambda )$. Then $\mathcal{T}_{n}(\lambda )$ $\equiv $ $\hat{%
z}_{n}^{2}(\lambda )/\hat{v}_{n}^{2}(\lambda )$ with $\hat{v}%
_{n}^{2}(\lambda )$ $\equiv $ $1/n\sum (y_{t}$ $-$ $\hat{\beta}_{n}x_{t})^{2}%
\hat{w}_{n,t}^{2}(\lambda )$, where $\hat{w}_{n,t}(\lambda )$ $\equiv $ $%
F_{t}(\lambda )$ $-$ $\hat{b}_{n}(\lambda )^{\prime }\hat{A}_{n}^{-1}x_{t}$,
$\hat{b}_{n}$ $\equiv $ $1/n\sum x_{t}F_{t}(\lambda )$ and $\hat{A}_{n}$ $%
\equiv $ $1/n\sum x_{t}x_{t}^{\prime }$
\citep[see][cf. Bierens,
1990]{White1989}. It is straightforward to show Assumption \ref%
{assum:locpow_suff}.a,b,c,e holds, and $\sup_{\lambda \in \Lambda }|\hat{v}%
_{n}^{2}(\lambda )$ $-$ $v^{2}(\lambda )|$ $\overset{p}{\rightarrow }$ $0$
by arguments used to prove Lemma C.1 in the SM. By Theorem \ref%
{lm:locpow_suff}, weak convergence (\ref{TZc}) therefore applies, and $%
\mathcal{T}_{n}(\lambda )$ is pointwise asymptotically $\chi ^{2}(1)$ under $%
H_{0}$.

\subparagraph{Tests}

We perform four tests. First, the PVOT over $\Lambda $ $=$ $[.0001,1]$ based
on the asymptotic p-value for $\mathcal{T}_{n}(\lambda )$. The discretized
set is $\Lambda _{n}$ $\equiv $ $\{.0001$ $+$ $1/(\varpi n)$, $.0001$ $+$ $%
2/(\varpi n),$ $...,$ $.0001$ $+$ $\bar{\imath}_{n}(\varpi )/(\varpi n)\}$
where $\bar{\imath}_{n}(\varpi )$ $\equiv $ $\argmax\{1$ $\leq $ $i$ $\leq $
$\varpi n$ $:$ $i$ $\leq $ $.9999\varpi n\}$, with a coarseness parameter $%
\varpi $ $=$ $100$. We can use a much smaller $\varpi $ if the sample size
is large enough (e.g. $\varpi $ $=$ $10$\ when $n$ $=$ $250$, or $\varpi $ $%
= $ $1$\ when $n$ $\geq $ $500$), but in general small $\varpi n$ leads to
over-rejection of $H_{0}$. Second, we use $\mathcal{T}_{n}(\lambda _{\ast })$
with a uniformly randomized $\lambda _{\ast }$ $\in $ $\Lambda $ and an
asymptotic p-value. Third, $\sup_{\lambda \in \Lambda _{n}}\mathcal{T}%
_{n}(\lambda )$ and $\int_{\Lambda _{n}}\mathcal{T}_{n}(\lambda )\mu
(d\lambda )$\ with uniform measure $\mu (\lambda )$, and wild bootstrapped
p-values. Fourth, Bierens and Ploberger's (\citeyear{BierensPloberger1997})
ICM $\widehat{\mathcal{I}}_{n}$ $\equiv $ $\int_{\Lambda _{n}}\hat{z}%
_{n}^{2}(\lambda )\mu (d\lambda )$ with uniform $\mu (\lambda )$, and the
critical value upper bound $c_{\alpha }$ $\int_{\Lambda }\hat{v}%
_{n}^{2}(\lambda )\mu (d\lambda )$, where $\{c_{.01},c_{.05},c_{.10}\}$ $=$ $%
\{6.81$, $4.26$, $3.23\}$ \citep[Section
6]{BierensPloberger1997}.

\subparagraph{Results}

Rejection frequencies for $\alpha $ $\in $ $\{.01,.05,.10\}$ are reported in
Table \ref{tbl:funcform}. The ICM test tends to be under sized, which is
expected due to the critical value upper bound. Randomized, average and
supremum tests have accurate empirical size for iid data, but exhibit size
distortions for time series data when $n$ $\in $ $\{100,250\}$. The PVOT
test has relatively sharp size in nearly every case, but is slightly
over-sized for time series data when $n$ $=$ $100$.

All tests except the supremum test have comparable power, while the ICM test
has low power at the $1\%$ level. The supremum test has the lowest power,
although its local power was essentially identical to the average and PVOT
tests for a similar test of omitted nonlinearity (see Section \ref%
{sec:local_num}). In the time series case, however, PVOT power when $n$ $=$ $%
100$ is lower than all other tests, except the supremum test in general and
the ICM test at level $\alpha $ $=$ $.01$. PVOT rejection frequencies are $%
\{.135,.206,.645\}$ for tests at levels $\{.01,.05,.10\}$, while randomized,
average, supremum and ICM power are $\{.135,.592,.846\}$, $%
\{.062,.412,.726\} $, $\{.021,.209,.561\}$ and $\{.004,.643,.866\}$\
respectively. These discrepancies, however, vanish when $n$ $\in $ $%
\{250,500\}$. The ICM test has dismal power at the $1\%$ level when $n$ $%
\leq $ $250$ and much lower power than all other tests when $n$ $=$ $500$,
but comparable or better power at levels $5\%$ and $10\%$. In summary,
across cases the various tests are comparable; supremum test power is
noticeably lower in many cases; and the PVOT test generally exhibits fewer
size distortions, and competitive or high power in nearly every case.

Of particular note, the accuracy of PVOT size provides further evidence that
the PVOT asymptotic critical value is identically $\alpha $. Finally, when $%
n $ $=$ $100$ the PVOT test took on average $.0085$ minutes ($.51$ seconds),
while the bootstrapped average or supremum test took $8.07$ minutes on
average. The 1000-fold increase is due to the number of bootstrap samples.
This demonstrates the PVOT test computational convenience, arising entirely
from its asymptotic critical value (upper bound) being the test level $%
\alpha $.

\subsection{Test of Functional Form with Weak Identification\label%
{sec:sim_weak}}

We now work with a Smooth Transition Autoregression [STAR], allowing for
weak identification. The following summarize the monte carlo study in \cite%
{Hill2021_weak}.\medskip

\subparagraph{Step-Up}

The data are drawn from:%
\begin{equation*}
y_{t}=\zeta _{0}y_{t-1}+\beta _{n}y_{t-1}\frac{1}{1+\exp \left\{ -10\left(
y_{t-1}-\pi _{0}\right) \right\} }+\varpi _{0}\frac{1}{1+y_{t-1}^{2}}%
+\epsilon _{t},
\end{equation*}%
where $\epsilon _{t}$ is iid $N(0,1)$. If $\varpi _{0}$ $=$ $0$ then $y_{t}$
is a Logistic STAR process and the null hypothesis is true. If $\beta _{n}$ $%
\rightarrow $ $0$ too quickly then $\pi _{0}$ cannot be identified and
estimation asymptotics are non-standard. We use $\zeta _{0}$ $=$ $.6$, $\pi
_{0}$ $=$ $0$ and $\varpi _{0}$ $\in $ $\{0,.03,.3\}$, the latter allowing
for weak and strong degrees of deviation from the null. We use $\beta _{n}$ $%
\in $ $\{.3,.3/\sqrt{n},0\}$ representing strong identification, weak
identification with $\sqrt{n}\beta _{n}$ $=$ $.3$ and $\beta _{n}$ $%
\rightarrow $ $\beta _{0}$ $=$ $0$, and non-identification with $\beta _{n}$
$=$ $\beta _{0}$ $=$ $0$.

Let $\iota $ $=$ $10^{-10}$. The estimated parameters satisfy $\beta _{n}$ $%
\in $ $\mathcal{B}^{\ast }$, $\zeta _{0}$ $\in $ $\mathcal{Z}^{\ast }(\beta
) $ and $\pi _{0}$ $\in $ $\Pi ^{\ast }$. The true parameter spaces are $%
\mathcal{B}^{\ast }$ $=$ $[-1$ $+$ $2\iota ,1$ $-$ $2\iota ]$, $\mathcal{Z}%
^{\ast }(\beta )$ $=$ $[-1-\beta $ $+$ $\iota <\zeta <1-\beta $ $-$ $\iota ]$%
, and $\Pi ^{\ast }$ $=$ $[-1,1]$. The estimation spaces are $\mathcal{B}$ $%
= $ $[-1$ $+$ $\iota ,1$ $-$ $\iota ]$, $\mathcal{Z}(\beta )$ $=$ $[-1-\beta
<\zeta <1-\beta ]$, and $\Pi $ $=$ $[-2,2]$. Thus $|\zeta $ $+$ $\beta |$ $<$
$1$ on $\Theta $ $\equiv $ $\mathcal{B}$ $\times $ $\mathcal{Z}(\beta )$ $%
\times $ $\Pi $, which ensures stationarity
\citep[see][Theorem
1]{BhattacharyaLee1995}.

We draw $100$ start values uniformly on $\Theta $ and estimate $\theta
_{0}=[\zeta _{0},\beta _{0},\pi _{0}]^{\prime }$ by least squares for each
start value, resulting in $\{\hat{\theta}_{n,i}\}_{i=1}^{100}$. The final $%
\hat{\theta}_{n}$ minimizes the least squares criterion over $\{\hat{\theta}%
_{n,i}\}_{i=1}^{100}$.\footnote{%
Computation is performed using Matlab R2016. An analytic gradient is used
for optimization. The criterion tolerance for ceasing iterations is $1e^{-8}$%
, and the maximum number of allowed iterations is $20,000$.} We also require
$\hat{\sigma}_{n}^{2}$ $=$ $1/n\sum_{t=2}^{n}(y_{t}-\hat{\zeta}_{n}y_{t-1}$ $%
-$ $\hat{\beta}_{n}y_{t-1}(1$ $+$ $\exp \left\{ -10\left( y_{t-1}-\hat{\pi}%
_{n}\right) \right\} )^{-1})^{2}$. Notice $\hat{\sigma}_{n}^{2}$ $\overset{p}%
{\rightarrow }$ $\sigma ^{2}$ under mild conditions and any degree of
(non)identification: if $\hat{\beta}_{n}$ $\overset{p}{\rightarrow }$ $0$
fast enough then the non-standard limit properties of $\hat{\pi}_{n}$\ are
irrelevant \citep[see][Theorem 4.1 and Remark 7]{Hill2021_weak}.

The test weight $\mathcal{F}(u)$ $=$ $1/(1$ $+$ $\exp \{u\})$, and $\mathcal{%
F}(\lambda ^{\prime }\Psi (x_{t}))$ uses the bounded one-to-one transform $%
\Psi (x)$ $=$ atan$(x)$ \citep[e.g.][p. 1445,
1453]{Bierens1990}. The parameter space is $\Lambda $ $=$ $[1,5]$,
discretized as $\Lambda _{n}$ with endpoints $\{1,5\}$ and equal increments
with $n$ elements (e.g. $\Lambda _{100}$ $=$ $\{1,$ $1.04,$ $1.08,...,$ $5$).

\subparagraph{Tests}

We perform eleven tests. The first five are not robust to weak
identification: $(i)$ uniformly randomize $\lambda ^{\ast }$ on $\Lambda $,
compute $\mathcal{T}_{n}(\lambda ^{\ast })$ and use $\chi ^{2}(1)$ for
p-value computation; $(ii)$ $\sup_{\lambda \in \Lambda _{n}}p_{n}(\lambda )$%
; $(iii)$ $\sup_{\lambda \in \Lambda _{n}}\mathcal{T}_{n}(\lambda )$ and $%
(iv)$ $\int_{\Lambda _{n}}\mathcal{T}_{n}(\lambda )\mu (d\lambda )$\ where $%
\mu $ is the uniform measure on $\Lambda $, and p-values are computed by
wild bootstrap; and $(v)$ the PVOT test using $\Lambda _{n}$, and a p-value
computed from the $\chi ^{2}(1)$ distribution.

The final six tests are robust based on the bootstrapped p-value procedure
in \cite{Hill2021_weak}. We compute $\mathcal{T}_{n}(\lambda ^{\ast })$
using ($vi$) the plug-in least-favorable [LF] and ($vii$) plug-in
Identification Category Selection Type 1 [ICS-1] p-values from
\citep[Sections 5 and
6]{Hill2021_weak}; $\sup_{\lambda \in \Lambda _{n}}p_{n}(\lambda )$ using ($%
vii$) the plug-in LF and ($ix$) plug-in ICS-1 p-values; and PVOT using ($x$%
)\ the plug-in LF and ($xi$) plug-in ICS-1 p-values. See
\citet[Section
7]{Hill2021_weak} for details on p-value computation for the present
experiment.

\subparagraph{Results}

Table \ref{tbl:starn100} contains rejection frequencies. All tests are
fairly comparable under strong identification $\beta _{n}$ $=$ $.3$. By
construction the LF p-values are larger than the ICS-1 p-values, which are
larger than the $\chi ^{2}$ p-values. This results in lower rejection rates
even under strong identification. The sup-p-value test is conservative by
construction, with comparatively smaller rejection rates.

Under weak and non-identification most non-robust tests over reject the null
hypothesis, and most distortions are comparatively large. Ironically, the
non-robust $\sup_{\lambda \in \Lambda _{n}}p_{n}(\lambda )$ is relatively
large, which pushes that test's rejection frequencies down. While this
inadvertently compensates for a potentially large size distortion, it leads
to lower empirical power.

The sole test that both controls for weak identification and obtains
relatively high power is the PVOT test with ICS-1 p-values. The PVOT test
with LF p-values also works well, but tends to have lower power than the
ICS-1 based PVOT test. This follows since the LF p-values are larger than
the ICS-1 p-values.

\subsection{Test of GARCH Effects\label{sec:sim_garch}}

\subparagraph{Setp-Up}

Samples $\{y_{t}\}_{t=1}^{n}$ are drawn from a GARCH process $y_{t}$ $=$ $%
\sigma _{t}\epsilon _{t}$ and $\sigma _{t}^{2}$ $=$ $\omega _{0}$ $+$ $%
\delta _{0}y_{t-1}^{2}$ $+$ $\lambda _{0}\sigma _{t-1}^{2}$ with parameter
values $\omega _{0}$ $=$ $1$, $\lambda _{0}$ $=$ $.6$, and $\delta _{0}$ $=0$
or $.3$, where $\epsilon _{t}$ is iid $N(0,1)$. The initial condition is $%
\sigma _{0}^{2}$ $=$ $\omega _{0}/(1$ $-$ $\lambda _{0})$ $=$ $2.5$.
Simulation results are qualitatively similar for other values $\lambda _{0}$
$\in $ $(0,1)$. Put $\Lambda $ $=$ $[.01,.99]$ with discretization $\Lambda
_{n}$ $\equiv $ $\{.01+1/(\varpi n),.01+2/(\varpi n),...,.01$ $+$ $\bar{%
\imath}_{n}(\varpi )/(\varpi n)\}$, where $\bar{\imath}_{n}(\varpi )$ $%
\equiv $ $\argmax\{1$ $\leq $ $i$ $\leq $ $\varpi n$ $:$ $i$ $\leq $ $%
.98\varpi n\}$, with coarseness $\varpi $ $=$ $1$. Hence there are $\mathcal{%
N}_{n}$ $\approx $ $n$ $-$ $1$ points in $\Lambda _{n}$. A finer grid based
on $\varpi $ $=$ $10$ or $100$, for example, leads to improved empirical
size at the 1\% level for the PVOT test, and more severe size distortions
for the supremum test. The cost, however, is computation time since a QML
estimator \textit{and} bootstrapped p-value are required for each sample. We
estimate $\pi _{0}$ $=$ $[\omega _{0},\delta _{0}]^{\prime }$ by QML for
fixed $\lambda $ $\in $ $\Lambda _{n}$, with criterion $Q_{n}(\pi ,\lambda )$
$=$ $\sum \{\ln \sigma _{t}^{2}(\pi ,\lambda )$ $+$ $y_{t}^{2}/\sigma
_{t}^{2}(\pi ,\lambda )\}$ where $\sigma _{t}^{2}(\pi ,\lambda )$ $=$ $%
\omega $ $+$ $\alpha y_{t-1}^{2}$ $+$ $\lambda \sigma _{t-1}^{2}(\pi
,\lambda )$, and $\sigma _{0}^{2}(\pi ,\lambda )$ $=$ $\omega /(1$ $-$ $%
\lambda )$. The estimator is $\hat{\pi}_{n}(\lambda )=[\hat{\omega}%
_{n}(\lambda ),\hat{\delta}_{n}(\lambda )]^{\prime }$ $=$ $\arg \min_{\pi
\in \Pi }Q_{n}(\pi ,\lambda )$ with space $\Pi $ $=$ $[.001,2]$ $\times $ $%
[0,.99]$.\footnote{%
We compute $\hat{\pi}_{n}(\lambda )$ using Matlab's built-in \textit{fmincon}
routine for constrained optimization, with numerical approximations for the
first and second derivatives. We cease computation iterations when the
numerical gradient, or the difference in the current and previous iteration
of $\hat{\pi}_{n}(\lambda )$, is less than $.0001$. The initial parameter
value is a uniform random uniform draw on $\Pi $.} The test statistic is $%
\mathcal{T}_{n}(\lambda )$ $=$ $n\hat{\delta}_{n}(\lambda )^{2}$, and the
p-value approximation $\hat{p}_{\widetilde{\mathcal{R}},\widetilde{\mathcal{M%
}},n}(\lambda )$ is computed by the method in Section \ref{sec:garch} with $%
\widetilde{\mathcal{M}}$ $=$ $10,000$ simulated samples of size $\widetilde{%
\mathcal{R}}$ $=$ $25,000$.

\subparagraph{Tests}

We handle the nuisance parameter $\lambda $ by uniformly randomizing it on $%
\Lambda $; computing the PVOT; and computing $\sup_{\lambda \in \Lambda }%
\mathcal{T}_{n}(\lambda )$ and $\int_{\Lambda }\mathcal{T}_{n}(\lambda )\mu
(d\lambda )$, along with corresponding simulation-based bootstrapped
p-values $\hat{p}_{\widetilde{\mathcal{R}},\widetilde{\mathcal{M}}%
,n}^{(\cdot )}$ detailed in Remark \ref{rm:p_boot_h}.

\subparagraph{Results}

Consult Table \ref{tbl:garch} for simulation results. The randomized test
under rejects the null, and has lower size adjusted power than the remaining
tests. Andrews' (2001)\nocite{Andrews2001} proposed supremum test is highly
over-sized, resulting in relatively low size adjusted power. The best tests
in terms of size and size adjusted power are the PVOT and average tests. The
average test tends to under reject the null at each level for sample sizes $%
n $ $\in $ $\{100,250\}$, and the PVOT test tends to over reject the null at
the $1\%$ level for $n$ $\in $ $\{100,250\}$. Recall the average test has
the highest weighted average power for alternatives near the null %
\citep{AndrewsPloberger1994}, hence the PVOT test performs on par with, or
is slightly better than, an optimal test (depending on $n$ and $\alpha $).
Finally, the PVOT size performance suggests the asymptotic critical value is
$\alpha $. The PVOT, average and supremum tests are roughly equal in terms
of computational cost due to the simulation procedure required for computing
the p-value. See Remark \ref{rm:p_boot_h}.

\section{Conclusion\label{sec:conclusion}}

\cite{HillAguilar13} and \cite{Hill_white_2012} develop the p-value
occupation time [PVOT] to smooth over a trimming tuning parameter. The idea
is extended here to tests when a nuisance parameter is present under the
alternative, and complete asymptotic theory is developed for the first time.
In the SM we show in a likelihood setting that the PVOT is a point estimate
of the weighted average rejection probability of the PV test, evaluated
under the null, making the PVOT a natural object of interest for hypothesis
testing when nuisance parameters are present. By construction, a critical
value upper bound for the PVOT test is the nominal significance level $%
\alpha $, making computation and interpretation very simple, and much easier
to perform than standard transforms like the average or supremum since these
typically require a bootstrapped p-value. If the original test is consistent
on a subset of $\Lambda $ with Lebesgue measure greater than $\alpha $ then
so is the PVOT test. Moreover, the PVOT form of smoothing naturally accepts
weak identification robust p-values, while conventionally smoothed test
statistics cannot be consistently bootstrapped under weak identification.
Indeed, evidently only the PVOT test with a weak identification robust
p-value achieves both accurate level and high power. We are not aware of any
other test statistic construction that allows for nuisance parameter
smoothing that is both robust to weak identification \textit{and} not
conservative. Interesting future work may include studying the PVOT test
when the data generating process is not encompassed under either hypothesis,
or looking at how pre-order selection, or the particular model
filter/estimator, may affect its performance.

\setcounter{equation}{0} \renewcommand{\theequation}{{\thesection}.%
\arabic{equation}} \appendix\onehalfspacing

\section{Appendix: Proofs\label{sec:proofs}}

\noindent \textbf{Proof of Theorem \ref{th:main}} Write $\{\mathcal{T}%
_{n}(\lambda )\}$ $=$ $\{\mathcal{T}_{n}(\lambda )$ $:$ $\lambda $ $\in $ $%
\Lambda \}$, etc. By Assumption \ref{assum:main}, $\{\mathcal{T}_{n}(\lambda
)\}$ $\Rightarrow ^{\ast }$ $\{\mathcal{T}(\lambda )\}$ under $H_{0}$, a
process with a version that has \emph{almost surely} bounded uniformly
continuous sample paths with respect to the sup-norm, where $\mathcal{T}%
(\lambda )$ has a continuous distribution function $F_{0}(c)$ $\equiv $ $P(%
\mathcal{T}(\lambda )$ $\leq $ $c)$ that is not a function of $\lambda $.
Hence by the continuous mapping theorem $\{\bar{F}_{0}(\mathcal{T}%
_{n}(\lambda ))\}$ $\Rightarrow ^{\ast }$ $\{\bar{F}_{0}(\mathcal{T}(\lambda
))\}$, where $\bar{F}_{0}(\cdot )$ $\equiv $ $1$ $-$ $F_{0}(\cdot )$, and $\{%
\bar{F}_{0}(\mathcal{T}(\lambda ))\}$ has a version with \emph{almost surely}
bounded uniformly continuous sample paths with respect to the sup-norm
\citep[e.g.][Theorem
2.7]{Billingsley1999}.

Furthermore, $\sup_{\lambda \in \Lambda }|p_{n}(\lambda )$ $-$ $\bar{F}_{0}(%
\mathcal{T}_{n}(\lambda ))|$ $\overset{p}{\rightarrow }$ $0$ by Assumption %
\ref{assum:main}.b, hence $\{p_{n}(\lambda ))\}$ $\Rightarrow ^{\ast }$ $\{%
\bar{F}_{0}(\mathcal{T}(\lambda ))\}$. By distribution continuity, $\mathcal{%
U}(\lambda )$ $\equiv $ $\bar{F}_{0}(\mathcal{T}(\lambda ))$ is for each $%
\lambda $ $\in $ $\Lambda $ uniformly distributed on $[0,1]$, and from above
$\{\mathcal{U}(\lambda )\}$ has a version with \emph{almost surely} bounded
uniformly continuous sample paths. Therefore the mapping $\mathcal{U}(\cdot
) $ $\mapsto $ $\int_{\Lambda }I(\mathcal{U}(\lambda )$ $<$ $\alpha
)d\lambda $ is continuous with probability one due to \emph{almost sure}
bounded continuity of the sample paths $\{\mathcal{U}(\lambda )\}$ and that
weak convergence is on $l_{\infty }(\Lambda )$ which is endowed with the
sup-norm. The continuous mapping theorem therefore yields $\mathcal{P}%
_{n}^{\ast }(\alpha )$ $=$ $\int_{\Lambda }I(p_{n}(\lambda )$ $<$ $\alpha
)d\lambda $ $\overset{d}{\rightarrow }$ $\int_{\Lambda }I(\mathcal{U}%
(\lambda )$ $<$ $\alpha )d\lambda $
\citep[Theorem IV.2.12, cf.
p.66-70]{Pollard1984}. Now use Lemma \ref{lm:U}, below, to yield $%
P(\int_{\Lambda }I(\mathcal{U}(\lambda )$ $<$ $\alpha )d\lambda $ $>$ $%
\alpha )$ $\leq $ $\alpha $ and each remaining claim. $\mathcal{QED}$.

\begin{lemma}
\label{lm:U}Let $\{\mathcal{U}(\lambda )$ $:$ $\lambda $ $\in $ $\Lambda \}$
be a stochastic process where $\mathcal{U}(\lambda )$ is distributed uniform
on $[0,1]$, and $\int_{\Lambda }d\lambda $ $=$ $1$. Then $(a)$ $%
P(\int_{\Lambda }I(\mathcal{U}(\lambda )$ $<$ $\alpha )d\lambda $ $>$ $%
\alpha )$ $\leq $ $\alpha $. In particular, $(b)$ if $\mathcal{U}(\lambda )$
$=$ $\mathcal{U}(\lambda ^{\ast })$ $=$ $a.s.$ $\forall \lambda $ $\in $ $%
\Lambda $ and some $\lambda ^{\ast }$ $\in $ $\Lambda $\ then $%
P(\int_{\Lambda }I(\mathcal{U}(\lambda )$ $<$ $\alpha )d\lambda $ $>$ $%
\alpha )$ $=$ $\alpha $. Finally, $(c)$ if $P(\mathcal{U}(\lambda )$ $<$ $%
\alpha ,\mathcal{U}(\tilde{\lambda})$ $<$ $\alpha )$ $>$ $\alpha ^{2}$ for
all couplets $(\lambda ,\tilde{\lambda})$ on a subset of $\Lambda $ $\times $
$\Lambda $ with positive measure, then $P(\int_{\Lambda }I(\mathcal{U}%
(\lambda )$ $<$ $\alpha )d\lambda $ $>$ $\alpha )$ $>0$.
\end{lemma}

\begin{remark}
\normalfont The key proof that $P(\int_{\Lambda }I(\mathcal{U}(\lambda )$ $<$
$\alpha )d\lambda $ $>$ $\alpha )$ $\leq $ $\alpha $ exploits a variation of
the Bernstein inequalities. If we know $\mathcal{U}(\lambda )$ is perfectly
dependent across $\lambda $ then the bound is exact.
\end{remark}

\noindent \textbf{Proof.\qquad }Let $\mathcal{P}$ $\equiv $ $\int_{\Lambda
}I(\mathcal{U}(\lambda )$ $<$ $\alpha )d\lambda $, where $\mathcal{P}$ $\in $
$[0,1]$ since $\int_{\Lambda }d\lambda $ $=$ $1$. In order to prove ($a$),
use Markov's inequality (cf. the Chermoff bound variation of the Bernstein
inequality) to yield
\begin{equation*}
P\left( \mathcal{P}>\alpha \right) \leq \inf_{k\geq 0}\left\{ e^{-k\alpha }E%
\left[ e^{k\mathcal{P}}\right] \right\} .
\end{equation*}%
Note that $E[\mathcal{P}^{i}]$ $\leq $ $E[\mathcal{P}]$ for all $i$ $\geq $ $%
1$ due to $\mathcal{P}$ $\in $ $[0,1]$. Now invoke Fubini's theorem, the
fact that $\mathcal{U}(\lambda )$\ is uniformly distributed on $[0,1]$, and $%
\int_{\Lambda }d\lambda $ $=$ $1$ to deduce:%
\begin{equation*}
E[\mathcal{P}]=E\left[ \int_{\Lambda }I(\mathcal{U}(\lambda )<\alpha
)d\lambda \right] =\int_{\Lambda }P(\mathcal{U}(\lambda )<\alpha )d\lambda
=\alpha \int_{\Lambda }d\lambda =\alpha \text{.}
\end{equation*}%
Expanding $E[e^{k\mathcal{P}}]$ around $k$ $=$ $0$, and exploiting $E[%
\mathcal{P}^{i}]$ $\leq $ $\alpha $, yields:
\begin{equation*}
P\left( \mathcal{P}>\alpha \right) \leq \inf_{k\geq 0}\left\{ e^{-k\alpha }E%
\left[ e^{k\mathcal{P}}\right] \right\} =\inf_{k\geq 0}\left\{ e^{-k\alpha
}\sum_{i=0}^{\infty }\frac{1}{i!}k^{i}E\left[ \mathcal{P}^{i}\right]
\right\} \leq \alpha \inf_{k\geq 0}\left\{ e^{-k\alpha }\sum_{i=0}^{\infty }%
\frac{1}{i!}k^{i}\right\} .
\end{equation*}%
Since $\alpha $ $\in $ $[0,1]$ and therefore $e^{k\left( 1-\alpha \right) }$
$\geq $ $1$ $\forall k$ $\geq $ $0$, trivially
\begin{equation*}
\inf_{k\in \mathcal{K}}\{e^{-k\alpha }\sum_{i=0}^{\infty
}k^{i}/i!\}=\inf_{k\geq 0}\{e^{-k\alpha }e^{k}\}=\inf_{k\geq 0}e^{k\left(
1-\alpha \right) }=1.
\end{equation*}%
This proves $P(\mathcal{P}$ $>$ $\alpha )$ $\leq $ $\alpha $\ as required.

Consider ($b$). If $P(\mathcal{U}(\lambda )$ $=$ $\mathcal{U}(\lambda ^{\ast
}))$ $=$ $1$ $\forall \lambda $ $\in $ $\Lambda $ and some $\lambda ^{\ast }$%
, then $P(\mathcal{P}$ $=$ $\int_{\Lambda }I(\mathcal{U}(\lambda ^{\ast
})<\alpha )d\lambda )$ $=$ $1$. Hence $P(\mathcal{P}$ $>$ $\alpha )$ $=$ $%
P(I(\mathcal{U}(\lambda ^{\ast })$ $<$ $\alpha )$ $>$ $\alpha )$ $=$ $P(%
\mathcal{U}(\lambda ^{\ast })<\alpha )$. The latter is identically $\alpha $
by uniform distributedness.

Finally, for ($c$) if $P(\mathcal{U}(\lambda )$ $<$ $\alpha ,\mathcal{U}(%
\tilde{\lambda})$ $<$ $\alpha )$ $>$ $\alpha ^{2}$ on a subset of $\Lambda $
$\times $ $\Lambda $ with positive measure, then $E[\mathcal{P}^{2}]$ $>$ $%
(E[\mathcal{P}])^{2}$ $=$ $\alpha ^{2}$. Since $E[\mathcal{P}^{2}]$ $=$ $E[%
\mathcal{P}^{2}I(\mathcal{P}^{2}$ $>$ $\alpha ^{2})]$ $+$ $E[\mathcal{P}%
^{2}I(\mathcal{P}^{2}$ $\leq $ $\alpha ^{2})]$, and $\mathcal{P}$\ is
bounded, by a variant of the second moment method $P(\mathcal{P}$ $>$ $%
\alpha )$ $\geq $ $(E[\mathcal{P}^{2}]$ $-$ $\alpha ^{2})^{2}/E[\mathcal{P}%
^{4}]$ $>$ $0$. $\mathcal{QED}$.\medskip \noindent \newline
\textbf{Proof of Theorem \ref{th:main_h1}.}\qquad \medskip \newline
\textbf{Claim (a).}\qquad Let $H_{0}$ be false, and define the set of $%
\lambda ^{\prime }s$ such that we reject the PV test for sample size $n$: $%
\Lambda _{n,\alpha }$ $\equiv $ $\{\lambda $ $\in $ $\Lambda $ $:$ $%
p_{n}(\lambda )$ $<$ $\alpha \}$. By assumption $\{p_{n}(\lambda )$ $:$ $%
\lambda $ $\in $ $\Lambda \}$ lies in a complete measure space, hence $%
\Lambda _{n,\alpha }$ is $\sigma (\mathcal{S}_{n})$-measurable %
\citep[see][p. 195-198]{Pollard1984}. Similarly, the Lebesgue measure $%
\int_{\Lambda _{n,\alpha }}d\lambda $ of $\Lambda _{n,\alpha }$ is $\sigma (%
\mathcal{S}_{n})$-measurable.

By construction
\begin{equation*}
\mathcal{P}_{n}^{\ast }(\alpha )\equiv \int_{\Lambda _{n,\alpha }}I\left(
p_{n}(\lambda )<\alpha \right) d\lambda +\int_{\Lambda /\Lambda _{n,\alpha
}}I\left( p_{n}(\lambda )<\alpha \right) d\lambda =\int_{\Lambda _{n,\alpha
}}d\lambda .
\end{equation*}%
Hence $\lim_{n\rightarrow \infty }P(\mathcal{P}_{n}^{\ast }(\alpha )$ $>$ $%
\alpha )$ $=$ $\lim_{n\rightarrow \infty }P(\int_{\Lambda _{n,\alpha
}}d\lambda $ $>$ $\alpha )$. Therefore $\lim_{n\rightarrow \infty }P(%
\mathcal{P}_{n}^{\ast }(\alpha )$ $>$ $\alpha )$ $>$ $0$ \textit{if and only
if} $\lim_{n\rightarrow \infty }P(\int_{\Lambda _{n,\alpha }}d\lambda $ $>$ $%
\alpha )$ $>$ $0$, \textit{if and only if} $\lim_{n\rightarrow \infty
}P(p_{n}(\lambda )$ $<$ $\alpha )$ $>$ $0$ on some subset with measure
greater than $\alpha $.\medskip \newline
\textbf{Claim (b).}\qquad Let $\Lambda _{\alpha }$ denote the set of $%
\lambda ^{\prime }s$ such that $\lim_{n\rightarrow \infty }P(p_{n}(\lambda )$
$<$ $\alpha )$ $=$ $1$, hence $\lim_{n\rightarrow \infty }P(p_{n}(\lambda )$
$<$ $\alpha )$ $<$ $1$ on $\Lambda /\Lambda _{\alpha }$. Then by dominated
convergence $\lim_{n\rightarrow \infty }P(\mathcal{P}_{n}^{\ast }(\alpha )$ $%
>$ $\alpha )$ $=$ $\lim_{n\rightarrow \infty }P(\int_{\Lambda _{\alpha
}}d\lambda $ $+$ $\int_{\Lambda /\Lambda _{\alpha }}I(p_{n}(\lambda )$ $<$ $%
\alpha )d\lambda $ $>$ $\alpha )$. If $\Lambda _{\alpha }$ has measure
greater than $\alpha $ then $\lim_{n\rightarrow \infty }P(\mathcal{P}%
_{n}^{\ast }(\alpha )$ $>$ $\alpha )$ $=$ $1$. $\mathcal{QED}$.\medskip
\newline
\textbf{Proof of Theorem \ref{th:local_pow_nl}.}\qquad Recall $F_{1}$ is a $%
\chi ^{2}(1)$ distribution, $\bar{F}_{1}$ $\equiv $ $1$ $-$ $F_{1}$, and $%
F_{1,v}$ is a noncentral chi-squared distribution with noncentrality $v$. By
construction $p_{n}(\lambda )$ $=$ $\bar{F}_{1}(\mathcal{T}_{n}(\lambda ))$.

In view of (\ref{TZc}), under $H_{1}^{L}$ it follows $p_{n}(\lambda )$ $%
\overset{d}{\rightarrow }$ $\bar{F}_{1}(\mathfrak{T}_{b})$, a law on $[0,1]$
where $\mathfrak{T}_{b}$ is distributed $F_{1,b^{2}c(\lambda )^{2}}$. Hence $%
\bar{F}_{1}(\mathfrak{T}_{b})$ is skewed left for $b$ $\neq $ $0$. Let $%
\mathcal{U}_{b}(\lambda )$ be distributed $\bar{F}_{0}(\mathfrak{T}_{b})$.
Then $\mathcal{U}_{0}(\lambda )$ is a uniform random variable, and in
general $P(\mathcal{U}_{b}(\lambda )$ $\leq $ $a)$ $-$ $P(\mathcal{U}%
_{0}(\lambda )$ $\leq $ $a)$ $>$ $0$ is monotonically increasing in $b$
since $P(\mathcal{U}_{b}(\lambda )$ $\leq $ $a)$ $\rightarrow $ $1$ is
monotonic as $|b|$ $\rightarrow $ $\infty $ for any $a$.

Now, by construction $\{\mathcal{U}_{b}(\lambda )\}$ has \emph{almost surely}
continuous sample paths with $\mathcal{U}_{b}(\lambda )$ distributed $F_{1}(%
\mathfrak{T}_{b})$. Hence under $H_{1}^{L}$ by (\ref{TZc}), and the
continuous mapping theorem:%
\begin{equation*}
\mathcal{P}_{n}^{\ast }(\alpha )=\int_{\Lambda }I\left( p_{n}(\lambda
)<\alpha \right) d\lambda \overset{d}{\rightarrow }\int_{\Lambda }I\left(
\mathcal{U}_{b}(\lambda )<\alpha \right) d\lambda .
\end{equation*}%
By construction $\int_{\Lambda }I(\mathcal{U}_{b}(\lambda )$ $<$ $\alpha
)d\lambda $ $\geq $ $\int_{\Lambda }I(\mathcal{U}_{0}(\lambda )$ $<$ $\alpha
)d\lambda $ with equality only if $b$ $=$ $0$: the asymptotic occupation
time of a p-value rejection $p_{n}(\lambda )$ $<$ $\alpha $ is higher under
any sequence of non-trivial local alternatives $H_{1}^{L}$ $:$ $\beta _{0}$ $%
=$ $b/n^{1/2}$, $b$ $\neq $ $0$. Further, $\int_{\Lambda }I(\mathcal{U}%
_{b}(\lambda )$ $<$ $\alpha )d\lambda $ $\rightarrow $ $1$ as $|b|$ $%
\rightarrow $ $\infty $. Hence as the local deviation from the null
increases, the probability of a PVOT test rejection of $H_{1}^{L}$
approaches one $\lim_{n\rightarrow \infty }P(\mathcal{P}_{n}^{\ast }(\alpha
) $ $>$ $\alpha )$ $\nearrow $ $1$ for any nominal level $\alpha $ $\in $ $%
[0,1)$. $\mathcal{QED}$.\medskip

\noindent \textbf{Proof of Theorem \ref{th:garch}.}\qquad The GARCH process
is stationary and has an iid error with a finite fourth moment. The claim
therefore follows from arguments in \citet[Section 4.1]{Andrews2001}. $%
\mathcal{QED}$.\medskip

\noindent \textbf{Proof of Theorem \ref{th:garch_p_val}.}\qquad By Theorem %
\ref{th:garch}, the limit process of $\{\mathcal{T}_{n}(\lambda )\}$ under $%
H_{0}$\ is $\{\mathcal{T}(\lambda )\}$, where $\mathcal{T}(\lambda )$ $=$ $%
(\max \{0,\mathcal{Z}(\lambda )\})^{2}$ and $\{\mathcal{Z}(\lambda )\}$ is
Gaussian with covariance $E[\mathcal{Z(}\lambda _{1})\mathcal{Z(}\lambda
_{2})]$ $=$ $(1$ $-$ $\lambda _{1}^{2})(1$ $-$ $\lambda _{2}^{2})/(1$ $-$ $%
\lambda _{1}\lambda _{2})$. Define $\bar{F}_{0}(c)=P(\mathcal{T}(\lambda
)\geq c)$ and $p_{n}(\lambda )\equiv \bar{F}_{0}(\mathcal{T}_{n}(\lambda ))$%
, the asymptotic p-value. Define $\mathcal{D}_{n}$ $\equiv $ $\sup_{\lambda
\in \Lambda }|\hat{p}_{\widetilde{\mathcal{R}}_{n},\widetilde{\mathcal{M}}%
_{n},n}(\lambda )$ $-$ $p_{n}(\lambda )|$. Theorems \ref{th:main} and \ref%
{th:main_h1} apply\ by Theorem \ref{th:garch}. Hence, by Lemma \ref%
{lm:boot_p_ulln}, below, and weak convergence arguments developed in the
proof of Theorem \ref{th:main}, under $H_{0}$ for some uniform process $\{%
\mathcal{U}(\lambda )\}$:%
\begin{eqnarray*}
&\int_{\Lambda }I\left( \mathcal{U}(\lambda )<\alpha \right) d\lambda &%
\overset{d}{\leftarrow }\int_{\Lambda }I\left( p_{n}(\lambda )-\mathcal{D}%
_{n}<\alpha \right) d\lambda \text{ }\geq \int_{\Lambda }I\left( \hat{p}_{%
\widetilde{\mathcal{R}}_{n},\widetilde{\mathcal{M}}_{n},n}(\lambda )<\alpha
\right) d\lambda \\
&&\text{ }\geq \int_{\Lambda }I\left( p_{n}(\lambda )+\mathcal{D}_{n}<\alpha
\right) d\lambda \overset{d}{\rightarrow }\int_{\Lambda }I\left( \mathcal{U}%
(\lambda )<\alpha \right) d\lambda .
\end{eqnarray*}%
Therefore $\int_{\Lambda }I(\hat{p}_{\widetilde{\mathcal{R}}_{n},\widetilde{%
\mathcal{M}}_{n},n}(\lambda )$ $<$ $\alpha )d\lambda $ $\overset{d}{%
\rightarrow }$ $\int_{\Lambda }I(\mathcal{U}(\lambda )$ $<$ $\alpha
)d\lambda $. The claim now follows from the proof of Theorem \ref{th:main}
and the fact that $\{\mathcal{T}(\lambda )\}$\ is weakly dependent in the
sense of Lemma \ref{lm:U}.c. $\mathcal{QED}$.

\begin{lemma}
\label{lm:boot_p_ulln}$\sup_{\lambda \in \Lambda }|\hat{p}_{\widetilde{%
\mathcal{R}}_{n},\widetilde{\mathcal{M}}_{n},n}(\lambda )$ $-$ $%
p_{n}(\lambda )|$ $\overset{p}{\rightarrow }$ $0$ where $\hat{p}_{\widetilde{%
\mathcal{R}}_{n},\widetilde{\mathcal{M}}_{n},n}(\lambda )$ $\equiv $ $1/%
\widetilde{\mathcal{M}}_{n}\sum_{i=1}^{\widetilde{\mathcal{M}}_{n}}I(%
\mathcal{T}_{\widetilde{\mathcal{R}}_{n},i}(\lambda )$ $\geq $ $\mathcal{T}%
_{n}(\lambda ))$.
\end{lemma}

\noindent \textbf{Proof.}\qquad We first state known properties and define
some terms. Assumption \ref{assum:main} applies to $\mathcal{T}_{n}(\lambda
) $ by Theorem \ref{th:garch}, where $\{\mathcal{T}_{n}(\lambda )\}$ $%
\Rightarrow ^{\ast }$ $\{\mathcal{T}(\lambda )\}$, $\mathcal{T}(\lambda )$ $%
= $ $(\max \{0,\mathcal{Z}(\lambda )\})^{2}$, and $\{\mathcal{Z}(\lambda )\}$
is a zero mean Gaussian process with a version that has \emph{almost surely}
continuous sample paths, and covariance function $(1$ $-$ $\lambda
_{1}^{2})(1$ $-$ $\lambda _{2}^{2})/(1$ $-$ $\lambda _{1}\lambda _{2})$ for $%
\lambda _{1},\lambda _{2}$ $\in $ $\Lambda $. Recall we have samples $%
\{Z_{j,i}\}_{j=1}^{\widetilde{\mathcal{R}}}$ where $Z_{j,i}$ $\overset{iid}{%
\sim }$ $N(0,1)$, and for $(\widetilde{\mathcal{R}},\widetilde{\mathcal{M}})$
$\in $ $\mathbb{N}$:
\begin{equation*}
\mathfrak{Z}_{\widetilde{\mathcal{R}},i}(\lambda )\equiv (1-\lambda
^{2})\sum_{j=1}^{\widetilde{\mathcal{R}}}\lambda ^{j}Z_{j,i}\text{ and }%
\mathcal{T}_{\widetilde{\mathcal{R}},i}(\lambda )\equiv \left( \max \{0,%
\mathfrak{Z}_{\widetilde{\mathcal{R}},i}(\lambda )\}\right) ^{2}\text{ for }%
i=1,...,\widetilde{\mathcal{M}}.
\end{equation*}%
$\mathfrak{Z}_{\infty }(\lambda )$\ has the same functional Gaussian
distribution as $\mathcal{Z}(\lambda )$, and therefore $(\max \{0,\mathfrak{Z%
}_{\infty }(\lambda )\})^{2}$ is a random draw from the distribution of $%
\mathcal{T}(\lambda )$. The distribution $\bar{F}_{0}(c)$ $\equiv $ $P(%
\mathcal{T}(\lambda )$ $\geq $ $c)$ is continuous and not a function of $%
\lambda $ under Assumption \ref{assum:main}. Hence, the p-value is
identically $p_{n}(\lambda )$ $=$ $\bar{F}_{0}(\mathcal{T}_{n}(\lambda ))$.
Let $\{\mathcal{T}_{1,i}(\lambda )\}_{i=1}^{\widetilde{\mathcal{M}}}$ and $%
\mathcal{T}_{2}(\lambda )$ be iid copies of $\mathcal{T}(\lambda )$, and
define%
\begin{equation*}
\mathcal{T}_{\widetilde{\mathcal{R}}}^{(\widetilde{\mathcal{M}})}(\lambda
)\equiv \left[ \mathcal{T}_{\widetilde{\mathcal{R}},i}(\lambda ),...,%
\mathcal{T}_{\widetilde{\mathcal{R}},\widetilde{\mathcal{M}}}(\lambda )%
\right] ^{\prime }\text{ \ and }\mathcal{T}_{1}^{(\widetilde{\mathcal{M}}%
)}(\lambda )\equiv \left[ \mathcal{T}_{1,i}(\lambda ),...,\mathcal{T}_{1,%
\widetilde{\mathcal{M}}}(\lambda )\right] ^{\prime }.
\end{equation*}

The arguments in \citet[Section 4.1]{Andrews2001} for weak convergence of $\{%
\mathcal{T}_{n}(\lambda )\}$ trivially extend to $[\mathcal{T}_{\widetilde{%
\mathcal{R}}_{n}}^{(\widetilde{\mathcal{M}})}(\lambda )^{\prime },\mathcal{T}%
_{n}(\lambda )]^{\prime }$ in view of independence of the individual
processes, and normality and smoothness of $\mathfrak{Z}_{\widetilde{%
\mathcal{R}}_{n},i}(\lambda )$. Specifically, there exist $\mathcal{T}_{1}^{(%
\widetilde{\mathcal{M}})}(\lambda )$ and $\mathcal{T}_{2}(\lambda )$ such
that:
\begin{equation*}
\left\{ \left[
\begin{array}{c}
\mathcal{T}_{\widetilde{\mathcal{R}}_{n}}^{(\widetilde{\mathcal{M}}%
)}(\lambda ) \\
\mathcal{T}_{n}(\lambda )%
\end{array}%
\right] :\lambda \in \Lambda \right\} \Rightarrow ^{\ast }\left\{ \left[
\begin{array}{c}
\mathcal{T}_{1}^{(\widetilde{\mathcal{M}})}(\lambda ) \\
\mathcal{T}_{2}(\lambda )%
\end{array}%
\right] :\lambda \in \Lambda \right\} \text{ as }n\rightarrow \infty \text{
for each }\widetilde{\mathcal{M}}\in \mathbb{N}.
\end{equation*}%
Hence, by two applications of the continuous mapping theorem, for each $%
\widetilde{\mathcal{M}}$ $\in $ $\mathbb{N}$ as $n$ $\rightarrow $ $\infty $:%
\begin{eqnarray*}
&&\left\{ \hat{p}_{\widetilde{\mathcal{R}}_{n},\widetilde{\mathcal{M}}%
,n}(\lambda )-\bar{F}_{0}(\mathcal{T}_{n}(\lambda )):\lambda \in \Lambda
\right\} =\left\{ \frac{1}{\widetilde{\mathcal{M}}}\sum_{i=1}^{\widetilde{%
\mathcal{M}}}I\left( \mathcal{T}_{\widetilde{\mathcal{R}}_{n},i}(\lambda
)\geq \mathcal{T}_{n}(\lambda )\right) -\bar{F}_{0}(\mathcal{T}_{n}(\lambda
)):\lambda \in \Lambda \right\} \\
&&\text{ \ \ \ \ \ \ \ \ \ \ \ \ \ \ \ \ \ \ \ \ \ \ \ \ \ \ \ \ \ \ \ \ \ \
\ \ \ \ \ \ \ \ \ \ \ }\Rightarrow ^{\ast }\left\{ \frac{1}{\widetilde{%
\mathcal{M}}}\sum_{i=1}^{\widetilde{\mathcal{M}}}I\left( \mathcal{T}%
_{1,i}(\lambda )\geq \mathcal{T}_{2}(\lambda )\right) -\bar{F}_{0}(\mathcal{T%
}_{2}(\lambda )):\lambda \in \Lambda \right\}
\end{eqnarray*}%
and%
\begin{equation*}
\sup_{\lambda \in \Lambda }\left\vert \hat{p}_{\widetilde{\mathcal{R}}_{n},%
\widetilde{\mathcal{M}},n}(\lambda )-\bar{F}_{0}(\mathcal{T}_{n}(\lambda
))\right\vert \overset{d}{\rightarrow }\sup_{\lambda \in \Lambda }\left\vert
\frac{1}{\widetilde{\mathcal{M}}}\sum_{i=1}^{\widetilde{\mathcal{M}}}I\left(
\mathcal{T}_{1,i}(\lambda )\geq \mathcal{T}_{2}(\lambda )\right) -\bar{F}%
_{0}(\mathcal{T}_{2}(\lambda ))\right\vert \text{ as }n\rightarrow \infty .
\end{equation*}

The proof is complete if we show%
\begin{equation}
\sup_{\lambda \in \Lambda }\left\vert \frac{1}{\widetilde{\mathcal{M}}}%
\sum_{i=1}^{\widetilde{\mathcal{M}}}I(\mathcal{T}_{1,i}(\lambda )\geq
\mathcal{T}_{2}(\lambda ))-\bar{F}_{0}(\mathcal{T}_{2}(\lambda ))\right\vert
\overset{p}{\rightarrow }0\text{ as }\widetilde{\mathcal{M}}\rightarrow
\infty ,  \label{ulln_I}
\end{equation}%
since this means $\sup_{\lambda \in \Lambda }|\hat{p}_{\widetilde{\mathcal{R}%
}_{n},\widetilde{\mathcal{M}},n}(\lambda )$ $-$ $\bar{F}_{0}(\mathcal{T}%
_{n}(\lambda ))|$ can be made arbitrarily close to zero in probability by
choice of $\widetilde{\mathcal{M}}$. Note that by construction $\bar{F}_{0}(%
\mathcal{T}_{2}(\lambda ))$ $=$ $E[I(\mathcal{T}_{1,i}(\lambda )$ $\geq $ $%
\mathcal{T}_{2}(\lambda ))|\mathcal{T}_{2}(\lambda )]$ since $\mathcal{T}%
_{1,i}(\lambda )$ and $\mathcal{T}_{2}(\lambda )$\ are iid copies of $%
\mathcal{T}(\lambda )$. We therefore derive a uniform LLN for
\begin{equation*}
\mathcal{I}_{i}(\lambda )\equiv I\left( \mathcal{T}_{1,i}(\lambda )\geq
\mathcal{T}_{2}(\lambda )\right) -E\left[ I\left( \mathcal{T}_{1,i}(\lambda
)\geq \mathcal{T}_{2}(\lambda )\right) |\mathcal{T}_{2}(\lambda )\right] .
\end{equation*}

Since $(\mathcal{T}_{1,i}(\lambda ),\mathcal{T}_{2}(\lambda ))$\ are iid
copies of $\mathcal{T}(\lambda )$, it follows $E[\bar{F}_{0}(\mathcal{T}%
_{2}(\lambda ))]$ $=$ $P(\mathcal{T}_{1,i}(\lambda )$ $\geq $ $\mathcal{T}%
_{2}(\lambda ))$ hence:
\begin{equation*}
E\left[ \mathcal{I}_{i}(\lambda )\right] =P\left( \mathcal{T}_{1,i}(\lambda
)\geq \mathcal{T}_{2}(\lambda )\right) -E\left[ \bar{F}_{0}(\mathcal{T}%
_{2}(\lambda ))\right] =P\left( \mathcal{T}_{1,i}(\lambda )\geq \mathcal{T}%
_{2}(\lambda )\right) -P\left( \mathcal{T}_{1,i}(\lambda )\geq \mathcal{T}%
_{2}(\lambda )\right) =0.
\end{equation*}%
Second, $1/\widetilde{\mathcal{M}}\sum_{i=1}^{\widetilde{\mathcal{M}}}%
\mathcal{I}_{i}(\lambda )$ $\overset{p}{\rightarrow }$ $0$ as $\widetilde{%
\mathcal{M}}$ $\rightarrow $ $\infty $ pointwise on $\Lambda $ since $%
\mathcal{I}_{i}(\lambda )$ is iid, and $E[\mathcal{I}_{i}(\lambda )]$ $=$ $0$%
.

It remains to demonstrate $\{\mathcal{I}_{i}(\lambda )$ $:$ $\lambda $ $\in $
$\Lambda \}$ is stochastically equicontinuous: $\forall (\epsilon ,\eta )$ $%
> $ $0$ there exists $\delta $ $>$ $0$ such that (see, e.g., Pollard %
\citeyear{Pollard1984}, and Billingsley \citeyear{Billingsley1999}, Chap. 7):%
\begin{equation*}
P\left( \sup_{\lambda ,\tilde{\lambda}\in \Lambda :||\lambda -\tilde{\lambda}%
||\leq \delta }\left\vert \frac{1}{\widetilde{\mathcal{M}}}\sum_{i=1}^{%
\widetilde{\mathcal{M}}}\left\{ \mathcal{I}_{i}(\lambda )-\mathcal{I}_{i}(%
\tilde{\lambda})\right\} \right\vert >\eta \right) <\varepsilon .
\end{equation*}%
The function $\mathcal{I}_{i}$ $:$ $\Lambda $ $\rightarrow $ $[-1,1]$ is not
continuous. We therefore adapt arguments developed in
\citet[proof of Theorem 2.1 and Lemma
2.1]{ArconesYu1994}, which requires the notion of the \textit{V-C subgraph }%
class of functions, denoted $\mathcal{V}(\mathcal{C})$. See \cite{VC1971}
and \citet[Section 7]{Dudley1978}, and see \citet[Chap.
II.4]{Pollard1984} for the closely related \textit{polynomial discrimination}
class. We use the following well known properties: $\mathcal{V}(\mathcal{C})$
contains continuous functions and the indicator function; $\mathcal{V}(%
\mathcal{C})$ contains linear combinations of $\mathcal{V}(\mathcal{C})$
functions; and $\mathcal{V}(\mathcal{C})$ transforms of $\mathcal{V}(%
\mathcal{C})$ functions are in $\mathcal{V}(\mathcal{C})$. Cf. \cite{VC1971}%
, \citet[Section
7]{Dudley1978} and \cite{Pollard1990}.

By using the approach of \cite{ArconesYu1994}, we may show that $1/%
\widetilde{\mathcal{M}}\sum_{i=1}^{\widetilde{\mathcal{M}}}\mathcal{I}%
_{i}(\lambda )$ is stochastically equicontinuous. $\mathcal{T}_{1,i}(\lambda
)$ and $\mathcal{T}_{2}(\lambda )$ are, respectively, versions of $(\max \{0,%
\mathfrak{Z}_{1,\infty ,i}(\lambda ))^{2}$ and \linebreak $(\max \{0,%
\mathfrak{Z}_{2,\infty }(\lambda ))^{2}$, where $\mathfrak{Z}_{1,\infty
,i}(\lambda )$ and $\mathfrak{Z}_{2,\infty }(\lambda )$ are independent
copies of $\mathfrak{Z}_{\infty }(\lambda )$, and $\mathfrak{Z}_{\infty
}(\lambda )$ $\equiv $ $(1$ $-$ $\lambda ^{2})\sum_{j=0}^{\infty }\lambda
^{j}Z_{j}$ is zero mean Gaussian with the same covariance function as $%
\mathcal{Z}(\lambda )$. By construction $\mathfrak{Z}_{\infty }(\lambda )$
is continuous in $\lambda $, hence it lies in $\mathcal{V}(\mathcal{C})$.
Further, $(\max \{0,\cdot )^{2}$ lies in $\mathcal{V}(\mathcal{C})$.
Therefore $(\max \{0,\mathfrak{Z}_{\infty }(\lambda ))^{2}$ lies in $%
\mathcal{V}(\mathcal{C})$, which implies $\mathcal{T}_{1,i}(\lambda )$ and $%
\mathcal{T}_{2}(\lambda )$ have versions that lie in $\mathcal{V}(\mathcal{C}%
)$. Hence $\mathcal{T}_{1,i}(\lambda )$ $-$ $\mathcal{T}_{2}(\lambda )$ has
a version in $\mathcal{V}(\mathcal{C})$. Therefore $I(\mathcal{T}%
_{1,i}(\lambda )$ $-$ $\mathcal{T}_{2}(\lambda )$ $\geq $ $0)$ has a version
in $\mathcal{V}(\mathcal{C})$. Moreover, the continuous transform $\bar{F}%
_{0}(\mathcal{T}_{2}(\lambda ))$ lies in $\mathcal{V}(\mathcal{C})$. Hence
the difference $\mathcal{I}_{i}(\lambda )$ $\equiv $ $I(\mathcal{T}%
_{1,i}(\lambda )$ $\geq $ $\mathcal{T}_{2}(\lambda ))$ $-$ $\bar{F}_{0}(%
\mathcal{T}_{2}(\lambda ))$ lies in $\mathcal{V}(\mathcal{C})$. This, and
boundedness of $\mathcal{I}_{i}(\lambda )$, imply that the covering numbers\
with respect to the $L_{p}$-metric satisfy, for any $p$ $>$ $2$, $\mathcal{N}%
(\varepsilon ,\Lambda ,||\cdot ||_{p})$ $<$ $a\varepsilon ^{-b}$ for all $%
\varepsilon $ $\in $ $(0,1)$ and some $a,b$ $>$ $0$ that may depend on $p$
(e.g. Lemma 7.13 in Dudley, \citeyear{Dudley1978}, and Lemma II.25 in
Pollard, \citeyear{Pollard1984}). Further, $\mathcal{I}_{i}(\lambda )$ is
uniformly bounded and iid. Therefore $\{\mathcal{I}_{i}(\lambda )$ $:$ $%
\lambda $ $\in $ $\Lambda \}$ is stochastically equicontinuous by adapting
the proof of Lemma 2.1 in \cite{ArconesYu1994}: see especially
\citet[eq.
(2.13)]{ArconesYu1994}. $\mathcal{QED}.$

\setstretch{1}
\bibliographystyle{econometrica}
\bibliography{refs_ot_test}

\setstretch{1} \clearpage
\begin{table}[h]
\caption{Functional Form Test Rejection Frequencies}
\label{tbl:funcform}
\begin{center}
\resizebox{\textwidth}{!}{
\begin{tabular}{lccccccccccccc}
\hline\hline
\multicolumn{14}{c}{iid data: linear vs. quadratic} \\ \hline\hline
& \multicolumn{1}{|c}{} & \multicolumn{1}{|c}{} & \multicolumn{3}{|c}{$n=100$} & \multicolumn{1}{|c}{} & \multicolumn{3}{|c}{$n=250$} &
\multicolumn{1}{|c}{} & \multicolumn{3}{|c}{$n=500$} \\ \hline\hline
Hyp$^{a}$ & \multicolumn{1}{|c}{Test} & \multicolumn{1}{|c}{} &
\multicolumn{1}{|c}{$1\%$} & $5\%$ & $10\%$ & \multicolumn{1}{|c}{} &
\multicolumn{1}{|c}{$1\%$} & $5\%$ & $10\%$ & \multicolumn{1}{|c}{} &
\multicolumn{1}{|c}{$1\%$} & $5\%$ & $10\%$ \\ \hline
& \multicolumn{1}{|l|}{sup-$p_{n}$ $^{b}$} & \multicolumn{1}{|c}{} &
\multicolumn{1}{|c}{.008$^{c}$} & .058 & .108 & \multicolumn{1}{|c}{} &
\multicolumn{1}{|c}{.000} & .039 & .094 & \multicolumn{1}{|c}{} &
\multicolumn{1}{|c}{.009} & .043 & .091 \\ \cline{2-14}
& \multicolumn{1}{|l|}{sup-$\mathcal{T}_{n}$ $^{d}$} & \multicolumn{1}{|c}{}
& \multicolumn{1}{|c}{.004} & .037 & .097 & \multicolumn{1}{|c}{} &
\multicolumn{1}{|c}{.008} & .041 & .083 & \multicolumn{1}{|c}{} &
\multicolumn{1}{|c}{.019} & .058 & .096 \\
$H_{0}$ & \multicolumn{1}{|l|}{aver-$\mathcal{T}_{n}$} & \multicolumn{1}{|c}{
} & \multicolumn{1}{|c}{.014} & .057 & .116 & \multicolumn{1}{|c}{} &
\multicolumn{1}{|c}{.007} & .040 & .088 & \multicolumn{1}{|c}{} &
\multicolumn{1}{|c}{.018} & .071 & .109 \\
& \multicolumn{1}{|l|}{rand-$\mathcal{T}_{n}$ $^{e}$} & \multicolumn{1}{|c}{}
& \multicolumn{1}{|c}{.014} & .056 & .117 & \multicolumn{1}{|c}{} &
\multicolumn{1}{|c}{.011} & .045 & .094 & \multicolumn{1}{|c}{} &
\multicolumn{1}{|c}{.021} & .059 & .109 \\ \cline{2-14}
& \multicolumn{1}{|l|}{ICM$^{f}$} & \multicolumn{1}{|c}{} &
\multicolumn{1}{|c}{.001} & .033 & .086 & \multicolumn{1}{|c}{} &
\multicolumn{1}{|c}{.001} & .014 & .075 & \multicolumn{1}{|c}{} &
\multicolumn{1}{|c}{.003} & .062 & .086 \\ \cline{2-14}
& \multicolumn{1}{|l|}{PVOT$^{g}$} & \multicolumn{1}{|c}{} &
\multicolumn{1}{|c}{.013} & .056 & .116 & \multicolumn{1}{|c}{} &
\multicolumn{1}{|c}{.010} & .044 & .092 & \multicolumn{1}{|c}{} &
\multicolumn{1}{|c}{.014} & .063 & .108 \\ \hline\hline
& \multicolumn{1}{|l|}{sup-$p_{n}$} & \multicolumn{1}{|c}{} &
\multicolumn{1}{|c}{.042} & .162 & .258 & \multicolumn{1}{|c}{} &
\multicolumn{1}{|c}{.137} & .337 & .473 & \multicolumn{1}{|c}{} &
\multicolumn{1}{|c}{.339} & .597 & .695 \\ \cline{2-14}
& \multicolumn{1}{|l|}{sup-$\mathcal{T}_{n}$} & \multicolumn{1}{|c}{} &
\multicolumn{1}{|c}{.051} & .156 & .251 & \multicolumn{1}{|c}{} &
\multicolumn{1}{|c}{.160} & .331 & .512 & \multicolumn{1}{|c}{} &
\multicolumn{1}{|c}{.354} & .539 & .743 \\
$H_{1}$ & \multicolumn{1}{|l|}{aver-$\mathcal{T}_{n}$} & \multicolumn{1}{|c}{
} & \multicolumn{1}{|c}{.051} & .211 & .316 & \multicolumn{1}{|c}{} &
\multicolumn{1}{|c}{.193} & .377 & .576 & \multicolumn{1}{|c}{} &
\multicolumn{1}{|c}{.412} & .643 & .776 \\
& \multicolumn{1}{|l|}{rand-$\mathcal{T}_{n}$} & \multicolumn{1}{|c}{} &
\multicolumn{1}{|c}{.051} & .221 & .316 & \multicolumn{1}{|c}{} &
\multicolumn{1}{|c}{.212} & .392 & .586 & \multicolumn{1}{|c}{} &
\multicolumn{1}{|c}{.404} & .668 & .798 \\ \cline{2-14}
& \multicolumn{1}{|l|}{ICM} & \multicolumn{1}{|c}{} &
\multicolumn{1}{|c}{.001} & .149 & .329 & \multicolumn{1}{|c}{} &
\multicolumn{1}{|c}{.043} & .330 & .606 & \multicolumn{1}{|c}{} &
\multicolumn{1}{|c}{.163} & .678 & .809 \\ \cline{2-14}
& \multicolumn{1}{|l|}{PVOT} & \multicolumn{1}{|c}{} &
\multicolumn{1}{|c}{.058} & .224 & .320 & \multicolumn{1}{|c}{} &
\multicolumn{1}{|c}{.232} & .391 & .604 & \multicolumn{1}{|c}{} &
\multicolumn{1}{|c}{.404} & .614 & .783 \\ \hline\hline
&  &  &  &  &  &  &  &  &  &  &  &  &  \\ \hline\hline
\multicolumn{14}{c}{time series data: AR vs. SETAR} \\ \hline\hline
& \multicolumn{1}{|c}{} & \multicolumn{1}{|c}{} & \multicolumn{3}{|c}{$n=100$} & \multicolumn{1}{|c}{} & \multicolumn{3}{|c}{$n=250$} &
\multicolumn{1}{|c}{} & \multicolumn{3}{|c}{$n=500$} \\ \hline\hline
Hyp & \multicolumn{1}{|c}{Test} & \multicolumn{1}{|c}{} &
\multicolumn{1}{|c}{$1\%$} & $5\%$ & $10\%$ & \multicolumn{1}{|c}{} &
\multicolumn{1}{|c}{$1\%$} & $5\%$ & $10\%$ & \multicolumn{1}{|c}{} &
\multicolumn{1}{|c}{$1\%$} & $5\%$ & $10\%$ \\ \hline
& \multicolumn{1}{|l|}{sup-$p_{n}$} & \multicolumn{1}{|c}{} &
\multicolumn{1}{|c}{.022} & .075 & .158 & \multicolumn{1}{|c}{} &
\multicolumn{1}{|c}{.008} & .052 & .113 & \multicolumn{1}{|c}{} &
\multicolumn{1}{|c}{.020} & .064 & .116 \\ \cline{2-14}
& \multicolumn{1}{|l|}{sup-$\mathcal{T}_{n}$} & \multicolumn{1}{|c}{} &
\multicolumn{1}{|c}{.001} & .003 & .039 & \multicolumn{1}{|c}{} &
\multicolumn{1}{|c}{.002} & .012 & .036 & \multicolumn{1}{|c}{} &
\multicolumn{1}{|c}{.003} & .052 & .124 \\
$H_{0}$ & \multicolumn{1}{|l|}{aver-$\mathcal{T}_{n}$} & \multicolumn{1}{|c}{
} & \multicolumn{1}{|c}{.002} & .022 & .082 & \multicolumn{1}{|c}{} &
\multicolumn{1}{|c}{.002} & .013 & .066 & \multicolumn{1}{|c}{} &
\multicolumn{1}{|c}{.008} & .072 & .132 \\
& \multicolumn{1}{|l|}{rand-$\mathcal{T}_{n}$} & \multicolumn{1}{|c}{} &
\multicolumn{1}{|c}{.021} & .113 & .193 & \multicolumn{1}{|c}{} &
\multicolumn{1}{|c}{.001} & .03 & .114 & \multicolumn{1}{|c}{} &
\multicolumn{1}{|c}{.018} & .082 & .143 \\ \cline{2-14}
& \multicolumn{1}{|l|}{ICM$^{f}$} & \multicolumn{1}{|c}{} &
\multicolumn{1}{|c}{.002} & .058 & .132 & \multicolumn{1}{|c}{} &
\multicolumn{1}{|c}{.000} & .030 & .066 & \multicolumn{1}{|c}{} &
\multicolumn{1}{|c}{.005} & .038 & .089 \\ \cline{2-14}
& \multicolumn{1}{|l|}{PVOT$^{g}$} & \multicolumn{1}{|c}{} &
\multicolumn{1}{|c}{.016} & .076 & .145 & \multicolumn{1}{|c}{} &
\multicolumn{1}{|c}{.011} & .047 & .115 & \multicolumn{1}{|c}{} &
\multicolumn{1}{|c}{.016} & .061 & .114 \\ \hline\hline
& \multicolumn{1}{|l|}{sup-$p_{n}$} & \multicolumn{1}{|c}{} &
\multicolumn{1}{|c}{.108} & .596 & .845 & \multicolumn{1}{|c}{} &
\multicolumn{1}{|c}{.925} & 1.00 & 1.00 & \multicolumn{1}{|c}{} &
\multicolumn{1}{|c}{1.00} & 1.00 & 1.00 \\ \cline{2-14}
& \multicolumn{1}{|l|}{sup-$\mathcal{T}_{n}$} & \multicolumn{1}{|c}{} &
\multicolumn{1}{|c}{.021} & .209 & .561 & \multicolumn{1}{|c}{} &
\multicolumn{1}{|c}{.685} & 1.00 & 1.00 & \multicolumn{1}{|c}{} &
\multicolumn{1}{|c}{1.00} & 1.00 & 1.00 \\
$H_{1}$ & \multicolumn{1}{|l|}{aver-$\mathcal{T}_{n}$} & \multicolumn{1}{|c}{
} & \multicolumn{1}{|c}{.062} & .412 & .726 & \multicolumn{1}{|c}{} &
\multicolumn{1}{|c}{.888} & 1.00 & 1.00 & \multicolumn{1}{|c}{} &
\multicolumn{1}{|c}{1.00} & 1.00 & 1.00 \\
& \multicolumn{1}{|l|}{rand-$\mathcal{T}_{n}$} & \multicolumn{1}{|c}{} &
\multicolumn{1}{|c}{.135} & .592 & .846 & \multicolumn{1}{|c}{} &
\multicolumn{1}{|c}{.960} & 1.00 & 1.00 & \multicolumn{1}{|c}{} &
\multicolumn{1}{|c}{1.00} & 1.00 & 1.00 \\ \cline{2-14}
& \multicolumn{1}{|l|}{ICM} & \multicolumn{1}{|c}{} &
\multicolumn{1}{|c}{.004} & .643 & .866 & \multicolumn{1}{|c}{} &
\multicolumn{1}{|c}{.108} & .928 & 1.00 & \multicolumn{1}{|c}{} &
\multicolumn{1}{|c}{.712} & 1.00 & 1.00 \\ \cline{2-14}
& \multicolumn{1}{|l|}{PVOT} & \multicolumn{1}{|c}{} &
\multicolumn{1}{|c}{.135} & .647 & .883 & \multicolumn{1}{|c}{} &
\multicolumn{1}{|c}{.957} & 1.00 & 1.00 & \multicolumn{1}{|c}{} &
\multicolumn{1}{|c}{1.00} & 1.00 & 1.00 \\ \hline\hline
\end{tabular}}
\end{center}
\par
{\small a. $H_{0}$ is $E[\epsilon |x]=0$. b. sup-$p_{n}$ is the $%
\sup_{\lambda \in \Lambda }p_{n}(\lambda )$ test. and ave-$\mathcal{T}_{n}$
tests are based on a wild bootstrapped p-value. c. Rejection frequency at
the given level. Empirical power is not size-adjusted. d. sup-$\mathcal{T}%
_{n}$ e. rand-$\mathcal{T}_{n}$ is an asymptotic $\chi^{2}$ test based on $%
\mathcal{T}_{n}(\lambda )$ with randomized $\lambda $\ on [0,1]. f. The ICM
test is based on critical value upper bounds in Bierens and Ploberger
(1997). g. PVOT: \textit{p-value occupation time} test.}
\end{table}

\clearpage

\begin{table}[h]
\caption{A. STAR Test Rejection Frequencies: Sample Size $n=100$}
\label{tbl:starn100}
\begin{center}
{\small
\begin{tabular}{l|ccc|c|ccc|c|ccc}
\hline\hline
& \multicolumn{3}{|c|}{$H_{0}$: LSTAR} &  & \multicolumn{3}{|c|}{$H_{1}$-weak
} &  & \multicolumn{3}{|c}{$H_{1}$-strong} \\ \hline
& 1\% & 5\% & 10\% &  & 1\% & 5\% & 10\% &  & 1\% & 5\% & 10\% \\
\hline\hline
& \multicolumn{11}{c}{Strong Identification: $\beta _{n}=.3$} \\ \hline\hline
sup $\mathcal{T}_{n}$ & \multicolumn{1}{|l}{.025} & \multicolumn{1}{l}{.094}
& \multicolumn{1}{l|}{.163} & \multicolumn{1}{|l|}{} & \multicolumn{1}{|l}{
.147} & \multicolumn{1}{l}{.280} & \multicolumn{1}{l|}{.365} &
\multicolumn{1}{|l|}{} & \multicolumn{1}{|l}{.757} & \multicolumn{1}{l}{.872}
& \multicolumn{1}{l}{.907} \\
aver $\mathcal{T}_{n}$ & \multicolumn{1}{|l}{.025} & \multicolumn{1}{l}{.078}
& \multicolumn{1}{l|}{.135} & \multicolumn{1}{|l|}{} & \multicolumn{1}{|l}{
.087} & \multicolumn{1}{l}{.209} & \multicolumn{1}{l|}{.289} &
\multicolumn{1}{|l|}{} & \multicolumn{1}{|l}{.552} & \multicolumn{1}{l}{.726}
& \multicolumn{1}{l}{.804} \\ \hline
rand $\mathcal{T}_{n}$ & \multicolumn{1}{|l}{.011} & \multicolumn{1}{l}{.052}
& \multicolumn{1}{l|}{.096} & \multicolumn{1}{|l|}{} & \multicolumn{1}{|l}{
.053} & \multicolumn{1}{l}{.143} & \multicolumn{1}{l|}{.232} &
\multicolumn{1}{|l|}{} & \multicolumn{1}{|l}{.446} & \multicolumn{1}{l}{.635}
& \multicolumn{1}{l}{.732} \\
rand LF & \multicolumn{1}{|l}{.007} & \multicolumn{1}{l}{.015} &
\multicolumn{1}{l|}{.038} & \multicolumn{1}{|l|}{} & \multicolumn{1}{|l}{.013
} & \multicolumn{1}{l}{.066} & \multicolumn{1}{l|}{.141} &
\multicolumn{1}{|l|}{} & \multicolumn{1}{|l}{.442} & \multicolumn{1}{l}{.553}
& \multicolumn{1}{l}{.661} \\
rand ICS-1 & \multicolumn{1}{|l}{.013} & \multicolumn{1}{l}{.050} &
\multicolumn{1}{l|}{.089} & \multicolumn{1}{|l|}{} & \multicolumn{1}{|l}{.028
} & \multicolumn{1}{l}{.089} & \multicolumn{1}{l|}{.170} &
\multicolumn{1}{|l|}{} & \multicolumn{1}{|l}{.379} & \multicolumn{1}{l}{.593}
& \multicolumn{1}{l}{.692} \\ \hline
sup $p_{n}$ & .009 & .039 & .068 &  & .036 & .118 & .209 &  & .378 & .554 &
.656 \\
sup $p_{n}$ LF & .006 & .009 & .032 &  & .012 & .057 & .120 &  & .262 & .457
& .572 \\
sup $p_{n}$ ICS-1 & .006 & .036 & .061 &  & .020 & .081 & .138 &  & .310 &
.506 & .617 \\ \hline
PVOT & \multicolumn{1}{|l}{.015} & \multicolumn{1}{l}{.065} &
\multicolumn{1}{l|}{.124} & \multicolumn{1}{|l|}{} & \multicolumn{1}{|l}{.101
} & \multicolumn{1}{l}{.257} & \multicolumn{1}{l|}{.335} &
\multicolumn{1}{|l|}{} & \multicolumn{1}{|l}{.727} & \multicolumn{1}{l}{.859}
& \multicolumn{1}{l}{.883} \\
PVOT LF & \multicolumn{1}{|l}{.007} & \multicolumn{1}{l}{.014} &
\multicolumn{1}{l|}{.052} & \multicolumn{1}{|l|}{} & \multicolumn{1}{|l}{.026
} & \multicolumn{1}{l}{.121} & \multicolumn{1}{l|}{.208} &
\multicolumn{1}{|l|}{} & \multicolumn{1}{|l}{.552} & \multicolumn{1}{l}{.781}
& \multicolumn{1}{l}{.817} \\
PVOT ICS-1 & \multicolumn{1}{|l}{.007} & \multicolumn{1}{l}{.043} &
\multicolumn{1}{l|}{.073} & \multicolumn{1}{|l|}{} & \multicolumn{1}{|l}{.042
} & \multicolumn{1}{l}{.153} & \multicolumn{1}{l|}{.237} &
\multicolumn{1}{|l|}{} & \multicolumn{1}{|l}{.622} & \multicolumn{1}{l}{.815}
& \multicolumn{1}{l}{.842} \\ \hline\hline
& \multicolumn{11}{c}{Weak Identification: $\beta _{n}=.3/\sqrt{n}$} \\
\hline\hline
sup $\mathcal{T}_{n}$ & \multicolumn{1}{|l}{.064} & \multicolumn{1}{l}{.155}
& \multicolumn{1}{l|}{.239} &  & \multicolumn{1}{|l}{.337} &
\multicolumn{1}{l}{.574} & \multicolumn{1}{l|}{.681} &  &
\multicolumn{1}{|l}{.929} & \multicolumn{1}{l}{.978} & \multicolumn{1}{l}{
.993} \\
aver $\mathcal{T}_{n}$ & \multicolumn{1}{|l}{.057} & \multicolumn{1}{l}{.146}
& \multicolumn{1}{l|}{.219} &  & \multicolumn{1}{|l}{.215} &
\multicolumn{1}{l}{.430} & \multicolumn{1}{l|}{.554} &  &
\multicolumn{1}{|l}{.739} & \multicolumn{1}{l}{.888} & \multicolumn{1}{l}{
.932} \\ \hline
rand $\mathcal{T}_{n}$ & \multicolumn{1}{|l}{.027} & \multicolumn{1}{l}{.083}
& \multicolumn{1}{l|}{.175} &  & \multicolumn{1}{|l}{.164} &
\multicolumn{1}{l}{.343} & \multicolumn{1}{l|}{.474} &  &
\multicolumn{1}{|l}{.604} & \multicolumn{1}{l}{.810} & \multicolumn{1}{l}{
.870} \\
rand LF & .012 & .042 & .093 &  & .060 & .161 & .308 &  & .467 & .685 & .794
\\
rand ICS-1 & .012 & .046 & .104 &  & .116 & .261 & .382 &  & .545 & .749 &
.841 \\ \hline
sup $p_{n}$ & .019 & .087 & .145 &  & .107 & .253 & .411 &  & .493 & .700 &
.785 \\
sup $p_{n}$ LF & .001 & .061 & .084 &  & .036 & .124 & .230 &  & .351 & .598
& .698 \\
sup $p_{n}$ ICS-1 & .001 & .065 & .085 &  & .088 & .193 & .335 &  & .454 &
.663 & .756 \\ \hline
PVOT & \multicolumn{1}{|l}{.038} & \multicolumn{1}{l}{.127} &
\multicolumn{1}{l|}{.196} &  & \multicolumn{1}{|l}{.328} &
\multicolumn{1}{l}{.542} & \multicolumn{1}{l|}{.591} &  &
\multicolumn{1}{|l}{.893} & \multicolumn{1}{l}{.968} & \multicolumn{1}{l}{
.950} \\
PVOT LF & .015 & .049 & .108 &  & \multicolumn{1}{|l}{.108} &
\multicolumn{1}{l}{.320} & \multicolumn{1}{l|}{.398} &  &
\multicolumn{1}{|l}{.710} & \multicolumn{1}{l}{.911} & \multicolumn{1}{l}{
.916} \\
PVOT ICS-1 & \multicolumn{1}{|l}{.014} & \multicolumn{1}{l}{.049} &
\multicolumn{1}{l|}{.107} &  & \multicolumn{1}{|l}{.221} &
\multicolumn{1}{l}{.435} & \multicolumn{1}{l|}{.486} &  &
\multicolumn{1}{|l}{.830} & \multicolumn{1}{l}{.942} & \multicolumn{1}{l}{
.932} \\ \hline\hline
& \multicolumn{11}{c}{Non-Identification: $\beta _{n}=\beta _{0}=0$} \\
\hline\hline
sup $\mathcal{T}_{n}$ & \multicolumn{1}{|l}{.066} & \multicolumn{1}{l}{.164}
& \multicolumn{1}{l|}{.249} &  & \multicolumn{1}{|l}{.358} &
\multicolumn{1}{l}{.584} & \multicolumn{1}{l|}{.696} &  &
\multicolumn{1}{|l}{.902} & \multicolumn{1}{l}{.970} & \multicolumn{1}{l}{
.983} \\
aver $\mathcal{T}_{n}$ & \multicolumn{1}{|l}{.062} & \multicolumn{1}{l}{.148}
& \multicolumn{1}{l|}{.226} &  & \multicolumn{1}{|l}{.233} &
\multicolumn{1}{l}{.438} & \multicolumn{1}{l|}{.548} &  &
\multicolumn{1}{|l}{.716} & \multicolumn{1}{l}{.872} & \multicolumn{1}{l}{
.911} \\ \hline
rand $\mathcal{T}_{n}$ & \multicolumn{1}{|l}{.044} & \multicolumn{1}{l}{.107}
& \multicolumn{1}{l|}{.186} &  & \multicolumn{1}{|l}{.184} &
\multicolumn{1}{l}{.380} & \multicolumn{1}{l|}{.505} &  &
\multicolumn{1}{|l}{.634} & \multicolumn{1}{l}{.793} & \multicolumn{1}{l}{
.864} \\
rand LF & .013 & .046 & .115 &  & .069 & .191 & .327 &  & .498 & .725 & .818
\\
rand ICS-1 & .013 & .047 & .116 &  & .137 & .298 & .481 &  & .583 & .769 &
.847 \\ \hline
sup $p_{n}$ & .018 & .080 & .167 &  & .117 & .272 & .363 &  & .514 & .710 &
.807 \\
sup $p_{n}$ LF & .011 & .043 & .083 &  & .042 & .122 & .221 &  & .383 & .612
& .740 \\
sup $p_{n}$ ICS-1 & .011 & .044 & .086 &  & .093 & .205 & .293 &  & .464 &
.683 & .783 \\ \hline
PVOT & \multicolumn{1}{|l}{.049} & \multicolumn{1}{l}{.134} &
\multicolumn{1}{l|}{.190} &  & \multicolumn{1}{|l}{.322} &
\multicolumn{1}{l}{.554} & \multicolumn{1}{l|}{.624} &  &
\multicolumn{1}{|l}{.890} & \multicolumn{1}{l}{.962} & \multicolumn{1}{l}{
.957} \\
PVOT LF & .015 & .061 & .117 &  & .122 & .322 & .415 &  & .740 & .911 & .936
\\
PVOT ICS-1 & .015 & .057 & .116 &  & .253 & .464 & .570 &  & .847 & .939 &
.954 \\ \hline\hline
\end{tabular}%
}
\end{center}
\par
{\small Numerical values are rejection frequency at the given level. LSTAR
is Logistic STAR. Empirical power is not size-adjusted. \textit{sup} $%
\mathcal{T}_{n}$ and \textit{ave} $\mathcal{T}_{n}$ tests are based on a
wild bootstrapped p-value. \textit{rand} $\mathcal{T}_{n}$: $\mathcal{T}%
_{n}(\lambda )$ with randomized $\lambda $\ on [1,5]. \textit{sup} ${p}_{n}$
is the supremum p-value test where p-values are computed from the
chi-squared distribution. PVOT uses the chi-squared distribution. LF implies
the least favorable p-value is used, and ICS-$1$ implies the type 1
identification category selection p-value is used with threshold $\kappa
_{n} $ $=$ $\ln (\ln (n))$.}
\end{table}

\clearpage
\addtocounter{table}{-1}

\begin{table}[h]
\caption{B. STAR Test Rejection Frequencies: Sample Size $n=250$}
\label{tbl:starn250}
\begin{center}
{\small
\begin{tabular}{l|ccc|c|ccc|c|ccc}
\hline\hline
& \multicolumn{3}{|c|}{$H_{0}$: LSTAR} &  & \multicolumn{3}{|c|}{$H_{1}$-weak
} &  & \multicolumn{3}{|c}{$H_{1}$-strong} \\ \hline
& 1\% & 5\% & 10\% &  & 1\% & 5\% & 10\% &  & 1\% & 5\% & 10\% \\
\hline\hline
& \multicolumn{11}{c}{Strong Identification: $\beta _{n}=.3$} \\ \hline\hline
\multicolumn{1}{l|}{sup $\mathcal{T}_{n}$} & \multicolumn{1}{|l}{.018} &
\multicolumn{1}{l}{.088} & \multicolumn{1}{l|}{.163} &  &
\multicolumn{1}{|l}{.359} & \multicolumn{1}{l}{.468} & \multicolumn{1}{l|}{
.551} &  & \multicolumn{1}{|l}{.953} & \multicolumn{1}{l}{.984} &
\multicolumn{1}{l}{.990} \\
\multicolumn{1}{l|}{aver $\mathcal{T}_{n}$} & \multicolumn{1}{|l}{.014} &
\multicolumn{1}{l}{.077} & \multicolumn{1}{l|}{.133} &  &
\multicolumn{1}{|l}{.262} & \multicolumn{1}{l}{.387} & \multicolumn{1}{l|}{
.468} &  & \multicolumn{1}{|l}{.873} & \multicolumn{1}{l}{.949} &
\multicolumn{1}{l}{.975} \\ \hline
\multicolumn{1}{l|}{rand $\mathcal{T}_{n}$} & \multicolumn{1}{|l}{.014} &
\multicolumn{1}{l}{.064} & \multicolumn{1}{l|}{.126} &  &
\multicolumn{1}{|l}{.165} & \multicolumn{1}{l}{.299} & \multicolumn{1}{l|}{
.396} &  & \multicolumn{1}{|l}{.793} & \multicolumn{1}{l}{.912} &
\multicolumn{1}{l}{.952} \\
\multicolumn{1}{l|}{rand LF} & .001 & .010 & .025 &  & .067 & .235 & .368 &
& .688 & .888 & .936 \\
\multicolumn{1}{l|}{rand ICS-1} & .008 & .031 & .077 &  & .076 & .244 & .375
&  & .762 & .902 & .947 \\ \hline
sup $p_{n}$ & .003 & .039 & .066 &  & .103 & .264 & .358 &  & .743 & .876 &
.917 \\
sup $p_{n}$ LF & .000 & .007 & .021 &  & .032 & .214 & .303 &  & .605 & .838
& .899 \\
\multicolumn{1}{l|}{sup $p_{n}$ ICS-1} & .003 & .035 & .063 &  & .038 & .217
& .316 &  & .714 & .870 & .912 \\ \hline
\multicolumn{1}{l|}{PVOT} & \multicolumn{1}{|l}{.016} & \multicolumn{1}{l}{
.067} & \multicolumn{1}{l|}{.125} &  & \multicolumn{1}{|l}{.328} &
\multicolumn{1}{l}{.437} & \multicolumn{1}{l|}{.517} &  &
\multicolumn{1}{|l}{.952} & \multicolumn{1}{l}{.983} & \multicolumn{1}{l}{
.991} \\
\multicolumn{1}{l|}{PVOT LF} & .004 & .020 & .041 &  & .132 & .348 & .417 &
& .938 & .972 & .976 \\
\multicolumn{1}{l|}{PVOT ICS-1} & .011 & .051 & .108 &  & .147 & .370 & .433
&  & .947 & .978 & .985 \\ \hline\hline
& \multicolumn{11}{c}{Weak Identification: $\beta _{n}=.3/\sqrt{n}$} \\
\hline\hline
\multicolumn{1}{l|}{sup $\mathcal{T}_{n}$} & \multicolumn{1}{|l}{.051} &
\multicolumn{1}{l}{.139} & \multicolumn{1}{l|}{.224} &  &
\multicolumn{1}{|l}{.764} & \multicolumn{1}{l}{.922} & \multicolumn{1}{l|}{
.957} &  & \multicolumn{1}{|l}{.992} & \multicolumn{1}{l}{1.00} &
\multicolumn{1}{l}{1.00} \\
\multicolumn{1}{l|}{aver $\mathcal{T}_{n}$} & \multicolumn{1}{|l}{.046} &
\multicolumn{1}{l}{.118} & \multicolumn{1}{l|}{.215} &  &
\multicolumn{1}{|l}{.539} & \multicolumn{1}{l}{.779} & \multicolumn{1}{l|}{
.853} &  & \multicolumn{1}{|l}{.969} & \multicolumn{1}{l}{.992} &
\multicolumn{1}{l}{.998} \\ \hline
\multicolumn{1}{l|}{rand $\mathcal{T}_{n}$} & \multicolumn{1}{|l}{.027} &
\multicolumn{1}{l}{.086} & \multicolumn{1}{l|}{.169} &  &
\multicolumn{1}{|l}{.451} & \multicolumn{1}{l}{.695} & \multicolumn{1}{l|}{
.785} &  & \multicolumn{1}{|l}{.911} & \multicolumn{1}{l}{.979} &
\multicolumn{1}{l}{.993} \\
\multicolumn{1}{l|}{rand LF} & .018 & .060 & .097 &  & .180 & .481 & .641 &
& .851 & .961 & .980 \\
\multicolumn{1}{l|}{rand ICS-1} & .018 & .058 & .098 &  & .298 & .633 & .770
&  & .926 & .975 & .991 \\ \hline
sup $p_{n}$ & .017 & .056 & .097 &  & .330 & .615 & .712 &  & .858 & .975 &
.991 \\
sup $p_{n}$ LF & .008 & .026 & .067 &  & .115 & .416 & .587 &  & .698 & .926
& .978 \\
\multicolumn{1}{l|}{sup $p_{n}$ ICS-1} & .008 & .030 & .072 &  & .294 & .580
& .687 &  & .852 & .975 & .991 \\ \hline
\multicolumn{1}{l|}{PVOT} & \multicolumn{1}{|l}{.051} & \multicolumn{1}{l}{
.122} & \multicolumn{1}{l|}{.201} &  & \multicolumn{1}{|l}{.740} &
\multicolumn{1}{l}{.894} & \multicolumn{1}{l|}{.934} &  &
\multicolumn{1}{|l}{1.00} & \multicolumn{1}{l}{1.00} & \multicolumn{1}{l}{
1.00} \\
\multicolumn{1}{l|}{PVOT LF} & .014 & .061 & .110 &  & .380 & .708 & .805 &
& .990 & 1.00 & 1.00 \\
\multicolumn{1}{l|}{PVOT ICS-1} & .015 & .060 & .111 &  & .618 & .848 & .878
&  & .999 & 1.00 & 1.00 \\ \hline\hline
& \multicolumn{11}{c}{Non-Identification: $\beta _{n}=\beta _{0}=0$} \\
\hline\hline
sup $\mathcal{T}_{n}$ & \multicolumn{1}{|l}{.061} & \multicolumn{1}{l}{.152}
& \multicolumn{1}{l|}{.223} &  & \multicolumn{1}{|l}{.751} &
\multicolumn{1}{l}{.922} & \multicolumn{1}{l|}{.956} &  &
\multicolumn{1}{|l}{1.00} & \multicolumn{1}{l}{1.00} & \multicolumn{1}{l}{
1.00} \\
aver $\mathcal{T}_{n}$ & \multicolumn{1}{|l}{.054} & \multicolumn{1}{l}{.145}
& \multicolumn{1}{l|}{.200} &  & \multicolumn{1}{|l}{.526} &
\multicolumn{1}{l}{.765} & \multicolumn{1}{l|}{.849} &  &
\multicolumn{1}{|l}{.975} & \multicolumn{1}{l}{.996} & \multicolumn{1}{l}{
.999} \\ \hline
rand $\mathcal{T}_{n}$ & \multicolumn{1}{|l}{.036} & \multicolumn{1}{l}{.123}
& \multicolumn{1}{l|}{.184} &  & \multicolumn{1}{|l}{.417} &
\multicolumn{1}{l}{.696} & \multicolumn{1}{l|}{.803} &  &
\multicolumn{1}{|l}{.025} & \multicolumn{1}{l}{.976} & \multicolumn{1}{l}{
.988} \\
rand LF & .008 & .047 & .108 &  & .205 & .504 & .655 &  & .838 & .955 & .973
\\
rand ICS-1 & .008 & .049 & .109 &  & .411 & .653 & .770 &  & .923 & .977 &
.989 \\ \hline
sup $p_{n}$ & .026 & .068 & .123 &  & .380 & .650 & .772 &  & .850 & .946 &
.968 \\
sup $p_{n}$ LF & .008 & .038 & .079 &  & .132 & .430 & .592 &  & .728 & .915
& .946 \\
sup $p_{n}$ ICS-1 & .008 & .004 & .081 &  & .340 & .629 & .750 &  & .842 &
.945 & .968 \\ \hline
PVOT & \multicolumn{1}{|l}{.036} & \multicolumn{1}{l}{.145} &
\multicolumn{1}{l|}{.211} &  & \multicolumn{1}{|l}{.732} &
\multicolumn{1}{l}{.885} & \multicolumn{1}{l|}{.930} &  &
\multicolumn{1}{|l}{1.00} & \multicolumn{1}{l}{1.00} & \multicolumn{1}{l}{
1.00} \\
PVOT LF & .010 & .058 & .114 &  & .373 & .717 & .806 &  & .990 & 1.00 & 1.00
\\
PVOT ICS-1 & .010 & .059 & .116 &  & .682 & .853 & .898 &  & 1.00 & 1.00 &
1.00 \\ \hline\hline
\end{tabular}%
}
\end{center}
\par
{\small Numerical values are rejection frequency at the given level. LSTAR
is Logistic STAR. Empirical power is not size-adjusted. \textit{sup} $%
\mathcal{T}_{n}$ and \textit{ave} $\mathcal{T}_{n}$ tests are based on a
wild bootstrapped p-value. \textit{rand} $\mathcal{T}_{n}$: $\mathcal{T}%
_{n}(\lambda )$ with randomized $\lambda $\ on [1,5]. \textit{sup} ${p}_{n}$
is the supremum p-value test where p-values are computed from the
chi-squared distribution. PVOT uses the chi-squared distribution. LF implies
the least favorable p-value is used, and ICS-$1$ implies the type 1
identification category selection p-value is used with threshold $\kappa
_{n} $ $=$ $\ln (\ln (n))$.}
\end{table}

\clearpage
\addtocounter{table}{-1}

\begin{table}[h]
\caption{C. STAR Test Rejection Frequencies: Sample Size $n=500$}
\label{tbl:starn500}
\begin{center}
{\small
\begin{tabular}{l|ccc|c|ccc|c|ccc}
\hline\hline
& \multicolumn{3}{|c|}{$H_{0}$: LSTAR} &  & \multicolumn{3}{|c|}{$H_{1}$-weak
} &  & \multicolumn{3}{|c}{$H_{1}$-strong} \\ \hline
& 1\% & 5\% & 10\% &  & 1\% & 5\% & 10\% &  & 1\% & 5\% & 10\% \\
\hline\hline
& \multicolumn{11}{c}{Strong Identification: $\beta _{n}=.3$} \\ \hline\hline
sup $\mathcal{T}_{n}$ & \multicolumn{1}{|l}{.029} & \multicolumn{1}{l}{.069}
& \multicolumn{1}{l|}{.153} &  & \multicolumn{1}{|l}{.441} &
\multicolumn{1}{l}{.590} & \multicolumn{1}{l|}{.676} &  &
\multicolumn{1}{|l}{.997} & \multicolumn{1}{l}{.999} & \multicolumn{1}{l}{
.999} \\
aver $\mathcal{T}_{n}$ & \multicolumn{1}{|l}{.022} & \multicolumn{1}{l}{.055}
& \multicolumn{1}{l|}{.120} &  & \multicolumn{1}{|l}{.382} &
\multicolumn{1}{l}{.546} & \multicolumn{1}{l|}{.624} &  &
\multicolumn{1}{|l}{.988} & \multicolumn{1}{l}{.996} & \multicolumn{1}{l}{
.997} \\ \hline
rand $\mathcal{T}_{n}$ & \multicolumn{1}{|l}{.008} & \multicolumn{1}{l}{.049}
& \multicolumn{1}{l|}{.098} &  & \multicolumn{1}{|l}{.328} &
\multicolumn{1}{l}{.488} & \multicolumn{1}{l|}{.598} &  &
\multicolumn{1}{|l}{.976} & \multicolumn{1}{l}{.999} & \multicolumn{1}{l}{
.996} \\
rand LF & .001 & .018 & .042 &  & .227 & .450 & .565 &  & .967 & .989 & .998
\\
rand ICS-1 & .009 & .046 & .096 &  & .230 & .449 & .565 &  & .974 & .990 &
.998 \\ \hline
sup $p_{n}$ & .005 & .039 & .078 &  & .295 & .457 & .536 &  & .961 & .990 &
.997 \\
sup $p_{n}$ LF & .002 & .010 & .033 &  & .223 & .427 & .528 &  & .949 & .985
& .997 \\
sup $p_{n}$ ICS-1 & .005 & .039 & .077 &  & .228 & .432 & .528 &  & .962 &
.990 & .997 \\ \hline
PVOT & \multicolumn{1}{|l}{.014} & \multicolumn{1}{l}{.055} &
\multicolumn{1}{l|}{.115} &  & \multicolumn{1}{|l}{.423} &
\multicolumn{1}{l}{.568} & \multicolumn{1}{l|}{.655} &  &
\multicolumn{1}{|l}{.996} & \multicolumn{1}{l}{.999} & \multicolumn{1}{l}{
.999} \\
PVOT LF & .002 & .023 & .051 &  & .311 & .509 & .618 &  & .995 & .998 & 1.00
\\
PVOT ICS-1 & .013 & .058 & .106 &  & .314 & .510 & .618 &  & .995 & .998 &
1.00 \\ \hline\hline
& \multicolumn{11}{c}{Weak Identification: $\beta _{n}=.3/\sqrt{n}$} \\
\hline\hline
sup $\mathcal{T}_{n}$ & \multicolumn{1}{|l}{.044} & \multicolumn{1}{l}{.134}
& \multicolumn{1}{l|}{.184} &  & \multicolumn{1}{|l}{.984} &
\multicolumn{1}{l}{.998} & \multicolumn{1}{l|}{1.00} &  &
\multicolumn{1}{|l}{1.00} & \multicolumn{1}{l}{1.00} & \multicolumn{1}{l}{
1.00} \\
aver $\mathcal{T}_{n}$ & \multicolumn{1}{|l}{.029} & \multicolumn{1}{l}{.125}
& \multicolumn{1}{l|}{.176} &  & \multicolumn{1}{|l}{.883} &
\multicolumn{1}{l}{.968} & \multicolumn{1}{l|}{/989} &  &
\multicolumn{1}{|l}{1.00} & \multicolumn{1}{l}{1.00} & \multicolumn{1}{l}{
1.00} \\ \hline
rand $\mathcal{T}_{n}$ & \multicolumn{1}{|l}{.032} & \multicolumn{1}{l}{.096}
& \multicolumn{1}{l|}{.162} &  & \multicolumn{1}{|l}{.817} &
\multicolumn{1}{l}{.929} & \multicolumn{1}{l|}{.970} &  &
\multicolumn{1}{|l}{.995} & \multicolumn{1}{l}{.998} & \multicolumn{1}{l}{
.998} \\
rand LF & .009 & .051 & .108 &  & .519 & .835 & .914 &  & .984 & .996 & .998
\\
rand ICS-1 & .009 & .051 & .120 &  & .785 & .921 & .954 &  & .990 & .998 &
1.00 \\ \hline
sup $p_{n}$ & .020 & .047 & .093 &  & .721 & .892 & .943 &  & .985 & .998 &
1.00 \\
sup $p_{n}$ LF & .015 & .025 & .054 &  & .451 & .772 & .883 &  & .961 & .992
& 1.00 \\
sup $p_{n}$ ICS-1 & .014 & .026 & .056 &  & .710 & .890 & .940 &  & .986 &
.998 & 1.00 \\ \hline
PVOT & \multicolumn{1}{|l}{.050} & \multicolumn{1}{l}{.118} &
\multicolumn{1}{l|}{.194} &  & \multicolumn{1}{|l}{.981} &
\multicolumn{1}{l}{.995} & \multicolumn{1}{l|}{1.00} &  &
\multicolumn{1}{|l}{1.00} & \multicolumn{1}{l}{1.00} & \multicolumn{1}{l}{
1.00} \\
PVOT LF & .012 & .053 & .109 &  & .823 & .965 & .975 &  & 1.00 & 1.00 & 1.00
\\
PVOT ICS-1 & .012 & .054 & .109 &  & .958 & .987 & .993 &  & 1.00 & 1.00 &
1.00 \\ \hline\hline
& \multicolumn{11}{c}{Non-Identification: $\beta _{n}=\beta _{0}=0$} \\
\hline\hline
sup $\mathcal{T}_{n}$ & \multicolumn{1}{|l}{.051} & \multicolumn{1}{l}{.151}
& \multicolumn{1}{l|}{.196} &  & \multicolumn{1}{|l}{.981} &
\multicolumn{1}{l}{.998} & \multicolumn{1}{l|}{.998} &  &
\multicolumn{1}{|l}{1.00} & \multicolumn{1}{l}{1.00} & \multicolumn{1}{l}{
1.00} \\
aver $\mathcal{T}_{n}$ & \multicolumn{1}{|l}{.043} & \multicolumn{1}{l}{.136}
& \multicolumn{1}{l|}{.189} &  & \multicolumn{1}{|l}{.886} &
\multicolumn{1}{l}{.968} & \multicolumn{1}{l|}{.984} &  &
\multicolumn{1}{|l}{1.00} & \multicolumn{1}{l}{1.00} & \multicolumn{1}{l}{
1.00} \\ \hline
rand $\mathcal{T}_{n}$ & \multicolumn{1}{|l}{.047} & \multicolumn{1}{l}{.111}
& \multicolumn{1}{l|}{.177} &  & \multicolumn{1}{|l}{.826} &
\multicolumn{1}{l}{.938} & \multicolumn{1}{l|}{.967} &  &
\multicolumn{1}{|l}{.997} & \multicolumn{1}{l}{1.00} & \multicolumn{1}{l}{
1.00} \\
rand LF & .006 & .058 & .110 &  & .549 & .859 & .926 &  & 1.00 & 1.00 & 1.00
\\
rand ICS-1 & .006 & .058 & .109 &  & .827 & .940 & .973 &  & 1.00 & 1.00 &
1.00 \\ \hline
sup $p_{n}$ & .032 & .081 & .126 &  & .718 & .904 & .934 &  & .995 & .999 &
.999 \\
sup $p_{n}$ LF & .013 & .051 & .085 &  & .414 & .778 & .875 &  & .965 & .999
& 1.00 \\
sup $p_{n}$ ICS-1 & .013 & .051 & .086 &  & .704 & .903 & .934 &  & .995 &
.999 & 1.00 \\ \hline
PVOT & \multicolumn{1}{|l}{.061} & \multicolumn{1}{l}{.148} &
\multicolumn{1}{l|}{.208} &  & \multicolumn{1}{|l}{.977} &
\multicolumn{1}{l}{.993} & \multicolumn{1}{l|}{.996} &  &
\multicolumn{1}{|l}{1.00} & \multicolumn{1}{l}{1.00} & \multicolumn{1}{l}{
1.00} \\
PVOT LF & .014 & .058 & .108 &  & .853 & .970 & .989 &  & 1.00 & 1.00 & 1.00
\\
PVOT ICS-1 & .013 & .057 & .107 &  & .978 & .996 & .998 &  & 1.00 & 1.00 &
1.00 \\ \hline\hline
\end{tabular}%
}
\end{center}
\par
{\small Numerical values are rejection frequency at the given level. LSTAR
is Logistic STAR. Empirical power is not size-adjusted. \textit{sup} $%
\mathcal{T}_{n}$ and \textit{ave} $\mathcal{T}_{n}$ tests are based on a
wild bootstrapped p-value. \textit{rand} $\mathcal{T}_{n}$: $\mathcal{T}%
_{n}(\lambda )$ with randomized $\lambda $\ on [1,5]. \textit{sup} ${p}_{n}$
is the supremum p-value test where p-values are computed from the
chi-squared distribution. PVOT uses the chi-squared distribution. LF implies
the least favorable p-value is used, and ICS-$1$ implies the type 1
identification category selection p-value is used with threshold $\kappa
_{n} $ $=$ $\ln (\ln (n))$.}
\end{table}

\clearpage
\begin{table}[t]
\caption{GARCH Effects Test Rejection Frequencies}
\label{tbl:garch}
\begin{center}
\resizebox{\textwidth}{!}{
\begin{tabular}{ccccccccccccc}
\hline\hline
& \multicolumn{1}{|c}{} & \multicolumn{3}{|c}{$n=100$} & \multicolumn{1}{|c}{
} & \multicolumn{3}{|c}{$n=250$} & \multicolumn{1}{|c}{} &
\multicolumn{3}{|c}{$n=500$} \\ \hline
Test & \multicolumn{1}{|c}{} & \multicolumn{1}{|c}{$1\%$} & $5\%$ & $10\%$ &
\multicolumn{1}{|c}{} & \multicolumn{1}{|c}{$1\%$} & $5\%$ & $10\%$ &
\multicolumn{1}{|c}{} & \multicolumn{1}{|c}{$1\%$} & $5\%$ & $10\%$ \\
\hline\hline
\multicolumn{13}{c}{No GARCH Effects (empirical size)$^{a}$} \\ \hline\hline
\multicolumn{1}{l}{sup-$p_{n}$ $^{b}$} & \multicolumn{1}{|c}{} &
\multicolumn{1}{|c}{.000$^{c}$} & .000 & .000 & \multicolumn{1}{|c}{} &
\multicolumn{1}{|c}{.000} & .000 & .000 & \multicolumn{1}{|c}{} &
\multicolumn{1}{|c}{.000} & .000 & .000 \\ \hline
\multicolumn{1}{l}{sup-$\mathcal{T}_{n}$ $^{d}$} & \multicolumn{1}{|c}{} &
\multicolumn{1}{|c}{.160} & .198 & .248 & \multicolumn{1}{|c}{} &
\multicolumn{1}{|c}{.148} & .188 & .224 & \multicolumn{1}{|c}{} &
\multicolumn{1}{|c}{.241} & .294 & .321 \\
\multicolumn{1}{l}{ave-$\mathcal{T}_{n}$} & \multicolumn{1}{|c}{} &
\multicolumn{1}{|c}{.004} & .032 & .052 & \multicolumn{1}{|c}{} &
\multicolumn{1}{|c}{.005} & .031 & .059 & \multicolumn{1}{|c}{} &
\multicolumn{1}{|c}{.008} & .053 & .107 \\
\multicolumn{1}{l}{rand-$\mathcal{T}_{n}$ $^{e}$} & \multicolumn{1}{|c}{} &
\multicolumn{1}{|c}{.004} & .004 & .012 & \multicolumn{1}{|c}{} &
\multicolumn{1}{|c}{.007} & .017 & .027 & \multicolumn{1}{|c}{} &
\multicolumn{1}{|c}{.003} & .028 & .038 \\ \hline
\multicolumn{1}{l}{PVOT$^{f}$} & \multicolumn{1}{|c}{} & \multicolumn{1}{|c}{
.015} & .059 & .096 & \multicolumn{1}{|c}{} & \multicolumn{1}{|c}{.019} &
.059 & .091 & \multicolumn{1}{|c}{} & \multicolumn{1}{|c}{.015} & .063 & .111
\\ \hline\hline
&  &  &  &  &  &  &  &  &  &  &  &  \\ \hline\hline
\multicolumn{13}{c}{GARCH Effects (empirical power)} \\ \hline\hline
\multicolumn{1}{l}{sup-$p_{n}$} & \multicolumn{1}{|c}{} &
\multicolumn{1}{|c}{.006} & .014 & .017 & \multicolumn{1}{|c}{} &
\multicolumn{1}{|c}{.000} & .010 & .017 & \multicolumn{1}{|c}{} &
\multicolumn{1}{|c}{.003} & .011 & .015 \\ \hline
\multicolumn{1}{l}{sup-$\mathcal{T}_{n}$} & \multicolumn{1}{|c}{} &
\multicolumn{1}{|c}{.848} & .934 & .934 & \multicolumn{1}{|c}{} &
\multicolumn{1}{|c}{.976} & .979 & .988 & \multicolumn{1}{|c}{} &
\multicolumn{1}{|c}{1.00} & 1.00 & 1.00 \\
\multicolumn{1}{l}{ave-$\mathcal{T}_{n}$} & \multicolumn{1}{|c}{} &
\multicolumn{1}{|c}{.733} & .891 & .904 & \multicolumn{1}{|c}{} &
\multicolumn{1}{|c}{.974} & .978 & .986 & \multicolumn{1}{|c}{} &
\multicolumn{1}{|c}{1.00} & 1.00 & 1.00 \\
\multicolumn{1}{l}{rand-$\mathcal{T}_{n}$} & \multicolumn{1}{|c}{} &
\multicolumn{1}{|c}{.446} & .555 & .633 & \multicolumn{1}{|c}{} &
\multicolumn{1}{|c}{.756} & .818 & .846 & \multicolumn{1}{|c}{} &
\multicolumn{1}{|c}{.873} & .923 & .935 \\ \hline
\multicolumn{1}{l}{PVOT} & \multicolumn{1}{|c}{} & \multicolumn{1}{|c}{.788}
& .914 & .914 & \multicolumn{1}{|c}{} & \multicolumn{1}{|c}{.975} & .988 &
.988 & \multicolumn{1}{|c}{} & \multicolumn{1}{|c}{1.00} & 1.00 & 1.00 \\
\hline\hline
&  &  &  &  &  &  &  &  &  &  &  &  \\ \hline\hline
\multicolumn{13}{c}{GARCH Effects (size adjusted power)} \\ \hline\hline
\multicolumn{1}{l}{sup-$p_{n}$} & \multicolumn{1}{|c}{} &
\multicolumn{1}{|c}{.006} & .014 & .017 & \multicolumn{1}{|c}{} &
\multicolumn{1}{|c}{.000} & .010 & .017 & \multicolumn{1}{|c}{} &
\multicolumn{1}{|c}{.003} & .011 & .015 \\ \hline
\multicolumn{1}{l}{sup-$\mathcal{T}_{n}$} & \multicolumn{1}{|c}{} &
\multicolumn{1}{|c}{.698} & .786 & .786 & \multicolumn{1}{|c}{} &
\multicolumn{1}{|c}{.838} & .841 & .864 & \multicolumn{1}{|c}{} &
\multicolumn{1}{|c}{.769} & .756 & .779 \\
\multicolumn{1}{l}{ave-$\mathcal{T}_{n}$} & \multicolumn{1}{|c}{} &
\multicolumn{1}{|c}{.739} & .909 & .952 & \multicolumn{1}{|c}{} &
\multicolumn{1}{|c}{.979} & .997 & 1.00 & \multicolumn{1}{|c}{} &
\multicolumn{1}{|c}{1.00} & .997 & .993 \\
\multicolumn{1}{l}{rand-$\mathcal{T}_{n}$} & \multicolumn{1}{|c}{} &
\multicolumn{1}{|c}{.452} & .601 & .721 & \multicolumn{1}{|c}{} &
\multicolumn{1}{|c}{.759} & .851 & .919 & \multicolumn{1}{|c}{} &
\multicolumn{1}{|c}{.880} & .945 & .997 \\ \hline
\multicolumn{1}{l}{PVOT} & \multicolumn{1}{|c}{} & \multicolumn{1}{|c}{.774}
& .902 & .902 & \multicolumn{1}{|c}{} & \multicolumn{1}{|c}{.966} & .979 &
.997 & \multicolumn{1}{|c}{} & \multicolumn{1}{|c}{.995} & .987 & .989 \\
\hline\hline
\end{tabular}}
\end{center}
\par
{\small a. The GARCH volatility process is $\sigma _{t}^{2}$ $=$ $\omega
_{0}+\delta _{0}y_{t-1}^{2}+\lambda _{0}\sigma _{t-1}^{2}$ with initial
condition $\sigma _{t}^{2}$ $=$ $\omega _{0}/(1-\lambda _{0}))$. The null
hypothesis is no GARCH effects $\delta _{0}=0$, and under the alternative $%
\delta _{0}=.3$. In all cases the true $\lambda _{0}=.6.$ b. sup-$p_{n}$ is
the $\sup_{\lambda \in \Lambda }p_{n}(\lambda )$ test. c. sup-$\mathcal{T}%
_{n}$ and ave-$\mathcal{T}_{n}$ tests are based on a wild bootstrapped
p-value. d. Rejection frequency at the given significance level. e. rand-$%
\mathcal{T}_{n}$ is an asymptotic $\chi ^{2}$ test based on $\mathcal{T}%
_{n}(\lambda )$ with randomized $\lambda $\ on [.01,.99]. f. PVOT: \textit{%
p-value occupation time} test. }
\end{table}

\clearpage\thispagestyle{empty}

\end{document}